\documentclass[twocolumn,superscriptaddress,floatfix,preprintnumbers, nofootinbib,hyperref]{revtex4} 
\pdfoutput=1
\usepackage[colorlinks=true,breaklinks=true]{hyperref}
\usepackage[normalem]{ulem}
\usepackage[utf8]{inputenc}
\hypersetup{allcolors=[rgb]{0.0 0.0 0.6},linkcolor=[rgb]{0.75 0.05 0.05}}
\usepackage{amsmath,amssymb}
\usepackage{epsfig}  
\usepackage{graphicx}   
\usepackage{slashed}             
\usepackage{url}
\usepackage{color}
\usepackage{comment}
\usepackage{subcaption}
\usepackage{caption}
\usepackage{multirow}
\usepackage[dvipsnames]{xcolor}
\usepackage{letltxmacro}
\LetLtxMacro{\oldcite}{\cite}
\renewcommand{\cite}[1]{\mbox{\oldcite{#1}}}
\usepackage{amssymb}
\usepackage{amsfonts}
\usepackage{amsmath}
\usepackage[toc,page]{appendix}
\usepackage{graphicx}
\usepackage{natbib}
\usepackage{url}
\usepackage{paralist}

\usepackage{textgreek}
\usepackage{braket}

\usepackage{xcolor}
\usepackage{mdframed}

\usepackage{braket}

\def\g{{\rm\thinspace g}}

\def\km{{\rm\thinspace km}}

\def\s{{\rm\thinspace s}}

\def\fourpanelwidth{0.45}

\def\kmps{\hbox{$\km\s^{-1}\,$}}

\def\h18{\hbox{H1821$+$643\,}}

\newcommand{\beq}{\begin{equation}}
\newcommand{\be}{\begin{equation}}
\newcommand{\ee}{\end{equation}}
\newcommand{\bea}{\begin{eqnarray}}
\newcommand{\eea}{\end{eqnarray}}
\newcommand{\eeq}{\end{equation}}
\newcommand{\bef}{\begin{figure}}
\newcommand{\eef}{\end{figure}}

\newcommand{\GeV}{{\rm GeV}}

\newcommand{\bB}{{\bf B}}

\def\z{\Theta}

\def\g{\ensuremath{g_{a\gamma}}}

\def\aap{\hbox{{\it Astron. \& Astrophys.}}}
\def\araa{\hbox{{\it Ann. Rev. Astron. Astrophys.}}}
\def\mnras{\hbox{{\it Mon. Not. Roy. Astron. Soc.}}}
\def\ssr{\hbox{{\it Space Sci. Rev.}}}
\definecolor{dm}{rgb}{0,0, 0}

\begin{document}

\title{The Fourier formalism for relativistic axion-photon conversion,\\ with astrophysical applications}

\author{M.C.~David Marsh}
\affiliation{The Oskar Klein Centre, Department of Physics, Stockholm University, Stockholm 106 91, Sweden}

\author{James H. Matthews}
\affiliation{Institute of Astronomy, University of Cambridge, Madingley Road, Cambridge, CB3 0HA, UK}

\author{Christopher Reynolds}
\affiliation{Institute of Astronomy, University of Cambridge, Madingley Road, Cambridge, CB3 0HA, UK}

\author{Pierluca Carenza}
\affiliation{Dipartimento Interateneo di Fisica “Michelangelo Merlin”, Via Amendola 173, 70126 Bari, Italy}
\affiliation{Istituto Nazionale di Fisica Nucleare - Sezione di Bari, Via Orabona 4, 70126 Bari, Italy}%

\date{\today}
\smallskip


\begin{abstract}
We study the weak mixing of photons and relativistic axion-like particles (axions) in plasmas with background magnetic fields, ${\bf B}$. We show that, to leading order in the axion-photon coupling, the conversion probability, $P_{\gamma \to a}$, is given by the one-dimensional power spectrum of the magnetic field components perpendicular to the particle trajectory. Equivalently, we express $P_{\gamma \to a}$ as the Fourier  transform of the magnetic field autocorrelation function, and establish a dictionary between properties of the real-space magnetic field and the energy-dependent conversion probability. For axions more massive than the plasma frequency, ($m_a>\omega_{\rm pl}$), we use this formalism to analytically solve the problem of perturbative axion-photon mixing in a general magnetic field. In the general case where {\color{dm} $\omega_{\rm pl}/m_a$} varies arbitrarily along the trajectory, we show that a naive application of the standard formalism for `resonant’ conversion can give highly inaccurate results, and that a careful calculation generically gives non-resonant contributions at least as large as the resonant contribution. Furthermore, we demonstrate how techniques based on the Fast Fourier Transform provide a new, highly efficient numerical method for calculating  axion-photon mixing. We briefly discuss magnetic field modelling in galaxy clusters in the light of our results and argue, in particular, that a recently proposed `regular' model used for studying axion-photon mixing (specifically applied to  the Perseus  cluster) is inconsistent with observations. Our formalism suggest new methods to search for imprints of axions, and will be important for spectrographs with  percent level sensitivity, which includes existing X-ray observations by \emph{Chandra} as well as the upcoming \emph{Athena} mission.
\end{abstract}

\maketitle

\tableofcontents
\section{Introduction}\label{sec:intro}

Determining the elementary particle content beyond the established Standard Model  is a central goal of contemporary high-energy physics. The QCD axion (cf.~\cite{Wilczek:1977pj}) and axion-like particles (axions) comprise a well-motivated class of hypothetical particles that frequently appear in extensions of the Standard Model, including effective theories derived from string theory~\cite{Svrcek:2006yi}. 
Both the QCD axion and axions can be understood as 
pseudo-Nambu-Goldstone bosons
of broken, approximate symmetries, and the QCD axion provides the leading candidate solution to the strong CP-problem. In this paper we will simply use `axions' to refer to the QCD axion and axion-like particles, as our results apply equally to these particles. 

Axions can naturally be  very light, with feeble couplings to matter and radiation, and provide an increasingly popular  candidate for explaining the nature of  dark matter \cite{Preskill:1982cy, Abbott:1982af, Dine:1982ah}. 
Characteristic of  axions is their  coupling to electromagnetism through the 
Lagrangian term
$$
{\cal L}_{\rm int} = \frac{1}{4} \g a F_{\mu \nu} \tilde F^{\mu \nu}
$$
where \g~denotes the axion-photon coupling and the axion, $a$, is assumed to have a mass $m_a$. This term induces a mixing between the photon and the axion in backgrounds with non-vanishing electromagnetic fields,  opening the possibility to interconvert axions and photons. Such interconversion underpins the majority of the experimental and observational efforts to search for axions (cf.~\cite{Raffelt:1996wa, PDG, Irastorza:2018dyq, Asztalos:2003px, Lawson:2019brd}). Particularly powerful are searches for axion-induced distortions in the spectra of luminous X-ray and gamma-ray sources located in galaxy clusters 
\cite{Hooper:2007bq, Chelouche:2008ta, Wouters:2013hua, Conlon:2013txa, Berg:2016ese,Marsh:2017yvc,Chen:2017mjf, Conlon:2017qcw, Reynolds:2019uqt,  HESS:2013udx, Cicoli:2014bfa, Meyer:2014epa,Meyer:2014gta, Fermi-LAT:2016nkz,  Zhang:2018wpc,Malyshev:2018rsh, Xia:2018xbt, Galanti:2018upl, Majumdar:2018sbv, Liang:2018mqm, Bu:2019qqg, Li:2020pcn,Guo:2020kiq, Cheng:2020bhr,Carenza:2021alz, Reynes:2021bpe}. State-of-the-art analyses using high-quality data from the \emph{Chandra X-ray observatory} have bounded axion-induced spectral distortions from the AGN at the centre of the Perseus cluster to the few percent level, leading to some of  the strongest limits on the axion-photon coupling to date ($\g \leq 7.9 \times 10^{-13}~{\rm GeV}^{-1}$ for $m_a < 10^{-12}~{\rm eV} $ \cite{Reynolds:2019uqt}).

The observational prospects are very good for even more sensitive searches for axions using  existing and planned X-ray and gamma-ray telescopes, such as \emph{Athena} \cite{nandra_athena_2013} and the \emph{Cerenkov Telescope Array} \cite{CTAConsortium:2017dvg}. However, such studies will be limited by modelling  uncertainties affecting the axion-photon mixing. The conversion probability is in general a complicated function of the mode energy, the axion parameters $m_a$ and $\g$, as well as the plasma density and magnetic field along the particle trajectory.  The robustness of the predictions to astrophysical modelling uncertainties
has been investigated by several groups over the past decade \cite{Angus:2013sua, Meyer:2014epa, Galanti:2018nvl, LT, Bu:2019qqg}, typically finding that limits on $\g$ can change by a factor of a few depending on the magnetic field model employed. A recent, unusual contribution to these studies is reference \cite{LT},  which investigated the robustness of certain gamma-ray constraints assuming --- as a limiting but ostensibly observationally consistent case --- that the cluster magnetic field in Perseus is highly regular, finding much weaker limits than those of  \cite{Fermi-LAT:2016nkz}.\footnote{We reexamine the astrophysical viability of this model in section \ref{sec:astrophys}.} Common to all these studies is the reliance on very simple magnetic field models to generate physical intuition (typically, a constant magnetic field), and numerical simulations to examine more involved models. This has left room for some confusion as to what properties of the magnetic field really drive the appearance of features in the conversion probability, and how robust the predictions really are.

In this paper, we revisit the theory of axion-photon conversion, and develop a powerful new method calculating and interpreting axion-photon mixing. Our results are equally valid for applications in the laboratory and space, but our main focus is on astrophysical applications involving conversion of high-energy photons into relativistic axions. {\color{dm} Our approach is inspired by --- and will be particularly useful for --- the emerging sub-field of precision X-ray searches for axions \cite{Conlon:2013txa, Berg:2016ese,Marsh:2017yvc,Chen:2017mjf, Conlon:2017qcw, Reynolds:2019uqt,Reynes:2021bpe}, which presently constrain the mixing probability to be no larger than a few percent. For such weak mixing, the system is in the perturbative regime, and the goal of this paper is to demonstrate the significant conceptual, calculational and methodological advances that can be gained using perturbation theory.  

}

It is well-known that the classical, linearised axion-photon system can be written as a Schrödinger-like equation, with time replaced by the spatial coordinate, say $z$, along the direction of propagation \cite{Raffelt:1987im}. In quantum mechanics, the asymptotic, perturbative transition amplitude can be expressed as a Fourier transform of the interaction Hamiltonian. In this paper, we show that a similar, but more subtle, statement also holds for classical axion-photon mixing. 

In the simplest case, which is directly relevant for gamma-ray searches, the mass of the axion is always larger than the plasma frequency. 
We find that, to leading order in $\g$, the axion-photon transition amplitude involving such massive axions 
is given by a sum of Fourier cosine and sine transforms of the (relevant component of the) magnetic field, $B$. The conversion probability is then given by the power spectrum of $B$. We  derive a version of the Wiener–Khintchine theorem that shows that the conversion probability is equal to the Fourier cosine transform of the magnetic autocorrelation function.   Strikingly, this means that questions about the spectrum of oscillations induced by axion-photon conversion map directly onto questions about the real-space properties of the magnetic autocorrelation. This constructively answers the questions of what properties of the magnetic field are reflected in the conversion probability. We demonstrate how to apply this formalism in a series of examples, ranging from the simple to the general. In particular, we analytically calculate the conversion probability and the magnetic autocorrelation function for a magnetic field expressed as a Fourier series. Since any physically relevant magnetic field can be expressed in such a way, this explicitly solves the problem of weak axion-photon mixing, for sufficiently massive axions.

Somewhat more subtle is the case of massless axions. In this case, the transition amplitude is no longer given by a Fourier transform obtained by integrating over the spatial coordinate $z$, as in the massive case. However, the amplitude can be written as a Fourier transform obtained  by integrating over {\color{dm} the} variable 
$$
\varphi(z) = \frac1 2 \int_0^z dz'\,\omega_{\rm pl}^2(z') \, .
$$
Again we find that the transition amplitude is given by cosine and sine transforms, but this time of the function $G=B/\omega_{\rm pl}^2$. The conversion probability  is given by the power spectrum of $G$, or equivalently by our version of the Wiener–Khintchine theorem, as the cosine transform of the autocorrelation function of $G$ (in {\color{dm} $\varphi$-space}).

In general, there may be some regions  where $m_a<\omega_{\rm pl}$, and others where $m_a> \omega_{\rm pl}$. In this case, it is not possible to express the transition amplitude as a single Fourier transform. However, we find that the transition amplitude reduces to a sum of `non-resonant' contributions that are given by Fourier transforms of $G$ (one for each region where $m_a- \omega_{\rm pl}$ has a definite sign), and a sum of `resonant' contributions (from points where $m_a=\omega_{\rm pl}$). 

An often used method to analytically evaluate resonant axion-photon mixing is the stationary phase approximation.  We show that, in the relativistic limit, the resonant amplitude calculated from the stationary phase approximation is enhanced by a factor of $\omega/m_a$, which is very large for light axions emerging from X-rays or gamma rays. However, we also show that a naive application of the stationary phase approximation to the resonant conversion gives a result that is highly inaccurate.  We address this issue by deriving a modified form of the stationary phase approximation that is relevant for relativistic axion-photon conversion: with this new formula, we see that in careful calculations of the amplitude, the non-resonant contributions are generically at least as large as the resonant contributions.

Fast Fourier Transforms (FFTs), and related methods, have revolutionised digital signal processing by providing highly effective ways to numerically evaluate the discrete Fourier transform. We show that with our formalism, axion-photon conversion can be evaluated  using methods based on FFTs. This drastically reduces the computational effort in determining the effects of axion-photon mixing, and allows for effective  marginalisation {\color{dm} or Monte Carlo} over astrophysical magnetic fields and plasma densities. See also \cite{Conlon:2018iwn} for a discussion of FFT techniques applied to axion searches. 

{\color{dm}
Since magnetic auto-correlations determine the predictions of perturbative axion-photon conversion, it is important to re-examine what is known about astrophysical magnetic fields in the environments most promising for axion searches.  To this end, we discuss the expected properties of magnetic fields in galaxy clusters. We review the arguments for turbulence in the intracluster medium (ICM), and the classes of models used in astrophysical axion searches. In particular, we critically examine the recently proposed `regular model' of \cite{LT}, finding it at odds with observations. }

We suggest that future studies of the magnetic field autocorrelation function in state-of-the-art magnetohydrodynamic simulations of the ICM  will be very useful in improving the sensitivity of axion searches. {\color{dm} Moreover, our formalism could lead to new methodologies for axion searches, e.g.~by generating the relevant mixing probabilities directly from an observationally inferred class of  autocorrelation functions, without ever explicitly solving the Schrödinger equation.}

This paper is organised as follows: in section \ref{sec:conv} we review the classical theory of axion-photon conversion, and how it can be understood through non-relativistic quantum mechanics. 
{\color{dm} We analytically develop the new formalism in sections \ref{sec:massive}--\ref{sec:general}, for the cases of a comparatively heavy axion (section \ref{sec:massive}), a very light axion (section \ref{sec:massless}), and the intermediate, general case (section \ref{sec:general}). We then test relevant aspects numerically in section \ref{sec:tests}, and discuss numerical implementations using discrete, Fourier-like transforms. In section \ref{sec:astrophys}, we discuss what is known about magnetic autocorrelations in the relevant astrophysical environments, and we indicate new possible directions for the methodology of axion searches. We conclude in  section~\ref{sec:conclusions}. 

The most important new results of this paper are (in no particular order) \emph{i)} the relation between the non-resonant conversion probability and the magnetic autocorrelation function given by equations \eqref{eq:Pautocorr}, \eqref{eq:Pmassless1} and \eqref{eq:Pgeneral}; \emph{ii)} accounting carefully for both resonant and non-resonant contributions when $\omega_{\rm pl} = m_a$ at one or more  points along the trajectory, cf.~section \ref{sec:res2}; \emph{iii)} the demonstration that one can use highly efficient FFT methods to calculate the conversion probabilities, cf.~section \ref{sec:numerical}; \emph{iv)} the identification of new, promising methods that are made possible by the new formalism, and which can open up new directions for axion searches, cf.~section \ref{sec:outlook}.
}


\section{Classical axion-photon conversion using `quantum' perturbation theory}\label{sec:conv}

In this section, we begin by reviewing the axion-photon mixing in magnetic fields following~ \cite{Raffelt:1987im, Battye:2019aco}. This leads to a Schrödinger-like equation from which the transition amplitudes can be calculated order-by-order in perturbation theory~\cite{Raffelt:1987im}{\color{dm}, in direct analogy with time-dependent perturbation theory in quantum mechanics}. The axion-photon interaction is described by the following Lagrangian
\begin{equation}
\mathcal{L}=-\frac{1}{4}F^{\mu\nu}F_{\mu\nu}+\frac{1}{2}\left(\partial_{\mu}\,a\partial^{\mu}a-m_{a}^{2}a^{2}\right)-\frac{g_{a\gamma}}{4}a\,F_{\mu\nu}\tilde{F}^{\mu\nu}\, , 
\end{equation}
where $a$ is the axion field, $A_{\mu}$ is the photon field, $F_{\mu\nu}$ is the electromagnetic tensor{\color{dm}, $\tilde F_{\mu \nu} = \tfrac{1}{2} \epsilon_{\mu\nu\lambda \rho} F^{\lambda \rho}$. We have }neglected the effects of Faraday rotations and QED birefringence at low energy.\footnote{The Faraday effect is inversely proportional to the energy, and completely negligible at X-ray energies or higher. Moreover, at X-ray energies, the QED birefringence contribution would in a typical galaxy cluster environments with $\mu{\rm G}$ magnetic fields and plasma frequencies of the order of $\omega_{\rm pl}\sim 10^{-12}$ eV lead to a fractional correction to $\Delta_\gamma$, that we define in \eqref{eq:Deltas}, of order $10^{-16}$. The QED contribution scales like $E^2 B^2$, and is less suppressed, and sometimes even non-negligible, in environments with strong magnetic fields probed at very high energies (cf.~\cite{Davies:2020uxn} for a recent example). }
This equation of motion are the Klein-Gordon equation for the axion field, and an extension of Maxwell's equations. The linearised equations around a static background magnetic field $\bB_{0}$ are given by
\begin{equation}
\begin{split}
(\Box + m_a^2)a &=- \g \dot {\bf A} \cdot {\bf B}_0 \, ,\\
(\Box + \omega_{\rm pl}^2) {\bf A} &= \g \dot a\, {\bf B}_0 \, , 
\end{split}
\label{eq:eom}
\end{equation}
where we included the plasma frequency $\omega_{\rm pl}$.
Most works on axion-photon mixing (from~\cite{Raffelt:1987im} to the more recent X-ray and gamma-ray studies, see e.g.~\cite{Horns:2012kw}) further simplify this system by assuming that all background quantities only vary along the $z$-direction, which is taken to be the axion or photon propagation direction.\footnote{For a recent discussions of extensions to anisotropic plasmas, see \cite{Battye:2019aco,Millar:2021}.} We consider a right-moving plane wave in the $z$-direction, and write the components of the background magnetic field respectively as  $B_{x}$ and $B_{y}$. Equations~\eqref{eq:eom} now become
\begin{equation}
\begin{split}
(\omega^{2}+\partial_{z}^{2} - m_a^2)a &=i \omega \g (A_{x}B_{x}+A_{y}B_{y}) \, ,\\
(\omega^{2}+\partial_{z}^{2}  - \omega_{\rm pl}^2)A_{j} &=-i \omega \g a B_{j}\quad j=x,y\, .
\end{split}
\end{equation}
In the relativistic limit, the equations of motion can be reduced to first order by appealing to the rotating wave approximation:  $\omega^{2}+\partial_{z}^{2}= (\omega+i \partial_{z})(\omega-i \partial_{z})\simeq2\omega(\omega-i \partial_{z})$.\footnote{An alternative method to arrive at the first-order equations is to use the WKB approximation, which is applicable also in the non-relativistic case~\cite{Battye:2019aco}.} 
After a shift of the axion field $a\rightarrow -i a$, the equations of motion are
\begin{equation}
\begin{split}
(\omega-i \partial_{z})a &=\frac{m_a^2}{2\omega}a-\frac{\g}{2} (A_{x}B_{x}+A_{y}B_{y}) \, ,\\
(\omega-i \partial_{z})A_{j} &=\frac{\omega_{\rm pl}^2}{2\omega}A_{j} -\frac{\g B_{j}}{2}a\quad j=x,y\, .
\end{split}
\label{eq:t1}
\end{equation}
These classical mixing equations are now of the form of the Schrödinger equation, with time replaced by the spatial coordinate $z$ \cite{Raffelt:1987im}:
\begin{equation}
i \frac{d}{dz} \Psi(z) = (H_{0}+H_{I})\Psi(z)\, .
\label{eq:EoM}
\end{equation}
Here the axion field and the components of the vector potential are components of the {\color{dm} `Schrödinger-picture state vector'}
\begin{equation}
\Psi(z) = \begin{pmatrix}
A_{x}\\
A_{y}\\
a
\end{pmatrix}\, ,
\end{equation}
{\color{dm} where we have suppressed the dependence on the mode energy $\omega$. The basis vectors are assumed to be $z$-independent, but the coefficients of the state vector evolve with $z$.
The} Hamiltonian is decomposed into free and interaction parts
\begin{equation}
    H_{0}= \omega \mathbb{I} + \begin{pmatrix}
\Delta_{\gamma}& 0&0\\
0& \Delta_{\gamma}&0\\
0&0&\Delta_{a}
\end{pmatrix} \quad H_{I}=\begin{pmatrix}
0&0&\Delta_{x}\\
0& 0&\Delta_{y}\\
\Delta_{x}&\Delta_{y}& 0
\end{pmatrix}\, , 
\end{equation}
and 
\begin{equation}
\begin{split}
\Delta_{a}&=-\frac{m_{a}^{2}}{2\omega}\, ,\\
\Delta_{\gamma}&=-\frac{\omega_{\rm pl}^{2}}{2\omega}\, ,\\
\Delta_{j}&=\frac{g_{a\gamma} B_{j}}{2} \quad\, .\\
\end{split}
\label{eq:Deltas}
\end{equation}
where $j=x,y$.\footnote{In the non-relativistic version of these identities, the factors of $\omega$ are replaced by the norm of the wavevector, which in general depends on $z$.}
{\color{dm}
The formal solution to the Schrödinger-like equation is 
$$\Psi(z) = U(z,0) \Psi(0)$$ where
\beq
U(z,0) = {\cal P}_z e^{i \int_0^z dz'\, (H_0 + H_I)} \, ,
\label{eq:formalsol}
\eeq
and where  ${\cal P}_z$ denotes the path-ordering operator. The $z$-evolution operator $U(z,0)$ is implicitly energy dependent.
Numerical solutions can be found by discretising the $z$-direction into a sufficiently large number of `cells' so that the Hamiltonian is approximately constant in each cell, and the total evolution operator $U$ is the product of the evolution operators for all cells, appropriately ordered. Such  a numerical approach is very common in astrophysical searches for axions, but becomes computationally costly when the mixing occurs over large distances with non-trivially varying magnetic fields, when the energy resolution needs to be finely sampled, and when a large number of possible magnetic field configurations must be considered. 

However, the structure of equation  \eqref{eq:EoM} is suggestive of different approach: perturbation theory in direct analogy  with time-dependent perturbation theory in quantum mechanics. This is most conveniently studied  in the interaction picture, where the equations of motion are given by} 
\begin{equation}
    i \frac{d}{dz} \Psi_{\rm int} = H_{\rm int}\Psi_{\rm int}\, , 
    \label{eq:intpic}
\end{equation}
where 
{\color{dm}
$H_{\rm int}=U_0^{\dagger}H_{I}U_0$ and $\Psi_{\rm int}=U_0^{\dagger}\Psi$ with the zeroth-order evolution operator $U_0={\rm exp}\left(-i\int_{0}^{z}dz'\,H_{0}\right)$.
 The general solution can be expressed as an expansion in the interaction Hamiltonian,
\begin{align}
&\Psi_{\rm int}(z) = \\
&\sum_{n=0}^\infty \left( -i\right)^n  \int_{0}^z dz_1' \ldots \int_{0}^{z_{n-1}} dz_n'\, H_{\rm int}(z_1') \ldots H_{\rm int}(z_n')\, \Psi_{\rm int}(0) , \nonumber 
\label{eq:1}
\end{align}
which equivalently corresponds to an expansion in the coupling constant, $\g$. To linear order in perturbation theory, the state in the interaction picture is given by 
\begin{equation}
    \Psi_{\rm int}(z)=\Psi_{\rm int}(0)-i\int_{0}^{z}dz'\, H_{\rm int}\Psi_{\rm int}(0)\, .
\end{equation}
In the original Schrödinger-picture basis, this corresponds to
\begin{equation}
\Psi(z) = \Big(U_0(z,0) +\int_0^z dz'\, U_0(z,z') H_I(z') U_0(z',0) \Big) \Psi(0) \, .
\end{equation}
The perturbative expansion can be expressed using Feynman diagrams, where each order in $H_I$ corresponds to an additional axion-photon mixing vertex, and where the zeroth-order evolution operator $U_0$ propagates the state between the vertices. 

We are interested in the transition amplitude of an initial photon state emerging as an axion. This corresponds to the initial condition  $\Psi(0) = (1,0,0)^{\rm T}$ if the photon is linearly $x$-polarised, and  $\Psi(0) = (0,1,0)^{\rm T}$ if its $y$-polarised.
 With  $j=x,y$ the leading order (LO) transition amplitude is then given by
\cite{Raffelt:1987im}
\begin{equation}
\label{eq:amplitude0}
\begin{split}
\mathcal{A}_{\gamma_{j}\rightarrow a}&= (0,0, 1) \cdot \Psi(z) =
-i\int_{0}^{z}dz'\,\Delta_{j}(z')\,e^{i\int_{0}^{z'} dz''\,(\Delta_{a}-\Delta_{\gamma})}\, . 
\end{split}
\end{equation}
 The conversion probability is obtained by squaring this transition amplitude: $P_{\gamma_{j}\rightarrow a}=|\mathcal{A}_{\gamma_{j}\rightarrow a}|^{2}$. This conversion probability calculated from the (classical) Schrödinger-like equation gives the ratio of the squared axion field to the squared electric field, which also corresponds to the observationally relevant flux ratio of axion and photons. 
 
 Next-to-leading-order (NLO) corrections appear at order ${\cal O}(\g^3)$ in the amplitude from the 3-vertex Feynman diagram: `photon $\to$ axion $\to$ photon $\to$ axion´. The LO conversion probability appear at ${\cal O}(\g^2)$, with higher-order corrections appearing at ${\cal O}(\g^4)$. Thus,  the perturbative expansion is generically a good approximation  when the conversion probability is small, with fractional corrections at the same order as the conversion probability (i.e.~a 1\% perturbative conversion probability generically receives corrections at the order of $0.01\%$, cf.~section \ref{sec:perturbativity} for a detailed discussion). \\

The above equations lead to unitary time evolution, and so neglects possible photon absorption. This is a good approximation when the medium is optically thin over the region at which axion-photon mixing occurs, which is the case for many astrophysical applications of practical relevance, cf.~e.g.~\cite{Marsh:2017yvc}.  However, including photon absorption is straightforward and leads to a modified  Schrödinger-like equation that features a  non-Hermitian   Hamiltonian:
$$
i \frac{d}{dz} \Psi(z) = (H_0 + H_I - i D) \Psi(z) \, ,
$$
where 
$$
D(z) =
\begin{pmatrix}
\beta(z) & 0 &0 \\
0 & \beta(z) &0 \\
0 & 0 & 0
\end{pmatrix}
$$
encodes the damping. The the real function $\beta$ depends on the absorption cross-section and astrophysical parameters, such as the free electron density. Since $[D(z), H_0(z')]=0$, it is straightforward to include damping as a zeroth-order modification to the perturbative expansion: 
equation \eqref{eq:intpic} in the interaction picture still holds, but with the unitary operator $U_0$ now replaced by the non-unitary transfer matrix 
$$
T_0(z_2, z_1) = e^{-i \int_{z_1}^{z_2}dz' \big(H_0 -iD \big) } \, ,
$$
so that $\Psi_{\rm int}(z) = T_0^{-1}(z,0) \Psi(0)$ and $H_{\rm int}  = T_0^{-1} H_I T_0$. This way, photon absorption modifies the zeroth-order propagator, and to linear order the Schrödinger-picture state is given by: 
\begin{equation}
\Psi(z) = \Big(T_0(z,0) +\int_0^z dz'\, T_0(z,z') H_I(z') T_0(z',0) \Big) \Psi(0) \, .
\end{equation}
In the following, we will neglect photon absorption.
\\}

In this paper, we will almost exclusively focus on polarised transition probabilities, involving a single component of the magnetic field. However, many bright astrophysical sources that are suitable for searches for axions are unpolarised. The conversion probability for an upolarised source of photons is $P_{\gamma \to a} = \frac{1}{2}\left( P_{\gamma_x \to a} + P_{\gamma_y \to a} \right)$, and the survival fraction of the unpolarised flux is 
\beq
P_{\gamma \to \gamma} = 1 - P_{\gamma \to a} = 1 - \frac{1}{2}\left( P_{\gamma_x \to a} + P_{\gamma_y \to a} \right) \, .
\label{eq:Punpol}
\eeq
In the following, we will primarily consider the conversion probability for a linearly polarised photon (with respect to a fixed, Cartesian coordinate system), with the understanding that the unpolarised probability can easily be obtained using formula \eqref{eq:Punpol}. For a discussion on polarised signals from axion-photon conversion, see \cite{Day:2018ckv}. \\

{\color{dm}
Finally, and for context, we note that solving the Schrödinger-like equation for axion photon mixing is equivalent to solving the von Neumann equation for the corresponding density matrix, $\rho$, as is done in much of the literature.  The equation of motion of $\rho$ is,
$$ 
i \frac{d \rho}{dz} = [H_0 + H_I, \rho] \, ,
$$
and the diagonal elements of $\rho$ are interpreted as the square of the wavefunctions of the two photon polarisation and the axion, respectively. The time-evolution of the density matrix is given by
$$
\rho(z) = U(z,0) \rho(0) U^\dagger(z,0) \, , 
$$
for the evolution operator of equation \eqref{eq:formalsol}. Numerical solutions in this formalism proceed  equivalently to the Schrödinger equation (i.e.~by finding the $z$-evolution operator), and there is no significant conceptual or calculational difference between the two formalisms. In particular, photon absorption is accounted for in an identical manner in the two formalisms, by replacing the unitary evolution operator by a non-unitary transfer matrix, and the perturbative expansion of $U$ can be applies equally to the two formalisms. For this reason, we will not discuss the density-matrix formulation of this problem any further in this paper.}


\section{Fourier transform formalism: the massive case}
\label{sec:massive}

\begin{figure}
\includegraphics[width=0.45\textwidth]{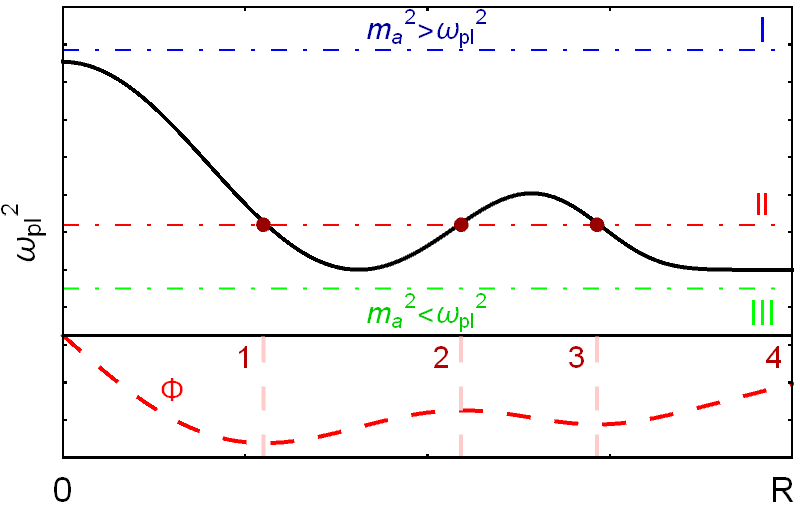} 
\caption{
 A hypothetical plasma frequency along the photon/axion trajectory. 
 In section \ref{sec:massive} we discuss case I (blue) where $m_a \gg \omega_{\rm pl}$, and in section \ref{sec:massless} we discuss case III (green) where $m_a\ll\omega_{\rm pl}$. Case II (red) is discussed in section \ref{sec:general} and corresponds to the general case, which can involve stationary points of the phase $\Phi$, plotted in the bottom panel for the example mass of this case. 
 }
\label{fig:SG}
\label{fig:coords}
\end{figure}

To leading order in the mixing, the amplitude for an initial photon state, linearly polarised along the $x$-direction, to transition into an axion is given by
\beq
{\cal A}_{\gamma_x \to a} = -i \int_0^z dz' \Delta_x(z') e^{-i\Phi(z')}
\label{eq:amp0} \, ,
\eeq
where the phase is given by
\beq
\Phi(z')  =\int_0^{z'} dz'' \big[\Delta_{\gamma}(z'') - \Delta_a \big] 
\, ,
\label{eq:Phi}
\eeq
with  $\Delta_\gamma$ and $\Delta_a$ as in equation \eqref{eq:Deltas}.
The phase $\Phi$ may both increase and decrease along the trajectory depending on the relative sizes of $\Delta_\gamma(z')$ and $\Delta_a$. 
The key property of $\Phi$ that we use in this paper is that  it can, in general, be factorised into a generalised spatial integration variable of equation \eqref{eq:amp0} (say $z'$ in the simplest case) multiplied  by an independent, `conjugate'  parameter (say $1/\omega$). In this section, we focus on the simplest case when $\Delta_\gamma$ can be neglected, and the links to Fourier analysis are most apparent.  We note that this factorisation is more subtle in the case of non-relativistic axions, which obey a similar Schrödinger-like equation but for which the factors of $1/\omega$ in \eqref{eq:Deltas} should be replaced by the inverse of a spatial wave-vector \cite{Battye:2019aco}, which depends on the spatial coordinate. In this paper, we focus on the simpler relativistic case.


\subsection{The cosine and sine transforms  for massive axions}
\label{sec:mass}
 We first consider relativistic axions with a mass larger than the plasma frequency 
 $$\omega_{\rm pl}^2 \ll m_a^2 \ll \omega^2 \, .$$ 
 This is the relevant case for gamma-ray searches for axions \cite{TheFermi-LAT:2016zue, LT}, and we will for brevity  refer to it as the case of `massive axions'.

In this case, we set $\Delta_{\gamma}=0$ and note that $\Delta_a = -m_a^2/(2\omega)$ is independent of $z$.   
The  transition amplitude becomes
\be
{\cal A}_{\gamma_x \to a} = -i \int_0^\infty dz' \Delta_x(z') e^{i z' \Delta_{a} } \, .
\label{eq:amp-c1}
\ee
Since we are interested in amplitudes evaluated far away from the transition region, we have extended the integral to infinity, assuming that  $\Delta_x$ is only non-vanishing in a finite region of space. 
This amplitude is now a half-sided Fourier transform of $\Delta_x$, with $\Delta_a$ being the conjugate `momentum'. This suggests that we may use Fourier analysis to analyse axion-photon mixing.

Functions defined on the real, positive line can be expressed independently using either sines or cosines. For sufficiently well-behaved functions that decay at infinity,\footnote{In all cases of physical interest that we will consider, the functions will be sufficiently well-behaved for the transforms to exist.} the relevant expansions are given by the Fourier sine and cosine transforms: 
\bea
{\cal F}_s(f)=
\hat f_s(\eta) &= \int_0^\infty dz\, \sin(\eta z) f(z)  
\\ 
{\cal F}_c(f)=
\hat f_c(\eta) &= \int_0^\infty  dz\,  \cos(\eta z) f(z) 
 \, .
\label{eq:f_c}
\eea
These transform have the convenient property of being, up to a constant, their own inverses:
\bea
f(z) = \frac{2}{\pi} \int_0^\infty d\eta\,  \hat f_c(\eta) \cos(\eta z) \label{eq:invcos} \\
f(z) = \frac{2}{\pi} \int_0^\infty d\eta\, \hat f_s(\eta) \sin(\eta z)  \label{eq:invsin} \, .
\eea
Moreover, the cosine and sine transforms are clearly real if $f(z)$ is real. 

The transition amplitude is now straightforwardly expressed through the cosine and sine  transforms as: 
\begin{equation}
\begin{split}
i{\cal A}_{\gamma_x \to a}(\eta) &= \int_0^\infty dz' \Delta_x(z') e^{-i z' \eta } 
= {\cal F}_c(\Delta_{x})-i{\cal F}_s( \Delta_{x})
\\
&=\frac{\g}{2} \Bigg( \hat{B}_{c}(\eta)-i\hat{B}_s(\eta) \bigg)\, ,
\label{eq:amp-c3}
\end{split}
\end{equation}
where we have dropped the spatial index on the magnetic field, and where the conjugate Fourier variable is given by 
\beq
\eta = - \Delta_a = m_a^2/(2\omega) \, ,
\eeq
 which is positive semi-definite. So, in the massive case, transition amplitudes are simple Fourier transforms of the magnetic field profile. Since these transforms are real, the conversion probability is conveniently given by
\beq
\begin{split}
P_{\gamma_x \to a}(\eta)& = \left| {\cal A}_{\gamma_x \to a} \right|^2={\cal F}_s( \Delta_{x})^2 + {\cal F}_c( \Delta_{x})^2 \\
&=\frac{\g^2}{4} \Bigg( 
\hat{B}^2_{c}(\eta)+\hat{B}^2_s(\eta) 
\Bigg)
\, ,
\end{split}
\label{eq:P1cs}
\eeq
without cross-terms between the cosine and sine transforms.
This expression means that the oscillatory pattern that axions induce on astrophysical spectra are directly given by the \emph{power spectrum} of the relevant magnetic fields. Moreover, for a fixed energy and mass, only a single wavelength of the magnetic field affects the conversion probability: the  conversion probability is localised in Fourier space.

The mathematical implications of equation \eqref{eq:P1cs}  can be further elucidated by relating the power spectra to the magnetic field autocorrelation function. To do so, we derive the Wiener–Khintchine theorem, as applied to the cosine transform.  
We begin by noting that the self-convolution of the standard, exponential Fourier transform implies the identities: 
\begin{equation}
\begin{split}
\Big({\cal F}_s& (\Delta_x) \Big)^2 = \\
&=\frac{1}{2}{\cal F}_c \Big(
\int_0^\infty dz\, \Delta_x(z)\left(\Delta_x(z+L) + \Delta^{\rm o}_x(z-L) \Big)
\right) 
\, ,\\
\Big( {\cal F}_c &(\Delta_x) \Big)^2 = \\
&\frac{1}{2}{\cal F}_c \Big(
\int_0^\infty dz\, \Delta_x(z)\left(\Delta_x(z+L) + \Delta^{\rm e}_x(z-L) \Big)
\right) \, ,
\end{split}
\end{equation}
where $\Delta^{\rm o}_x$ and $\Delta^{\rm e}_x$  respectively denote the odd and even extensions of $\Delta_x$ to negative values of the spatial coordinate. 
We now use that 
\beq
\begin{split}
&\int_0^\infty dz\, \Delta_x(z)\Big(\Delta^{\rm e}_x(z-L) + \Delta^{\rm o}_x(z-L) \Big) \\
&=2 \int_L^\infty dz\, \Delta_x(z)\Delta_x(z-L) \\
&=2 \int_0^\infty dz\, \Delta_x(z+L)\Delta_x(z) \, ,
\end{split}
\eeq
and identify 
\beq
c_{\Delta_x}(L) \equiv \int_0^\infty dz\, \Delta_x(z+L)\Delta_x(z) \, ,
\eeq
as the auto-correlation function of $\Delta_x$. We now have that
\beq
P_{\gamma_x \to a}(\eta) = 2 {\cal F}_c \Big(c_{\Delta_x}(L) \Big) = \frac{\g^2}{2} {\cal F}_c \Big( c_{B_x}(L)\Big)\, .
\label{eq:Pautocorr}
\eeq
This compact equation is one of the main results of this paper.  It says that the form of axion-induced modulations in high-energy spectra is  encoded in the magnetic auto-correlation function, through a simple transform. 
Since the cosine transform is its own inverse (up to a constant), the spectrum of the probability as defined by the cosine transform, i.e.~${\cal F}_c(P_{\gamma_x \to a})$, is directly given by the magnetic autocorrelation at a given spatial length scale. This means that questions about the spectrum of oscillations in the conversion probability map  onto sharp questions about the real-space properties of the magnetic field. 

Through equation \eqref{eq:Pautocorr}, we can translate  established properties of deterministic autocorrelation functions into \emph{general} statements about the perturbative conversion probability. For example, the autocorrelation function  peaks at zero: $c_B(0)>c_B(L)$ for $L>0$, and decays at large $L$ (at least when  the magnetic field is of physical origin, or when the Fourier transform is applicable). This means that the zero mode of $P_{\gamma \to a}(\eta)$ is always the largest Fourier (cosine) component, and the conversion probability decays as $\eta \to \infty$.

We close this section by commenting on two additional new relations that follow from this formalism. First, we note that equation \eqref{eq:Pautocorr} implies a new equation for extrema of the conversion probability. We have that $P'_{\gamma_x \to a}(\eta)=0$ whenever
\beq
{\cal F}_s \Big( L\, c_{B_x}(L) \Big) = 0 \, .
\eeq
This equation can be of phenomenological interest as the distribution of extreme probability points is a measure of how featured the probability curve is. 

Second, we note  a non-trivial equality for the integrated conversion probability.   Plancherel's formulas imply the  identities:
\beq
\int_0^\infty dz\, \Delta_x(z)^2  = \frac{2}{\pi} \int_0^\infty d\eta\,   {\cal F}_c( \Delta_{x})^2  = \frac{2}{\pi} \int_0^\infty d\eta\,  {\cal F}_s( \Delta_{x})^2 \, .
\eeq
It follows that the integrated probability over all wavelengths ($\eta$) equals the real-space integral over the squared magnetic field:
\be
\int_0^\infty d\eta\, P_{\gamma_x \to a}  = \frac{\g \pi}{2}  \int_0^\infty dz\, B_x^2  \, .
\ee

\subsection{Examples}
\label{sec:massive_examples}
We now apply the general formalism described in section \ref{sec:mass} to a series of examples. We find that all types of models that have previously been considered in the literature on axion-photon conversion (i.e.~`cell models', turbulent models defined in Fourier space, and regular models defined from Taylor expansions) can  be solved analytically.  Indeed, since any magnetic field with a finite extent can be expressed as a Fourier series for which we find the general solution in Example 4, this analysis constructively solves the problem of perturbative axion-photon mixing in the massive case. {\color{dm} We note that all figures presented in this section (and similar figures in sections  \ref{sec:massless} and \ref{sec:general}) correspond to analytic solutions evaluated in the perturbative regime, so agree excellently with the full solution.} 

\subsubsection{Example 1: A single cell of constant $B$}
\label{sec:single_cell}
The simplest example of axion-photon conversion corresponds to a constant magnetic field over a finite range. We take $B$ constant for $0\leq z \leq z_{\rm max}$, while being zero elsewhere. 
The real and imaginary parts of the transition amplitude are given by
\begin{equation}
\begin{split}
   {\rm Re}\left({\cal A}_{\gamma_x \to a} \right) &= {\cal F}_s\Big( \Delta_x\Big) =\frac{\g B}{2} \int_0^{z_{\rm max}} dz\, \sin(\eta z) \\
   &=\frac{\g B}{2 \eta}\Big(1-\cos(\eta z_{\rm max})\Big) \, ,\\
   {\rm Im}\left({\cal A}_{\gamma_x \to a} \right) &= -{\cal F}_c\Big( \Delta_x\Big) =-\frac{\g B}{2} \int_0^{z_{\rm max}} dz\,  \cos(\eta z) \\
   &=-\frac{\g B}{2 \eta}\sin(\eta z_{\rm max}) \, .
 \end{split}
\end{equation}
where $\eta=m^2_a/2\omega$. 
Squaring the amplitude, the conversion probability is
\begin{equation}
P_{\gamma \to a} =\frac{g_{a\gamma}^{2}B^{2}}{\eta^{2}}\sin^{2}\left(\frac{\eta z_{\rm max}}{2}\right)\, .
\label{eq:Psingledomain}
\end{equation}
This result (of course) agrees with the standard single-domain formula to leading order in $\g$
\begin{equation}
\begin{split}
P_{\gamma \to a}&=\frac{4\Delta_{x}^{2}}{\Delta_{\rm osc}^{2}}\sin^{2}\left(\frac{\Delta_{\rm osc}L}{2}\right)\simeq\frac{g_{a\gamma}^{2}B^{2}}{\eta^{2}}\sin^{2}\left(\frac{\eta z_{\rm max}}{2}\right)\, , 
\end{split}
\end{equation}
where $\Delta_{\rm osc}=\sqrt{\eta^{2}+4\Delta_{x}^{2}}$.

The alternate way to calculate the transition probability is to compute the magnetic autocorrelation function, and then take its cosine transform, cf.~equation \eqref{eq:Pautocorr}. The autocorrelation function of a rectangular box is a simple, linear function:
\beq
\begin{split}
 c_B(L) &= \int_0^\infty dz B_x(z+L) B_x(z) = \\
 &=B^2  \int_0^\infty dz W_{[0, z_{\rm max}]}(z+L) W_{[0,z_{\rm max}]}(z) = \\
 &=  B^2  (z_{\rm max}-L)\, \Theta(z_{\rm max}-L) \, .
\end{split}
\eeq\
Here, we have introduced the step functions
 \beq
 W_{[z_{\rm min},\, z_{\rm max }]}(z) =
 \begin{cases}
1 & {\rm if~} z_{\rm min} \leq z\leq z_{\rm max} \\
0 &~{\rm otherwise} \, ,
 \end{cases}
 \label{eq:step}
 \eeq 
 and 
 \beq
 \Theta(z)= W_{[0,\infty]}(z) =
 \begin{cases}
1 & {\rm if~} z\geq0 \\
 0
& {\rm otherwise.} \end{cases}
 \eeq

 The cosine transform of the magnetic autocorrelation  is then given by:
 \beq
 \begin{split}
 {\cal F}_c(c_B(L) )& = \int_0^\infty dL \cos(L \eta) c_B(L) \\
 &=B^2 \int_0^{z_{\rm max}} dL \cos(L \eta)   (z_{\rm max}-L)   \\
 &=2 B^2 \frac{\sin^2(z_{\rm max}\eta/2)}{\eta^2} \, .
 \end{split}
 \label{eq:singledomainPc}
 \eeq
Equation \eqref{eq:Pautocorr} then reproduces the correct conversion probability: 
\beq
P_{\gamma \to a} = \frac{g_{a\gamma}^{2}B^{2}}{\eta^{2}}\sin^{2}\left(\frac{\eta z_{\rm max}}{2}\right)\,  .
\eeq
It is interesting to note how the conversion probability changes with $z_{\rm max}$: the larger the $z_{\rm max}$, the more rapid the probability oscillations in $\eta$-space. From equation \eqref{eq:singledomainPc} we can infer that this is a general property following from the relation between the conversion probability and the magnetic autocorrelation function: long-ranged autocorrelations map to rapidly oscillating modes in  $P_{\gamma \to a}(\eta)$.

    \begin{figure}[t]
        \centering
        \includegraphics[width=\linewidth]{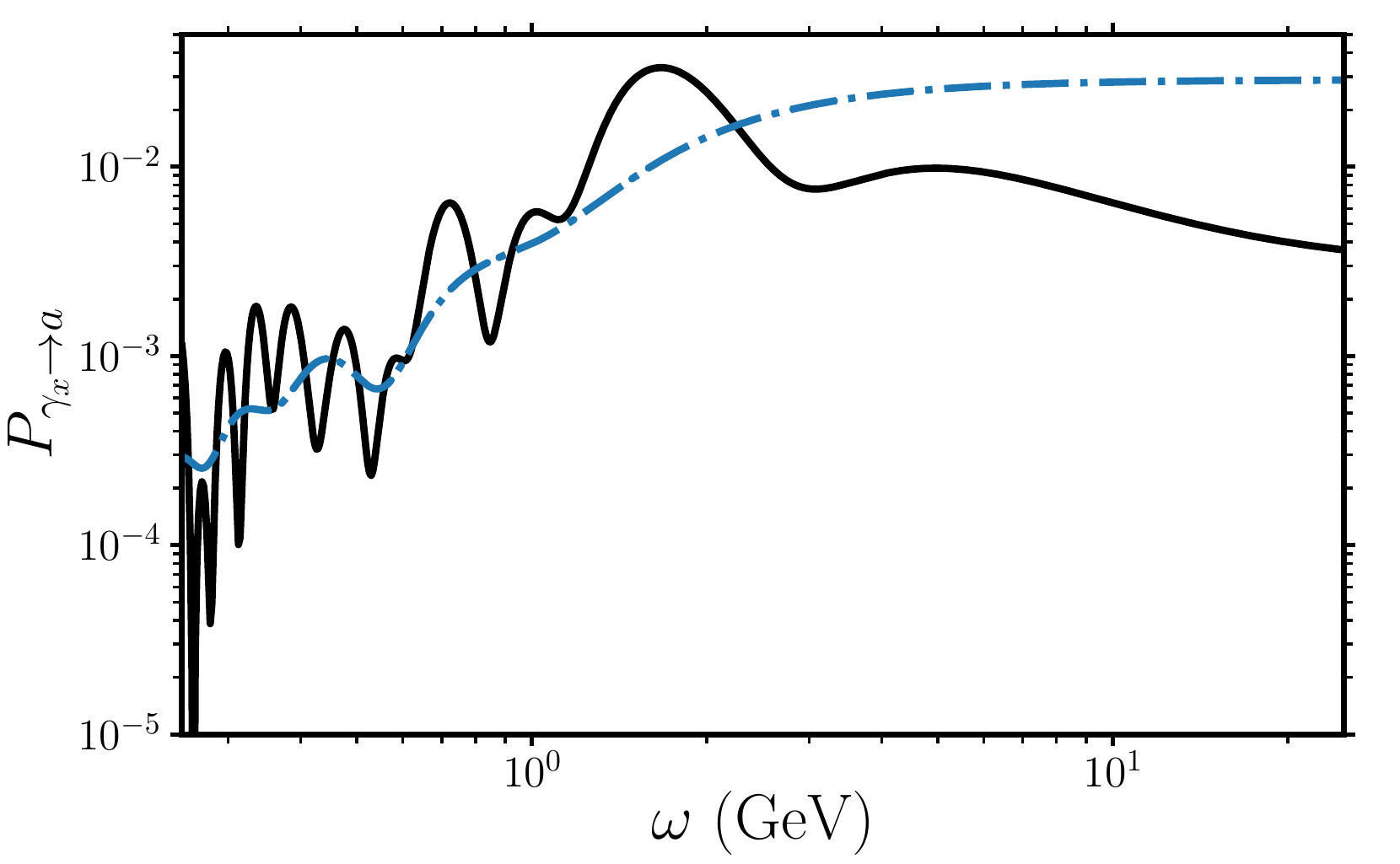} 
        \caption{Conversion probability for $g_{a\gamma}=10^{-11} \GeV^{-1}$, $m_{a} =5\times10^{-9}$~eV and  $\omega_{\rm pl}$ negligible, including (black solid line) or excluding (blue dot-dash line) the interference terms in equation \eqref{eq:cellmodel}. The magnetic field model is a cell model of the type described in Example 2 (section \ref{sec:general_cell}); the specific model is plotted with a solid line in the top left panel of Fig.~\ref{fig:fig_comparison}.
        }
        \label{fig:fig2_comp}
    \end{figure}

\begin{figure*}[t!] 
  \begin{subfigure}[t]{\fourpanelwidth\linewidth}

    \includegraphics[width=\linewidth]{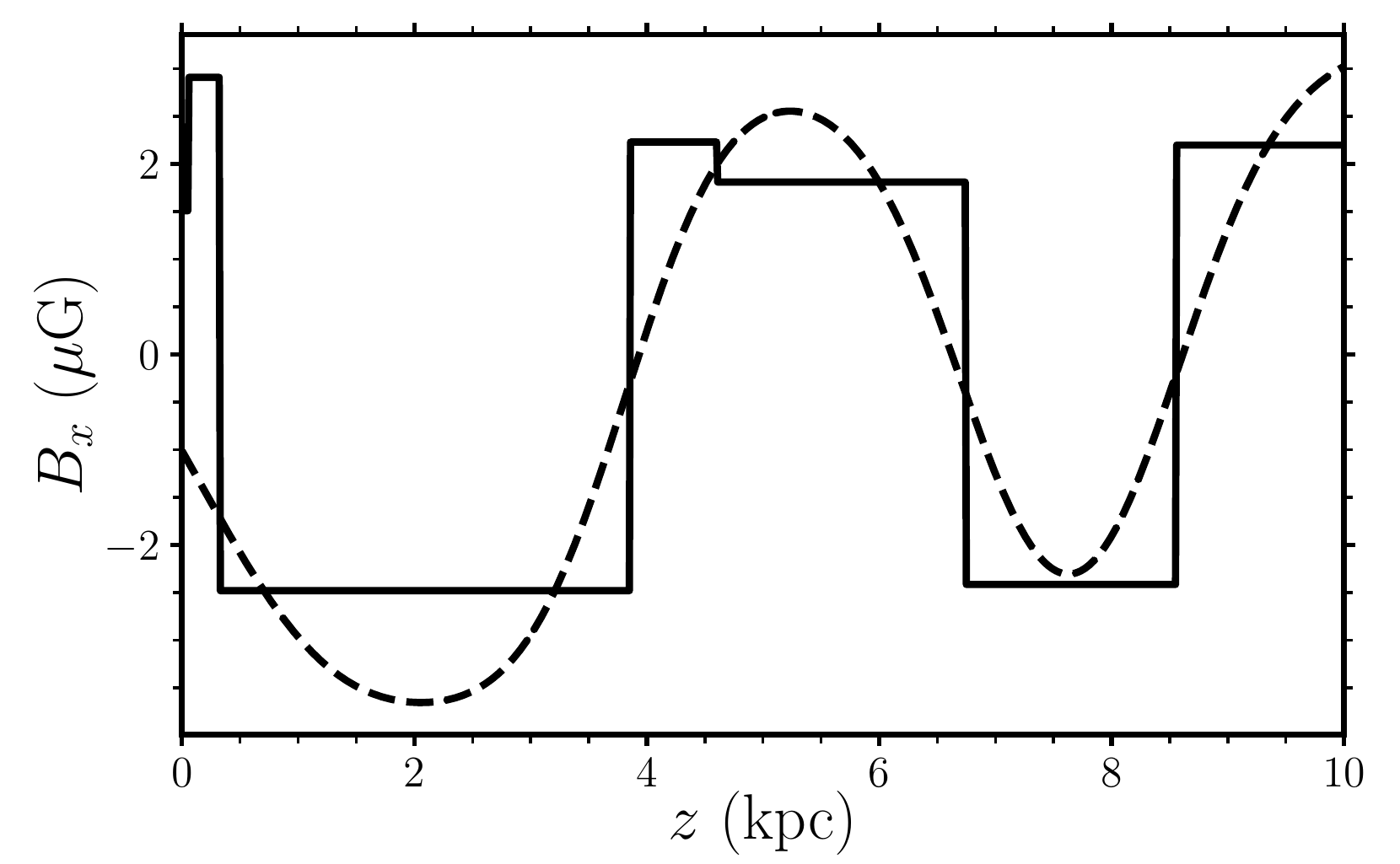} 
    \caption{Magnetic field profiles for a cell model (solid) and a smoothed version of the same field (dashed).} \label{fig:fig2_B}
    \vspace{4ex}
  \end{subfigure}
  \begin{subfigure}[t]{\fourpanelwidth\linewidth}
    
    \includegraphics[width=\linewidth]{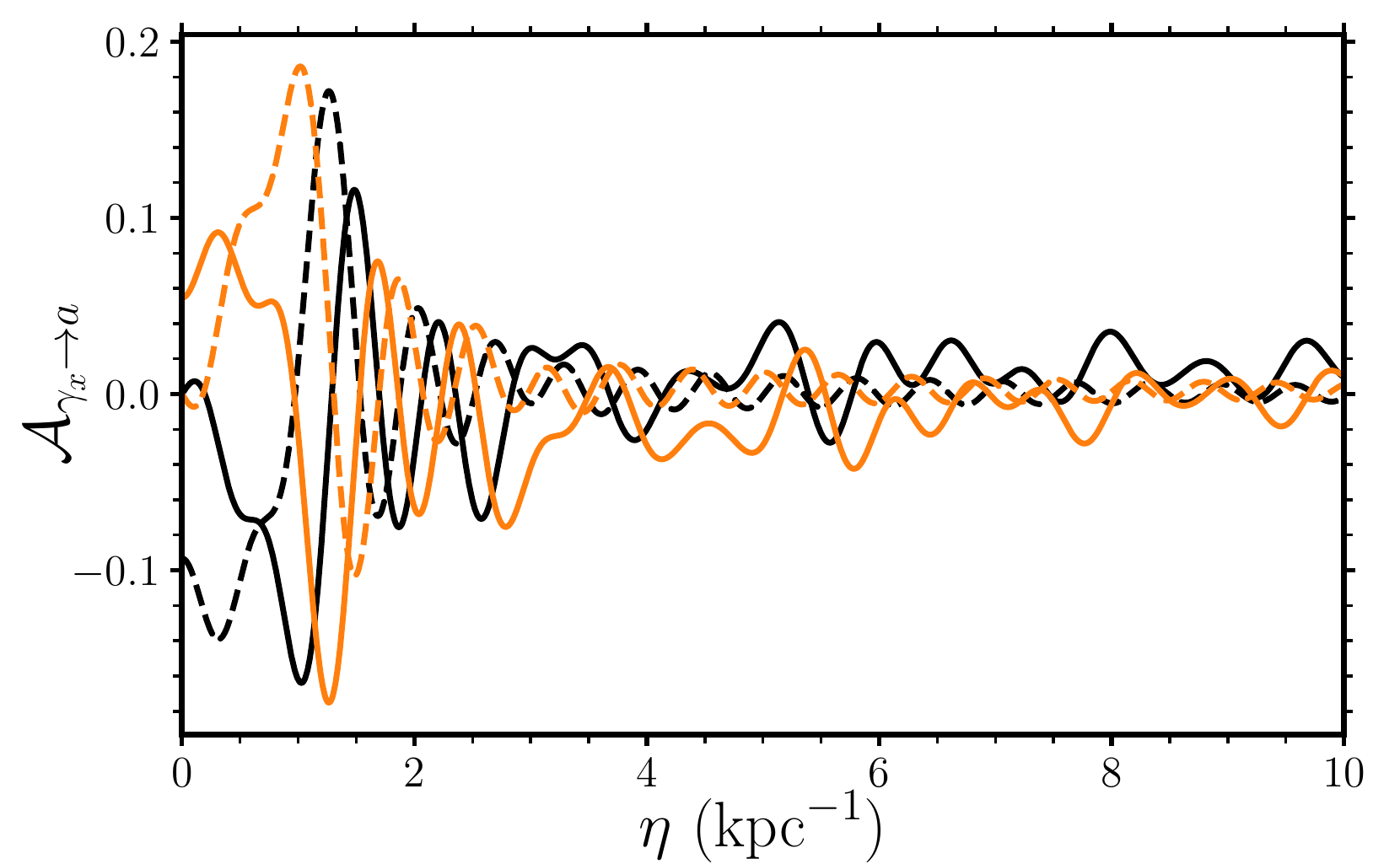} 
    \caption{The real, Re(${\cal A}_{\gamma_x \to a})={\cal F}_s(\Delta_x)$ (black), and imaginary, Im(${\cal A}_{\gamma_x \to a})=-{\cal F}_c(\Delta_x)$ (orange), parts of the amplitude ${\cal A}_{\gamma_x \to a}$ for the cell model (solid) and the smooth (dashed) magnetic fields} \label{fig:fig2_amp}
    \vspace{4ex}
  \end{subfigure} 
  \hfill
  \begin{subfigure}[t]{\fourpanelwidth\linewidth}
    \includegraphics[width=\linewidth]{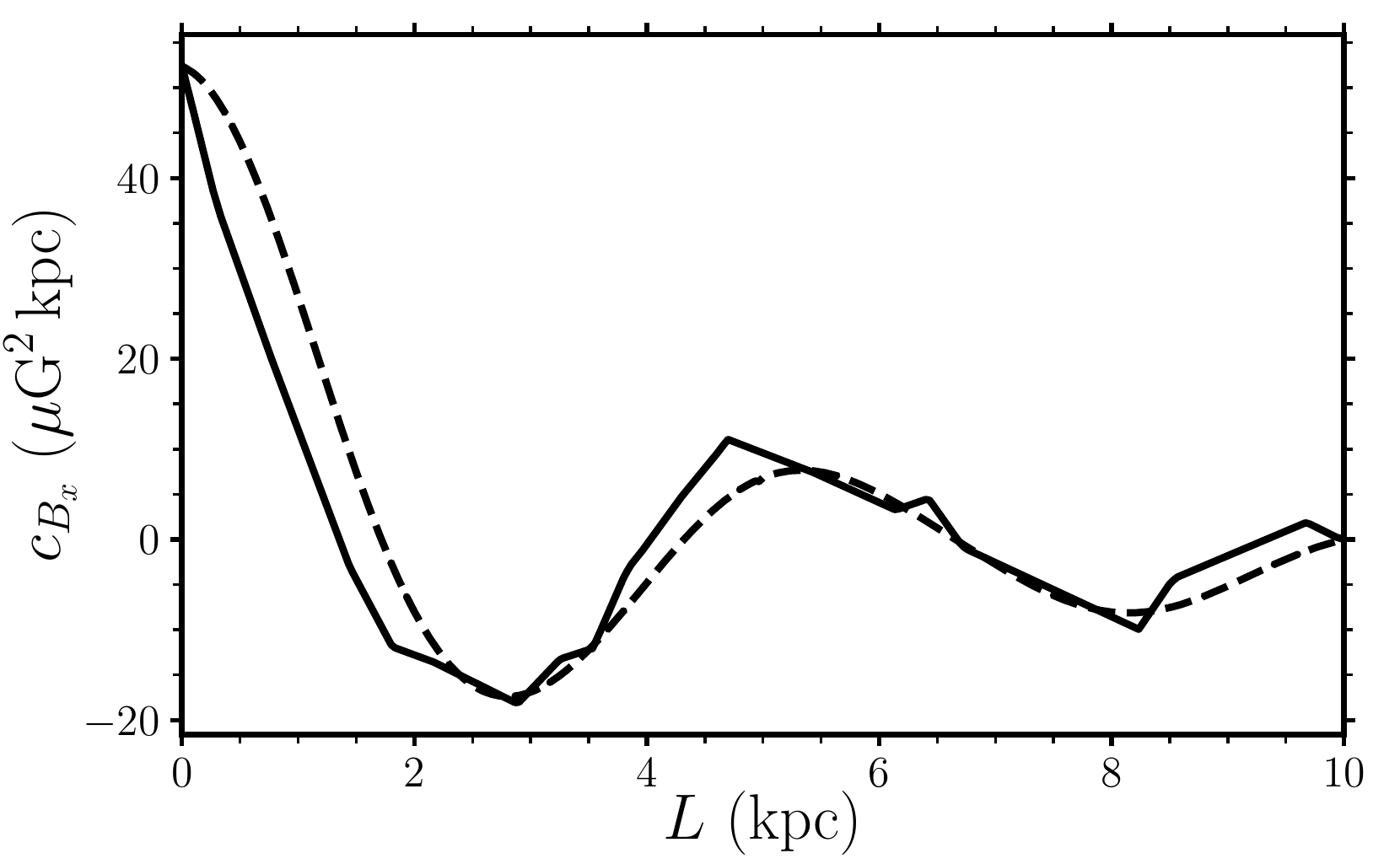} 
    \caption{Magnetic field autocorrelation function for the cell model (solid) and the smooth field (dashed).} \label{fig:fig2_corr}
    \vspace{4ex}
  \end{subfigure}
  \begin{subfigure}[t]{\fourpanelwidth\linewidth}

    \includegraphics[width=\linewidth]{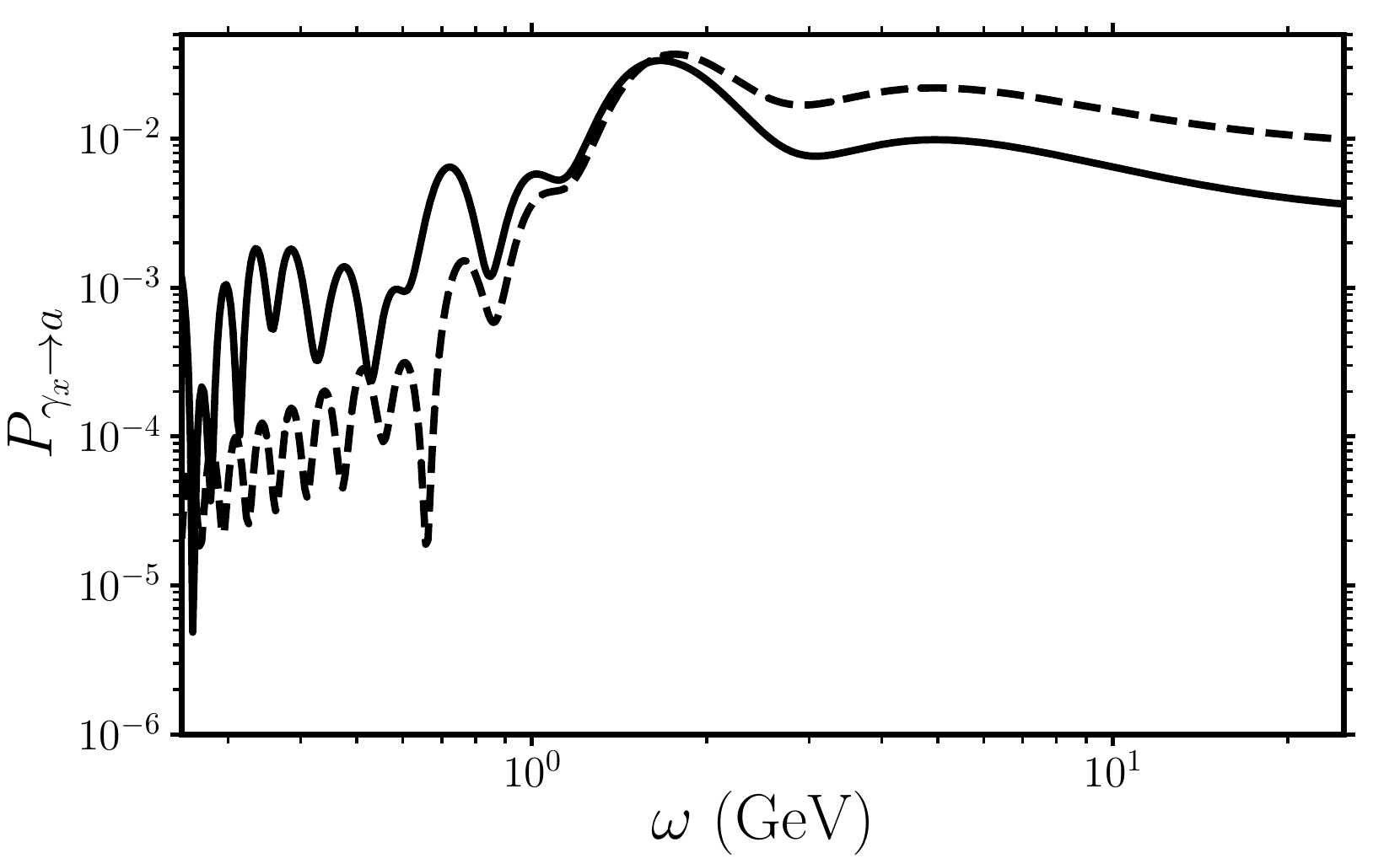} 
    \caption{Conversion probability as function of the energy for the cell model (solid) and the smoothed  field (dashed).} \label{fig:fig2_P}
    \vspace{4ex}
  \end{subfigure} 
    \caption{
   The axion-photon conversion probability for $g_{a\gamma}=10^{-11} \GeV^{-1}$, $m_{a} =5\times10^{-9}$~eV and  $\omega_{\rm pl}$ negligible (bottom right) is calculated for two magnetic field models: a cell model (Example 2, cf.~section \ref{sec:general_cell}) and a smoothed version of the same field (top left). The sine and cosine transforms of these fields (top right) show significant differences of the resulting amplitudes.
   The autocorrelation functions (bottom left) are qualitatively similar however, and the features of the conversion probability (bottom right) are in qualitative agreement at high energies, but differ at low energies where the cell model produces larger amplitude oscillations (still small relative to the maximum of $P_{\gamma_x \to a}$).
    }
    \label{fig:fig_comparison}
\end{figure*}

\subsubsection{Example 2: General cell models}
\label{sec:general_cell}
In an often considered class of `cell models', the magnetic field is  taken to be constant within a series of domains along the particle trajectory. In general cell models, the magnitudes and directions of the magnetic field and the size of each cell are treated as independent variables, that may e.g.~be generated through some specified probability distributions (cf.~e.g.~\cite{Conlon:2015uwa, Marsh:2017yvc, Reynolds:2019uqt,Galanti:2018nvl,Conlon:2018iwn}). The benefits of cell models are that they are quick to generate and  enable rather efficient numerical evaluation of multi-scale magnetic fields, which is important when marginalising over magnetic field realisations to understand the astrophysical variability of the axion predictions. Due to their close resemblance with the constant domain model, cell models have also been used to gain intuition about general properties of axion-photon conversion by simply summing the probabilities from each cell (as opposed to the amplitudes), neglecting the interference terms \cite{Conlon:2018iwn}.  

In this section, we determine the  conversion probability analytically in general cell models. This allows us to address the questions of the importance of the interference terms, and how the predictions from a a cell-model magnetic field differs from a smoothed version.

The ($x$-component of the) magnetic field in a general cell model can be defined as
\begin{equation}
B(z)=\sum_{i}B_{i} W_{[z_{i},z_{i+1}]}(z)\, ,
\end{equation}
with the constant parameters $B_i$ and where $z_i < z_{i+1}$. The real and imaginary parts of the transition amplitude are given by
\begin{equation}
    \begin{split}
   {\rm Re}&\left({\cal A}_{\gamma_x \to a} \right) = {\cal F}_s\Big( \Delta_x\Big)  \\
   %
&=\sum_i \frac{\g B_i}{\eta} \sin\left(\frac{\eta}{2} (z_{i+1}- z_i) \right)\sin\left(\frac{\eta}{2} (z_{i+1}+ z_i) \right)\, ,
\\
   {\rm Im}&\left({\cal A}_{\gamma_x \to a} \right) = 
     -{\cal F}_c\Big( \Delta_x\Big) 
     =\\
     &=-\sum_i \frac{\g B_i}{\eta} \sin\left(\frac{\eta}{2} (z_{i+1}- z_i) \right)\cos\left(\frac{\eta}{2} (z_{i+1}+ z_i) \right)
     \, .
\end{split}
\end{equation}
To find the conversion probability, one may of course  sum the squares of the real and imaginary parts of the amplitude. However, in this case, it is arguably simpler to directly compute the autocorrelation function and its cosine transform, as we now show.  

The magnetic field auto-correlation function is given by,
\begin{equation}
\begin{split}
c_{B}(L)&=\sum_{i, j}B_{i}B_{j}\int_{0}^{\infty}dz\,W_{[z_{i},z_{i+1}]}(z+L)W_{[z_{j},z_{j+1}]}(z)=\\
&=\sum_{i, j}B_{i}B_{j}\, \theta(z_{i+1}-L-z_{j})\theta(z_{j+1}-z_{i}+L)\\
&\quad\quad\Big[{\rm Min}(z_{j+1},z_{i+1}-L)-{\rm Max}(z_{j},z_{i}-L)\Big]\, .\\
\end{split}
\end{equation}
This equation means that in a general cell model, the magnetic autocorrelation function is a continuous and piece-wise linear function of $L$, with discontinuities in its first derivative. This generalises the well-known result that the autocorrelation function of a step-like signal is a triangle function.
\pagebreak

The conversion probability is given by the cosine transform of this function, according to equation \eqref{eq:Pautocorr}. Explicitly, we obtain
\begin{widetext}
\begin{equation}
\begin{split}
&\mathcal{F}_{c}(c_B)=\sum_{i}B_{i}^{2}\, \frac{1- \cos(\eta d_i)}{\eta^2}+
\sum_{i>j}B_{i}B_{j}
\Bigg[\theta(d_{i}-d_{j})
\Bigg(
F_{-}
\begin{pmatrix}
z_{i+1}-z_{j+1} \\
z_{i+1}-z_{j} \\ 
z_{i+1}-z_{j} 
\end{pmatrix} 
+ F_{+}
\begin{pmatrix}
z_{i}-z_{j+1} \\ 
z_{i}-z_{j} \\ z_{j+1}-z_{i}
\end{pmatrix}
+\\
&+
d_j
G
\begin{pmatrix}
z_{i}-z_{j} \\ 
z_{i+1}-z_{j+1}
\end{pmatrix}
\Bigg)
+\theta(d_{j}-d_{i})
\Bigg(F_{-}
\begin{pmatrix}
z_{i}-z_{j} \\ z_{i+1}-z_{j} \\ z_{i+1}-z_{j})
\end{pmatrix}
+
%
F_{+}
\begin{pmatrix}
z_{i}-z_{j+1} \\
z_{i+1}-z_{j+1} \\
z_{j+1}-z_{i})
\end{pmatrix}
+
%
d_i G
\begin{pmatrix}
z_{i+1}-z_{j+1} \\ z_{i}-z_{j}
\end{pmatrix}
\Bigg)
\Bigg]
\, ,\\
\end{split}
\label{eq:cellmodel}
\end{equation}
where $d_{i}=z_{i+1}-z_{i}$ is the cell length and
\begin{equation}
\begin{split}
F_{+}
\begin{pmatrix}
a \\ b \\z
\end{pmatrix}
&=\int_{a}^{b} dL\,\cos(\eta L)(z+ L)=\frac{\cos(b\eta)-\cos(a\eta)+(z+b)\eta\sin(b\eta)-(z+a)\eta\sin(a\eta)}{\eta^{2}}\, ,\\
F_{-}
\begin{pmatrix}
a \\ b \\z
\end{pmatrix}
&=\int_{a}^{b} dL\,\cos(\eta L)(z- L)=\frac{\cos(a\eta)-\cos(b\eta)-(z-a)\eta\sin(a\eta)-(z-b)\eta\sin(b\eta)}{\eta^{2}}\, ,\\
G
\begin{pmatrix}
a \\ b 
\end{pmatrix}
&=\int_{a}^{b} dL\,\cos(\eta L)=\frac{\sin(b\eta)-\sin(a\eta)}{\eta}\, .\\
\end{split}
\end{equation}
\end{widetext}
The conversion probability is obtained as $P_{\gamma_x \to a} = \tfrac{\g^2}{2} {\cal F}_c\big( c_B\big)$, and setting $\eta= m_a^2/(2\omega)$. 

We identify the first term in equation \eqref{eq:cellmodel} as an incoherent sum of oscillatory transition probabilities from each cell. For each term, the oscillation frequency in $\eta$-space is simply set by the cell size, $d_i$.

All other terms are due to interference effects, and are also given by simple oscillatory functions (possibly with a prefactor of $\eta$). For the interference terms, the frequency of oscillation in $\eta$-space is set by the separations between domain boundaries, which need not be adjacent. Thus, interference terms can contribute with a wide range of oscillation frequencies to $P_{\gamma_x\to a}(\eta)$, including rapid oscillations. Moreover, unless there is a hierarchy of magnetic field strengths, the interference terms are unsuppressed. 
Figure \ref{fig:fig2_comp} shows the effects of the interference terms in equation \eqref{eq:cellmodel} 
are particular apparent at low energies (large $\eta$), where they 
generate the fast oscillations  in the conversion probability. The magnitude of this discrepancy depends on the number of cells, and here we illustrate just one simple example. However, we believe that an inclusion of the interference effects may further improve interesting analyses that have neglected them, such as \cite{Conlon:2018iwn}.

In figure \ref{fig:fig_comparison} we compare the results of  equation \eqref{eq:cellmodel} for a cell model and the complete numerical solution for the same field smoothed out on the $\sim 1$~kpc scale. The corresponding magnetic autocorrelation functions are qualitatively similar, but the auto-correlation of the cell-model is only piece-wise linear, with clear discontinuities in its first derivative. These jagged features of $c_{B_x}(L)$ translate into additional support at large $\eta$ for $P_{\gamma_x \to a}(\eta)$. This is visible also in figure \ref{fig:fig2_P}: at high energies (small $\eta$) the shape of the conversion probabilities are  similar, and mostly differ due to the different magnetic field strengths of the cell model and the smoothed field. At lower energies (large $\eta$), both the conversion probabilities are oscillatory, but that of the  cell model is more `featured', and decays more slowly. We conclude that the differences between cell models and smoothed versions of the magnetic field are mostly confined to comparatively low energies. Observations that are only sensitive to axions in the high-energy region where the conversion probability is the largest are likely to be rather insensitive to the differences between cell models and smoothed versions.


\subsubsection{Example 3: Single mode oscillatory magnetic fields}
A well-known result of \cite{Raffelt:1987im} is that an oscillating magnetic field can lead to an enhanced axion-photon conversion probability, analogously to magnetic  resonance  in quantum mechanics, where an oscillating magnetic field provides the additional energy for rapid transition between Zeeman split energy levels. For axion-photon conversion, the basic observation is that an oscillating magnetic field of the form
\beq
B = B_0 \Theta(R-z) \cos (k z) \, ,
\label{eq:Bsinglemode}
\eeq
leads to an amplitude involving the terms $e^{i( \eta + k)R} $ and $e^{i(\eta-k)R}$. For $\eta \approx k$ and $\eta R \gg 1$, these correspond, respectively, to a rapidly and a slowly oscillating term. Neglecting the former by appealing to the rotating phase approximation, one finds a conversion probability of the form \cite{Raffelt:1987im}
\beq
P_{\gamma \to a} \approx \frac{(\g B_0 R)^2}{4}\,  \frac{\sin^2\left(\frac{(\eta-k)R}{2}\right)}{(\eta - k)^2 R^2} \, ,
\label{eq:Pflopping}
\eeq
which peaks at $\eta=k$. We will now discuss the relation between our more general result of equation \eqref{eq:Pautocorr} and equation \eqref{eq:Pflopping}. 

Assuming a magnetic field of the form of equation \eqref{eq:Bsinglemode}, the 
real and imaginary parts of the transition amplitude are given by
\begin{equation}
\begin{split}
   {\rm Re}\left({\cal A}_{\gamma_x \to a} \right) &= {\cal F}_s\Big( \Delta_x\Big) =\\
   &=\frac{g_{a\gamma}B_{0}R}{4}\left[\frac{\sin^2\xi_{+}}{\xi_{+}}+\frac{\sin^2\xi_{-}}{\xi_{-}}\right]\,  ,\\
   {\rm Im}\left({\cal A}_{\gamma_x \to a} \right) &= -
     {\cal F}_c\Big( \Delta_x\Big) 
 =\\
 &=-\frac{g_{a\gamma}B_{0}R}{4}\left[\frac{\sin\xi_{+} \cos\xi_{+}}{\xi_{+}}+\frac{\sin\xi_{-}\cos\xi_{-}}{\xi_{-}}\right]\, , 
    \label{eq:amplit}
\end{split}
\end{equation}
where $\xi_{\pm}=(\eta\pm k)R/2$ and $\eta=m_{a}^{2}/2\omega$. The conversion probability is calculated by squaring this amplitude 
\begin{widetext}
\begin{equation}
\begin{split}
    P_{\gamma\rightarrow a}=\frac{g_{a\gamma}^{2}B_{0}^{2}R^{2}}{16}&\bigg[\frac{\sin^{2}\xi_{+}}{\xi_{+}^{2}}+\frac{\sin^{2}\xi_{-}}{\xi_{-}^{2}}+2\cos(\xi_{+}-\xi_{-})\frac{\sin\xi_{+}\sin\xi_{-}}{\xi_{+}\xi_{-}}\bigg]\, .
    \label{eq:Psinglemode}
    \end{split}
\end{equation}
\end{widetext}
Note that as $R$ increases (keeping $k$ fixed), the interference term becomes negligible and the probability approaches $P_{a\gamma}=g_{a\gamma}^{2}B_{0}^{2}R \,\delta(k-\eta)/16$.

Alternatively, one can find the conversion probability by computing the autocorrelation function,
\begin{equation}
    c_{B}(L)=\frac{B_{0}^{2}}{2}\left[(R-L)\cos(kL)+
    \frac{\sin(k(R-L))\cos(kR)}{k}
    \right]\,  .
    \label{eq:cBsingle}
\end{equation}
and taking its cosine transform. The first term of equation \eqref{eq:cBsingle} then gives the first two terms of equation \eqref{eq:Psinglemode}, including the  `resonant' term for $\xi_- \to 0$. This provides a new perspective on the resonance:  its characteristic $\sin^2 \xi/\xi^2$ behaviour  emerges when the autocorrelation function includes  a cosine mode with linearly changing amplitude $\sim L \cos(kL) $

\subsubsection{Example 4: General, `turbulent' magnetic fields}
\label{sec:turbulent}
In astrophysical environments, the components of the magnetic field inevitably involve more than one Fourier mode which leads to interference effects that are absent in the single-mode case. Indeed, `turbulent' magnetic fields are often modelled as Gaussian random fields, by drawing the amplitudes of a set of modes from some specified power spectrum.\footnote{To ensure that the resulting magnetic field is divergence free, it's convenient to generate the Fourier components of the gauge potential, rather than the magnetic field as we do in two numerical examples in section \ref{sec:numerical}. To make the link between the properties of the magnetic field and the resulting conversion probability apparent however, it's more convenient to work with the Fourier components of the magnetic field directly, as we do in this section.} 

We consider the most general magnetic field defined for $0<z<R$ as expressed through its Fourier series
\beq
B = \Theta(R-z) \sum_n (B^{c}_n \cos(k_n z) + B^{s}_n \sin(k_n z) )\, .
\label{eq:Bturb}
\eeq
The wavenumbers are $k_n = 2\pi n /R$ with $n=0, 1, \ldots$. 

\begin{figure*}[t!] 
  \begin{subfigure}[t]{\fourpanelwidth\linewidth}

    \includegraphics[width=\linewidth]{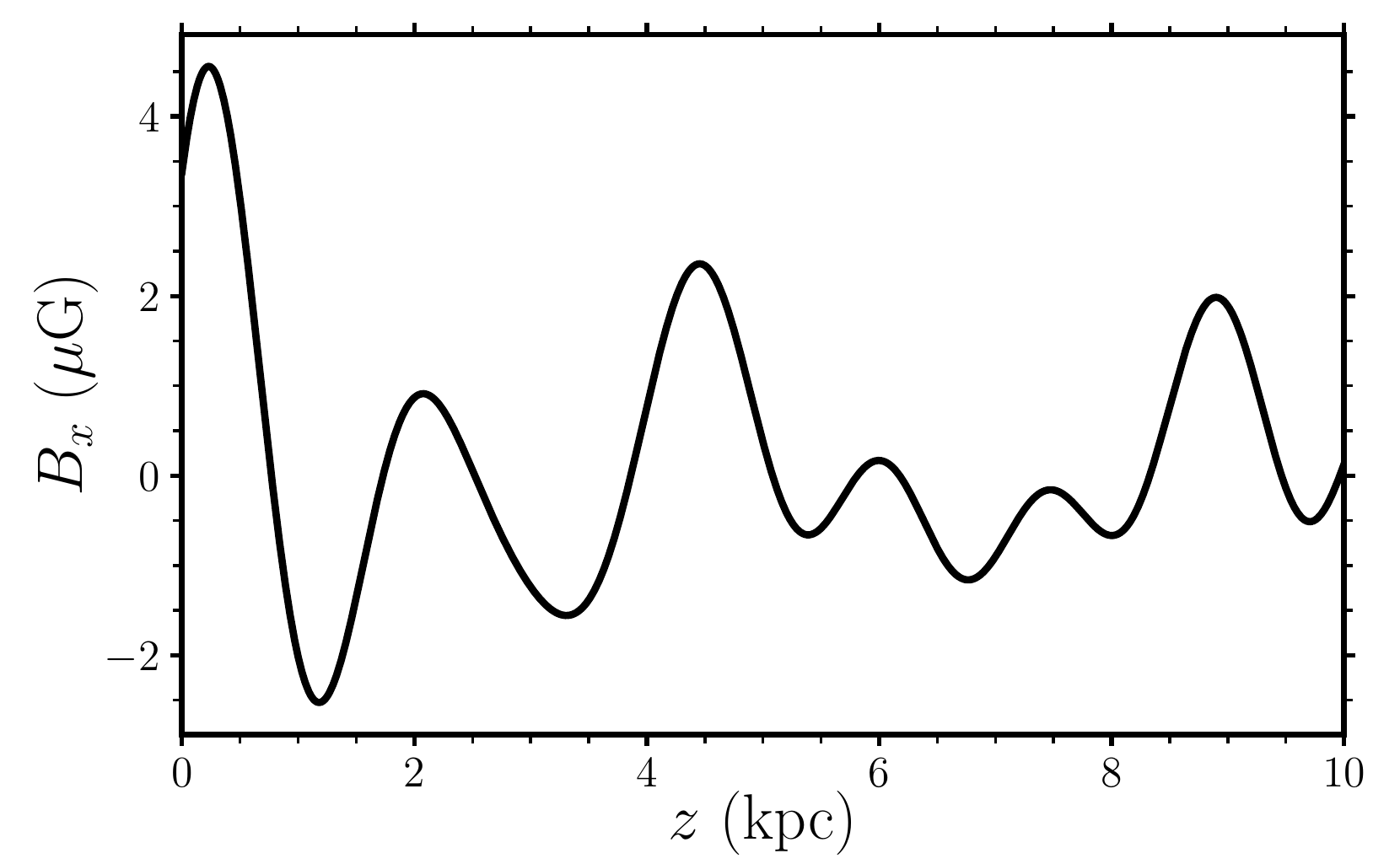} 
    \caption{Magnetic field profile.} \label{fig:fig4_B}
    \vspace{4ex}
  \end{subfigure}
  \begin{subfigure}[t]{\fourpanelwidth\linewidth}
    
    \includegraphics[width=\linewidth]{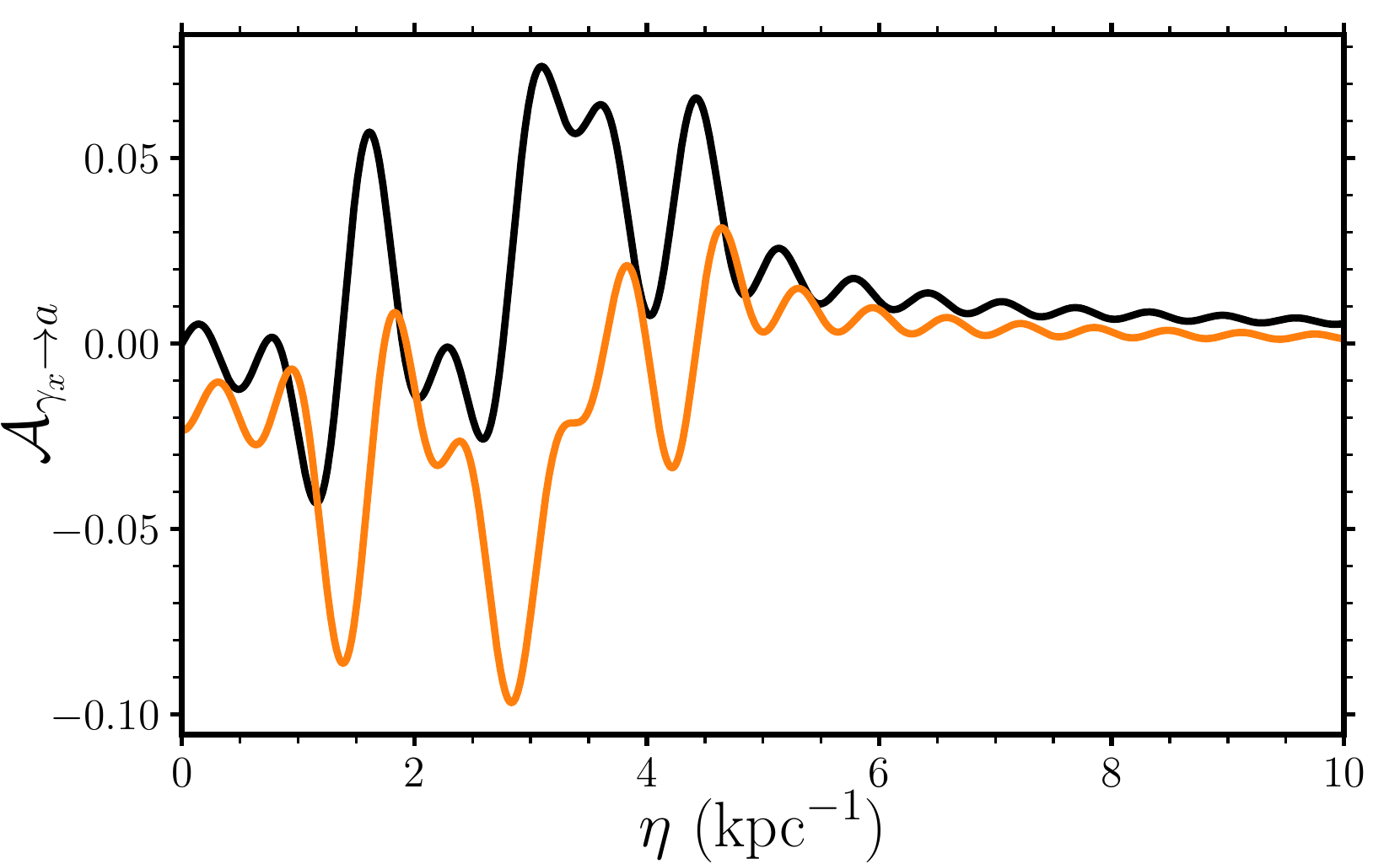} 
   \caption{The real, Re(${\cal A}_{\gamma_x \to a})={\cal F}_s(\Delta_x)$ (black), and imaginary, Im(${\cal A}_{\gamma_x \to a})=-{\cal F}_c(\Delta_x)$ (orange), parts of the amplitude ${\cal A}_{\gamma_x \to a}$.} \label{fig:fig4_transf}
    \vspace{4ex}
  \end{subfigure} 
  \begin{subfigure}[t]{\fourpanelwidth\linewidth}
    \includegraphics[width=\linewidth]{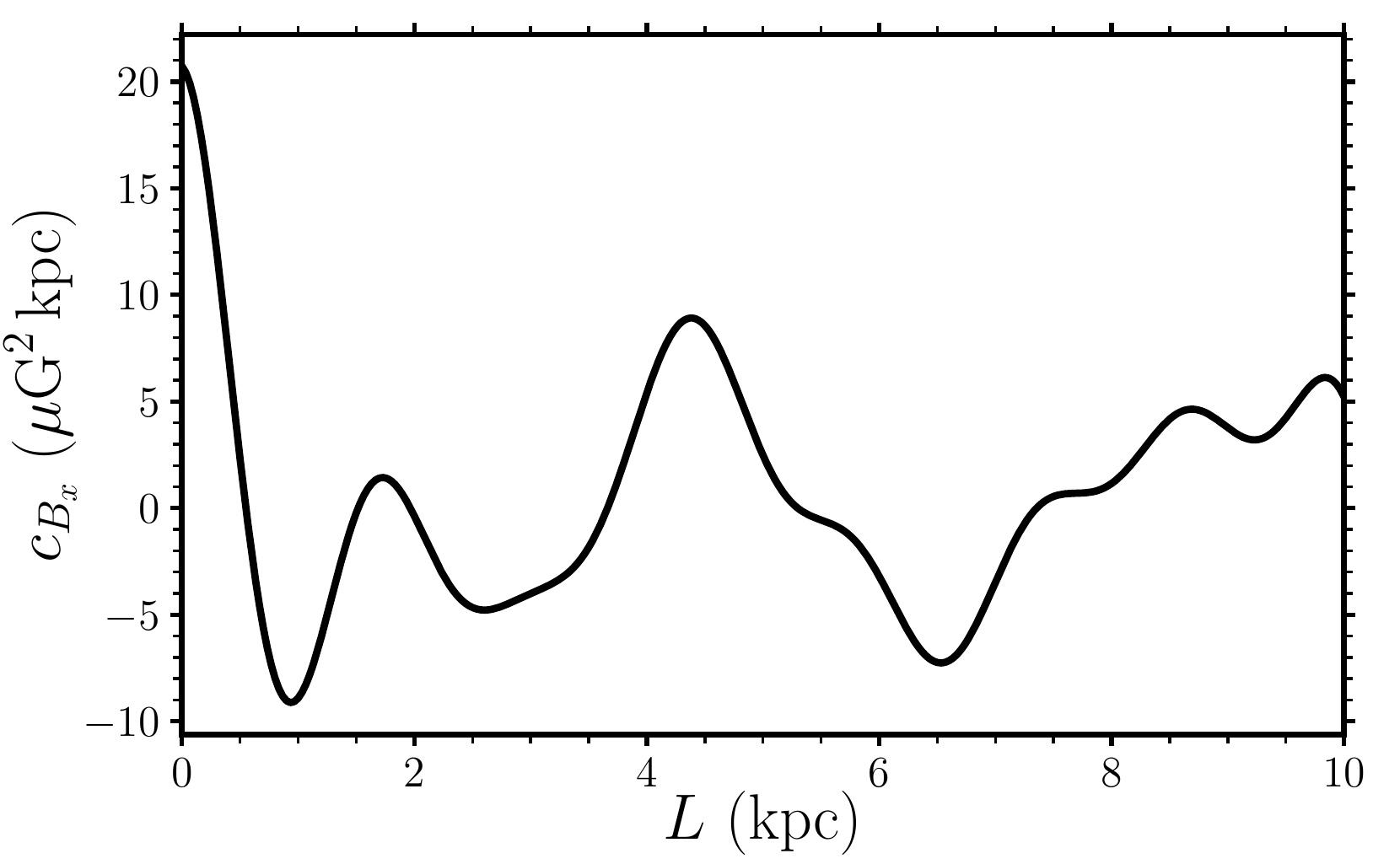} 
      \caption{Magnetic field autocorrelation function.} \label{fig:fig4_corr}
    \vspace{4ex}
  \end{subfigure}
  \begin{subfigure}[t]{\fourpanelwidth\linewidth}

    \includegraphics[width=\linewidth]{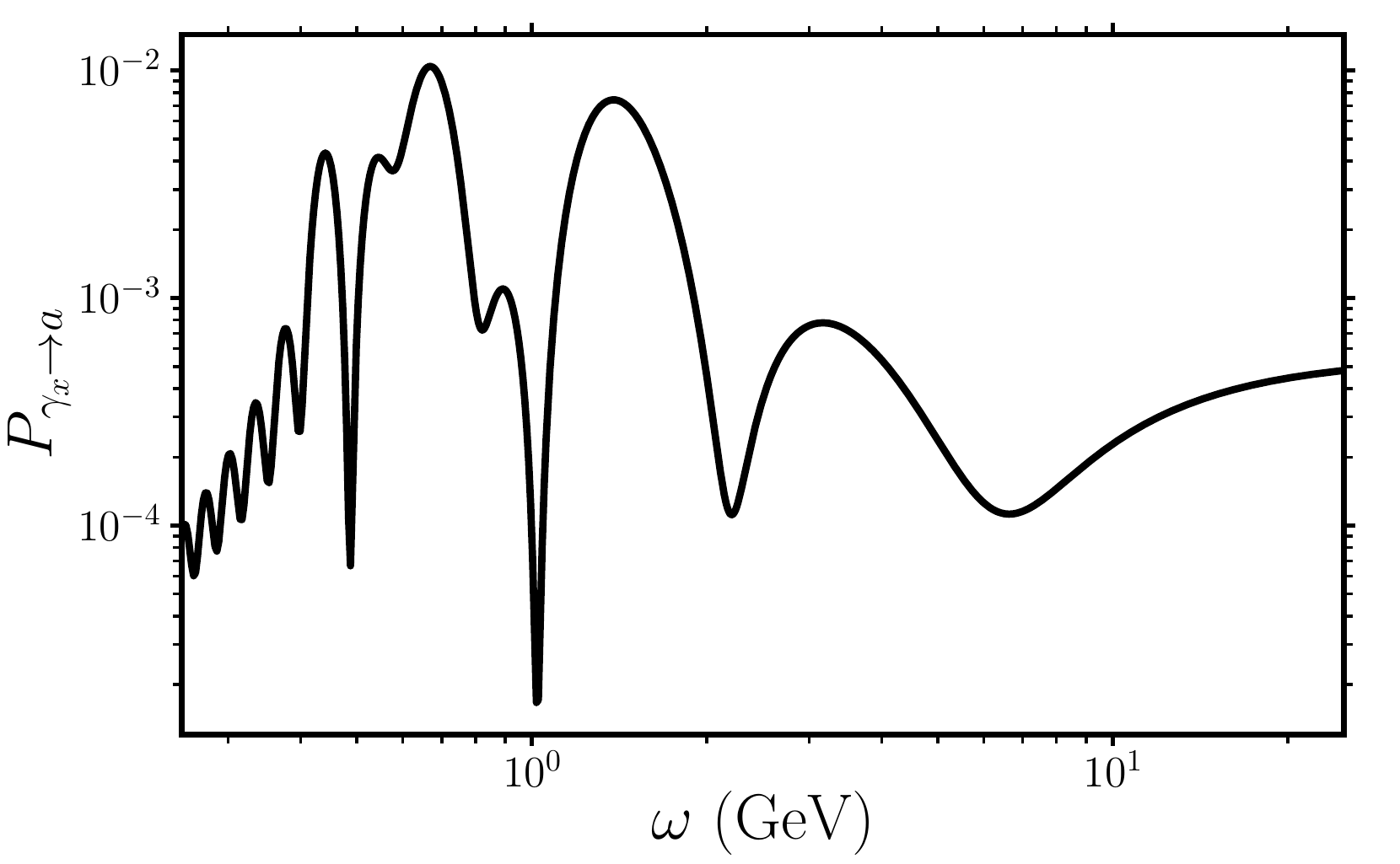} 
      \caption{Conversion probability as function of the axion energy.} \label{fig:fig4_P}
    \vspace{4ex}
  \end{subfigure} 
    \caption{
The axion-photon conversion probability for $g_{a\gamma}=10^{-11} \GeV^{-1}$, $m_{a} =5\times10^{-9}$~eV and  $\omega_{\rm pl}$ negligible (bottom right) calculated for the magnetic field model in equation \eqref{eq:Bturb} for 7 modes (top left), cf.~section \ref{sec:turbulent} (Example 4). The real and imaginary parts of the amplitude are shown in the top right panel and the autocorrelation function in the bottom left panel.
    }
  \label{fig:fig_osc}
\end{figure*}

\begin{widetext}
The real and imaginary parts of the amplitude for the $n$-th mode is simply obtained by direct integration, or equivalently by taking the cosine and sine transforms as in \eqref{eq:amp-c1}:
\begin{align}
   {\rm Re}\left({\cal A}_{\gamma_x \to a} \right) &= {\cal F}_s\Big( \Delta_x\Big) = \frac{g_{a\gamma}R}{4}\left[B_{n}^{c}\left(\frac{\sin^2\xi^{+}_{n}}{\xi^{+}_{n}}+\frac{\sin^2\xi^{-}_{n}}{\xi^{-}_{n}}\right)+B_{n}^{s}\left(\frac{\sin\xi^{-}_{n}\cos\xi^{-}_{n}}{\xi^{-}_{n}}-\frac{\sin\xi^{+}_{n} \cos\xi^{+}_{n}}{\xi^{+}_{n}}\right)\right]\,  ,
   \nonumber
\end{align}
\begin{align}
   {\rm Im}\left({\cal A}_{\gamma_x \to a} \right) &= -
     {\cal F}_c\Big( \Delta_x\Big) 
 =-\frac{g_{a\gamma}R}{4}\left[B_{n}^{c}\left(\frac{\sin\xi^{+}_{n} \cos\xi^{+}_{n}}{\xi^{+}_{n}}+\frac{\sin\xi^{-}_{n}\cos\xi^{-}_{n}}{\xi^{-}_{n}}\right)+B_{n}^{s}\left(\frac{\sin^2\xi^{+}_{n}}{\xi^{+}_{n}}-\frac{\sin^2\xi^{-}_{n}}{\xi^{-}_{n}}\right)\right]\, , 
    \label{eq:amplit2}
\end{align}
where $\xi^{\pm}_{n}=(\eta\pm k_{n})R/2$ and $\eta=m_{a}^{2}/2\omega$. The conversion probability is then given by 
\begin{equation}
\begin{split}
    P_{\gamma\rightarrow a}&=\frac{g_{a\gamma}^{2}R^{2}}{16}\sum_{n}\Bigg[(B_{n}^{+})^{2}\left(\frac{\sin^{2}\xi^{+}_{n}}{(\xi^{+}_{n})^{2}}+\frac{\sin^{2}\xi^{-}_{n}}{(\xi^{-}_{n})^{2}}\right)+2(B_{n}^{-})^{2}\cos(\xi^{+}_{n}-\xi^{-}_{n})\frac{\sin\xi^{+}_{n}\sin\xi^{-}_{n}}{\xi^{+}_{n}\xi^{-}_{n}}+\\
   &+2\sum_{m< n}(B_{n}^{c}B_{m}^{c}+B_{n}^{s}B_{m}^{s})\left(\frac{\sin(\xi_m^+)\sin(\xi_n^+)}{\xi_m^+ \xi_n^+} \cos(\xi_m^+- \xi_n^+)+\frac{\sin(\xi_m^-)\sin(\xi_n^-)}{\xi_m^- \xi_n^-} \cos(\xi_m^-- \xi_n^-)\right)+\\
   & +2\sum_{m< n}(B_{n}^{c}B_{m}^{c}-B_{n}^{s}B_{m}^{s})\left(\frac{\sin(\xi_m^+)\sin(\xi_n^-)}{\xi_m^+ \xi_n^-} \cos(\xi_m^+- \xi_n^-)+\frac{\sin(\xi_m^-)\sin(\xi_n^+)}{\xi_m^- \xi_n^+} \cos(\xi_m^- - \xi_n^+)\right)+\\
   &+4B_{n}^{c}B_{n}^{s}\frac{\sin(\xi_{n}^{+})\sin(\xi_{n}^{-})}{\xi_{n}^{+}\xi_{n}^{-}}\sin(\xi_{n}^{+}-\xi_{n}^{-})+\\
   &+\sum_{m< n}(B_{n}^{c}B_{m}^{s}+B_{m}^{c}B_{n}^{s})\left(\frac{\sin\xi^{+}_{n}\sin\xi^{-}_{m}\sin(\xi^{+}_{n}-\xi^{-}_{m})}{\xi^{+}_{n}\xi^{-}_{m}}-\frac{\sin\xi^{-}_{n}\sin\xi^{+}_{m}\sin(\xi^{-}_{n}-\xi^{+}_{m})}{\xi^{-}_{n}\xi^{+}_{m}}\right)+\\
  & -\sum_{m< n}(B_{n}^{c}B_{m}^{s}-B_{m}^{c}B_{n}^{s})\left(\frac{\sin\xi^{+}_{n}\sin\xi^{+}_{m}\sin(\xi^{+}_{n}-\xi^{+}_{m})}{\xi^{+}_{n}\xi^{+}_{m}}-\frac{\sin\xi^{-}_{n}\sin\xi^{-}_{m}\sin(\xi^{-}_{n}-\xi^{-}_{m})}{\xi^{-}_{n}\xi^{-}_{m}}\right)\Bigg]\, , 
    \end{split}
    \label{eq:Pmanymodes}
\end{equation}
\end{widetext}
where $(B_{n}^{\pm})^{2}=(B_{n}^{c})^{2}\pm(B_{n}^{s})^{2}$. The terms proportional to $\sin^2(\xi_n^\pm)/(\xi_n^\pm)^2$, corresponds to an incoherent addition of  contributions of the form \eqref{eq:Pflopping} and would be the only term present in a strict application of the rotating phase approximation.

The magnetic autocorrelation function of  \eqref{eq:Bturb} has the general form
\begin{widetext}
\beq    \begin{split}
c_{B}(L)&=\sum_{n}\left[(B_{n}^{c})^{2}C_{1,n}(L)+(B_{n}^{s})^{2}C_{2,n}(L)+B_{n}^{c}B_{n}^{s}(C_{3,n}(L)+C_{4,n}(L))\right]+\\
&+\sum_{m\neq n}\left[B_{n}^{c}B_{m}^{c}C_{1,nm}(L)+B_{n}^{s}B_{m}^{s}C_{2,nm}(L)+B_{n}^{c}B_{m}^{s}C_{3,nm}(L)+B_{n}^{s}B_{m}^{c}C_{4,nm}(L)\right]
\label{eq:cBmanymodes}
\end{split}
    \eeq
where the functions $C$ are defined as
\begin{equation}
    \begin{split}
C_{1,n}(L)&=\int_{0}^{R-L}dz\, \cos(k_{n}z)\cos(k_{n}(z+L))=\\  
&=\frac{2k_{n}(R-L)\cos(k_{n}L)-\sin(k_{n}L)-\sin(k_{n}(L-2R))}{4k_{n}}\\
C_{2,n}(L)&=\int_{0}^{R-L}dz\, \sin(k_{n}z)\sin(k_{n}(z+L))=\\  
&=\frac{2k_{n}(R-L)\cos(k_{n}L)+\sin(k_{n}L)+\sin(k_{n}(L-2R))}{4k_{n}}\\
C_{3,n}(L)&=\int_{0}^{R-L}dz\, \sin(k_{n}z)\cos(k_{n}(z+L))=\\
&=\frac{2k_{n}(L-R)\sin(k_{n}L)-\cos(k_{n}L)-\cos(k_{n}(L-2R))}{4k_{n}}\\
C_{4,n}(L)&=\int_{0}^{R-L}dz\, \cos(k_{n}z)\sin(k_{n}(z+L))=\\
&=\frac{2k_{n}(R-L)\sin(k_{n}L)+\cos(k_{n}L)-\cos(k_{n}(L-2R))}{4k_{n}}    \end{split}
\end{equation}
and 
\begin{equation}
    \begin{split}
C_{1,nm}(L)&=\int_{0}^{R-L}dz\, \cos(k_{n}z)\cos(k_{m}(z+L))=\\
        &=\frac{k_{n}\cos(k_{m}R)\sin(k_{n}(L-R))+k_{m}\sin(k_{m}R)\cos(k_{n}(L-R))-k_{m}\sin(k_{m}L)}{k_{m}^{2}-k_{n}^{2}}\, \\
 C_{2,nm}(L)&=\int_{0}^{R-L}dz\, \sin(k_{n}z)\sin(k_{m}(z+L))=\\
        &=\frac{k_{n}\cos(k_{n}(L-R))\sin(k_{m}R)+k_{m}\cos(k_{m}R)\sin(k_{n}(L-R))-k_{n}\sin(k_{m}L)}{k_{n}^{2}-k_{m}^{2}}
        \end{split}
        \end{equation}
        \begin{equation}
    \begin{split}
 C_{3,nm}(L)&=\int_{0}^{R-L}dz\, \sin(k_{n}z)\cos(k_{m}(z+L))=\\
        &=\frac{k_{m}\sin(k_{n}(L-R))\sin(k_{m}R)-k_{n}\cos(k_{m}R)\sin(k_{n}(L-R))+k_{n}\cos(k_{m}L)}{k_{n}^{2}-k_{m}^{2}}\, \\
C_{4,nm}(L)&=\int_{0}^{R-L}dz\, \cos(k_{n}z)\sin(k_{m}(z+L))=\\
        &=\frac{k_{n}\sin(k_{m}R)\sin(k_{n}(L-R))-k_{m}\cos(k_{n}(L-R))\cos(k_{m}R)+k_{m}\cos(k_{m}L)}{k_{m}^{2}-k_{n}^{2}} 
        \label{eq:csmanymodeslast}\end{split}
\end{equation}
\end{widetext}
Equations \eqref{eq:Pmanymodes}-\eqref{eq:cBmanymodes} provide the most general solution for relativistic axion-photon conversion of axions with $m_a \gg \omega_{\rm pl}$, and equations \eqref{eq:cBmanymodes}-\eqref{eq:csmanymodeslast} provide the corresponding cosine transforms. In figure \ref{fig:fig_osc}, we show a simple, 7-mode example of the magnetic field, and its solution. We have tested our analytical solution against numerical simulations, and found perfect agreement.


\subsubsection{Example 5: Monomial magnetic fields}  
Sufficiently slowly varying fields are often conveniently Taylor expanded to some finite order in $z/R$. In this example, we consider the simplest case of a linear dependence on the radius, and in section \ref{sec:regular} we determine the conversion probability for a general polynomial.

The simplest monomial magnetic field is linear 
\begin{equation}
    B_x(z)=B_{0}\frac{z}{R}\Theta(R-z)\, , 
\end{equation}
and the transition amplitude is 
\begin{equation}
\begin{split}
   {\rm Re}\left({\cal A}_{\gamma_x \to a} \right) &= {\cal F}_s\Big( \Delta_x\Big) = \\
   &=   \frac{g_{a\gamma}B_{0}}{2 \eta^{2}R } \Big( 
   \sin(\eta R)-\eta R \cos(\eta R)  \Big)
   \,  ,
\\
   {\rm Im}\left({\cal A}_{\gamma_x \to a} \right) &= -
 {\cal F}_c\Big( \Delta_x\Big) 
 =\\
 &= \frac{g_{a\gamma}B_{0}}{2 \eta^{2}R} \Big( 
 1-\cos(\eta R) -\eta R \sin(\eta R)
 \Big)
 \, , 
 \end{split}
    \label{eq:ampmonomial}
\end{equation}
and the conversion probability is
\begin{equation}
\begin{split}
    P_{\gamma_x\rightarrow a}&=\frac{g_{a\gamma}^{2}B_{0}^{2}R^{2}}{4}\frac{1}{\eta^{4}R^{4}}\\
    &\quad\left[2+\eta^{2}R^{2}-2\cos(\eta R)-2\eta R\sin(\eta R)\right]\, .
    \end{split}
\end{equation}

\subsubsection{Example 6: `Regular' magnetic fields}
\label{sec:regular}
\label{sec:LTconv}

In this  section, we consider a broad class of magnetic fields that can be modelled by finite-order polynomials within a finite radius, i.e.~as
\beq
 B_x(z) = B_0\, b\left(\frac{z}{R}\right) \Theta(R-z) \, ,
\label{eq:poly}
\eeq
where  $b(z)$ can be expressed as a polynomial in $z/R$. 
 The purpose of this section is to provide an algorithmic  prescription for finding the conversion probability of such models analytically, using our Fourier analysis approach. 
 As an example, we also calculate the conversion probability explicitly in the regular magnetic field model proposed in \cite{LT} (however, see our discussion about the astrophysical consistency of this model in section \ref{sec:magneticmodels}).

\begin{figure*}[t!] 
  \begin{subfigure}[t]{\fourpanelwidth\linewidth}

    \includegraphics[width=\linewidth]{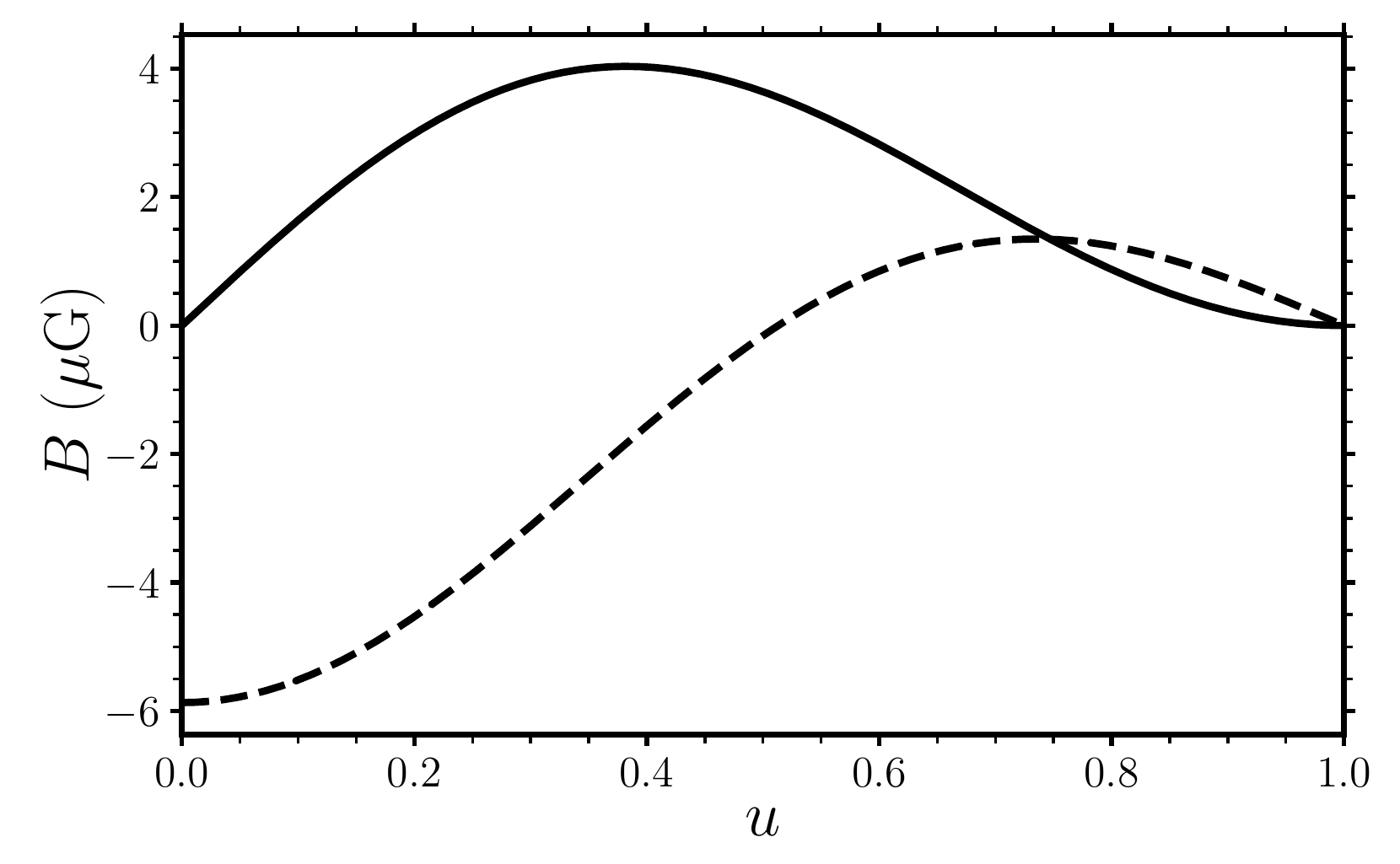} 
     \caption{Components of the magnetic field as a function of $u=z/R$} \label{fig:bLT}
    \vspace{4ex}
  \end{subfigure}
  \begin{subfigure}[t]{\fourpanelwidth\linewidth}
        \includegraphics[width=\linewidth]{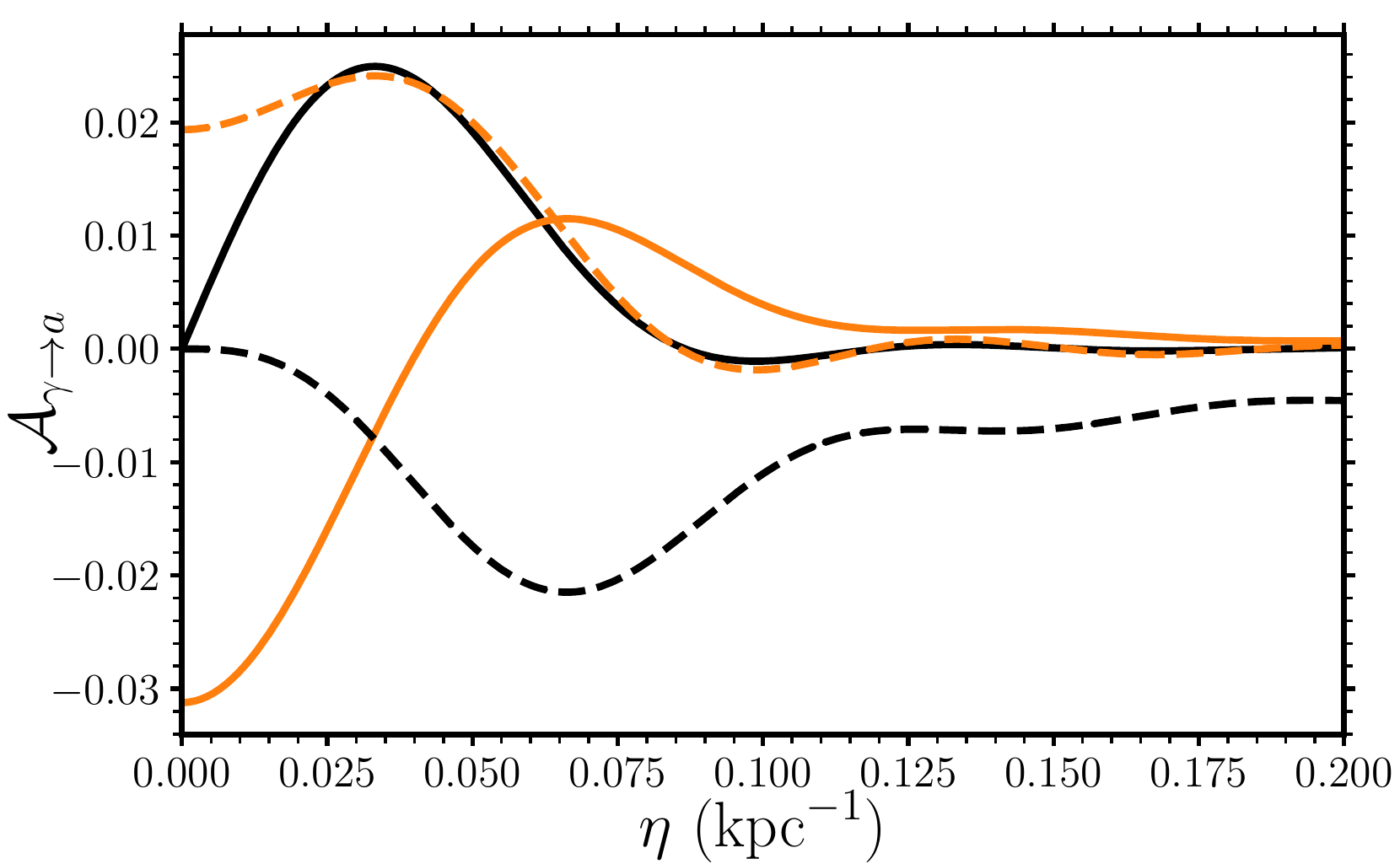} 
        \caption{The real, Re(${\cal A}_{\gamma \to a})={\cal F}_s(\Delta_x)$ (black), and imaginary, Im(${\cal A}_{\gamma \to a})=-{\cal F}_c(\Delta_x)$ (orange), parts of the amplitude ${\cal A}_{\gamma \to a}$ for the two field components. } \label{fig:bcsLT}
    \vspace{4ex}
  \end{subfigure} 
  \begin{subfigure}[t]{\fourpanelwidth\linewidth}
        \includegraphics[width=\linewidth]{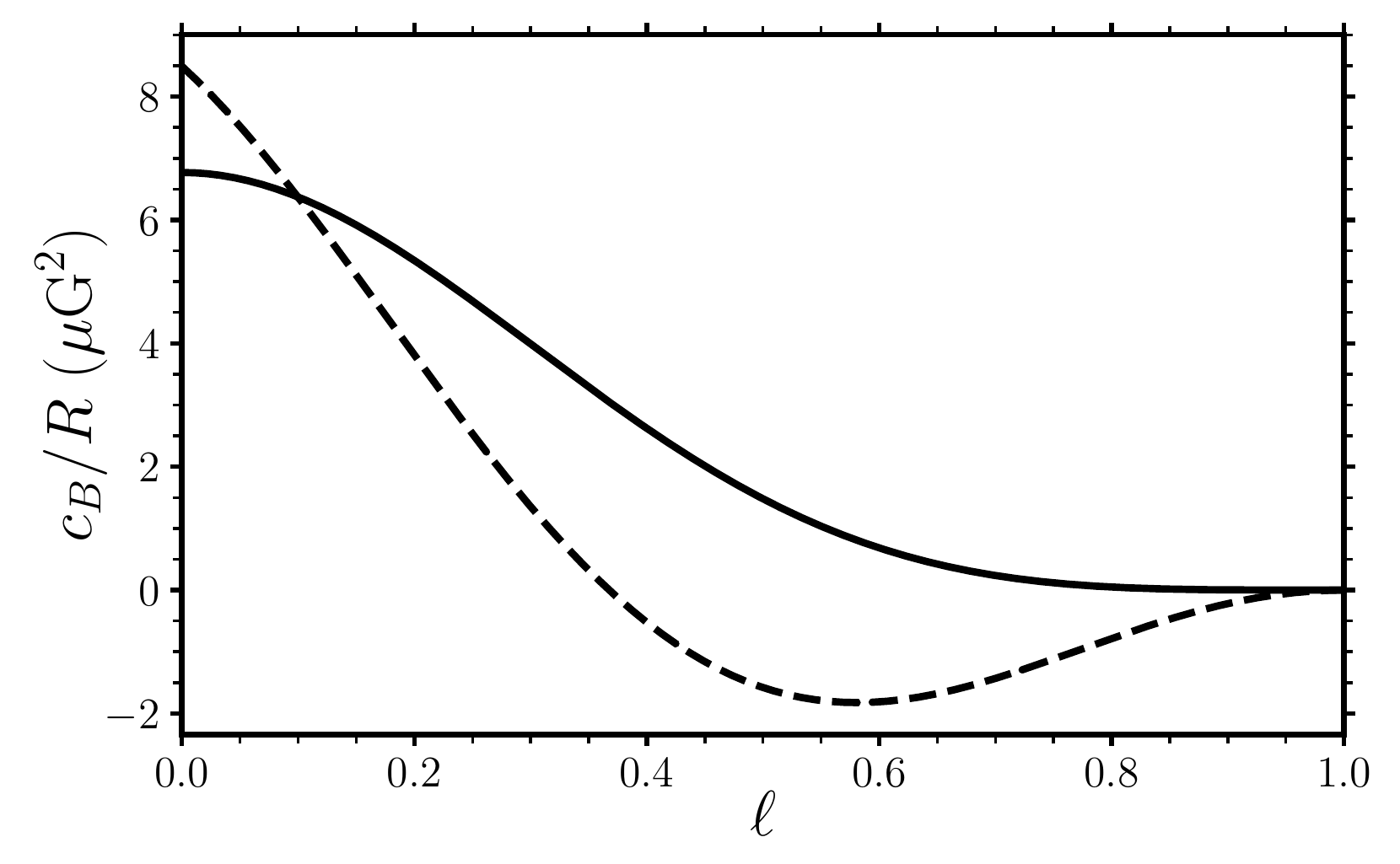} 
        \caption{Magnetic field autocorrelation function as a function of $\ell=L/R$.} \label{fig:cbLT}
    \vspace{4ex}
  \end{subfigure}
  \begin{subfigure}[t]{\fourpanelwidth\linewidth}
        \includegraphics[width=\linewidth]{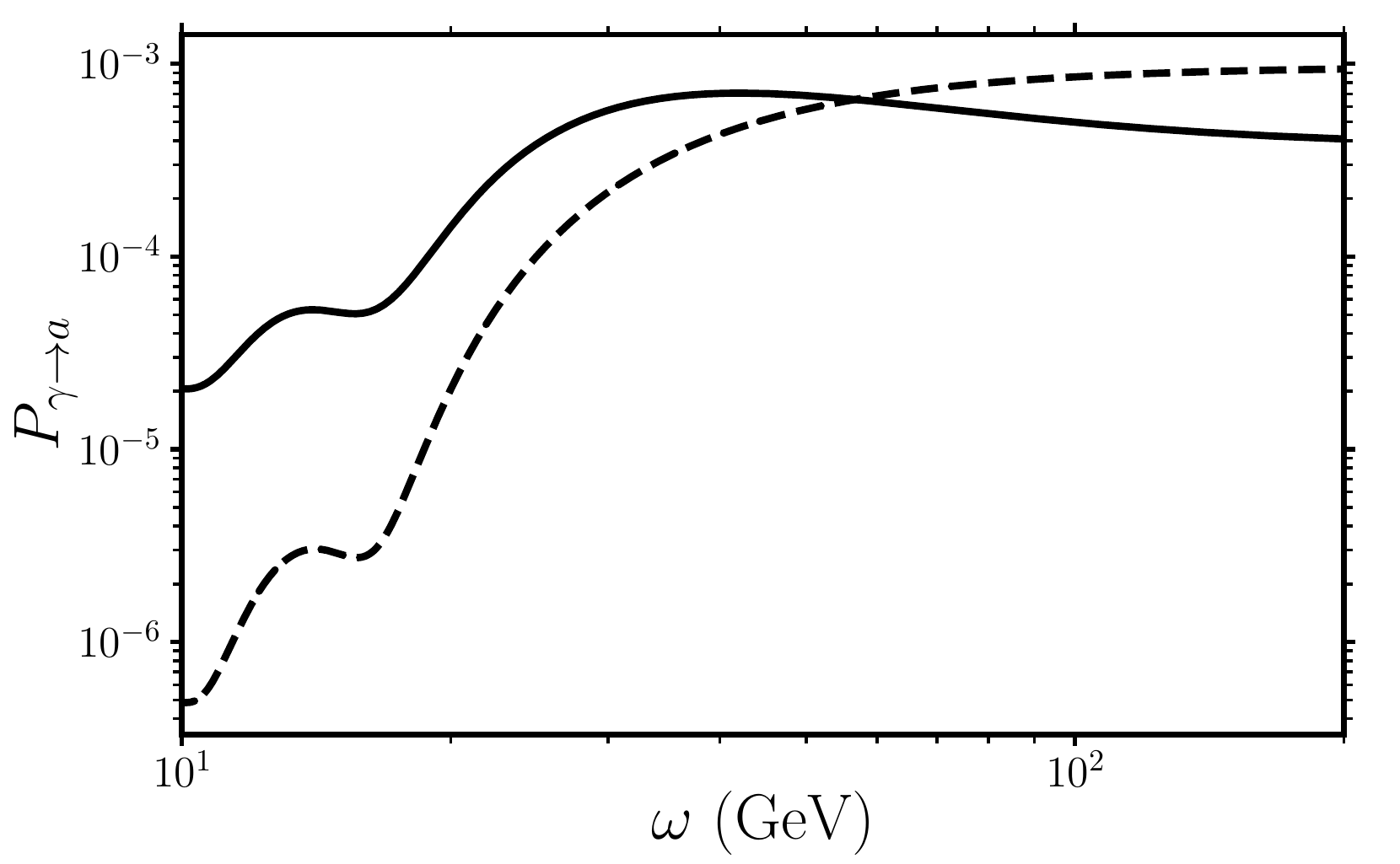}
        \caption{Conversion probability as a function of energy for $\g=10^{-13}\, {\rm GeV}^{-1}$, $m_a=5\times10^{-9}\, {\rm eV}$ and $R=93$\,kpc.} \label{fig:PLT}
    \vspace{4ex}
  \end{subfigure} 
    \caption{The conversion probability (bottom right) for the regular magnetic field (top left) discussed in \cite{LT, Gourgouliatos:2010nu} and section \ref{sec:regular} (Example 6). 
     The solid and dashed curves respectively correspond to the $\phi$ and $\theta$ components of the regular magnetic field, cf.~equations \eqref{eq:BphiLT} and \eqref{eq:BthetaLT}. See however section \ref{sec:magneticmodels} for a discussion of the interpretation of these results. 
    }
    \label{fig:LT}
\end{figure*}

The transition amplitude involves the sine and cosine transforms of $b\left(\frac{z}{R}\right) \Theta(R-z)$.  
It is convenient to define $u=z/R$ and write $b(u) = \sum_{n=0}^{n_{\rm max}} b_n u^n$. The cosine and sine transforms then involve terms like
\beq
\int_0^1du\, u^n \cos(\tilde \eta u)~~~\text{and}~~~\int_0^1du\,  u^n \sin(\tilde \eta u)\, ,
\eeq
where $\tilde \eta = \eta R$. These integrals respectively, and rather  formally, evaluate to instances of the regular and generalised hypergeometric functions. For our purposes, it will simpler to evaluate them by repeated integration by parts, i.e.~by using
\begin{align}
\int_0^1du\, u^n \cos(\tilde \eta u) &=\frac{1}{\tilde \eta} \sin(\tilde \eta) - \frac{n}{\tilde \eta} \int_0^1 du\, u^{n-1} \sin(\tilde \eta u)
\label{eq:cosintparts}\\
\int_0^1du\, u^n \sin(\tilde \eta u) &=-\frac{1}{\tilde \eta} \cos(\tilde \eta) + \frac{n}{\tilde \eta} \int_0^1 du\, u^{n-1} \cos(\tilde \eta u)~.
\end{align}
The resulting amplitude then has a simple form
\begin{equation}
\begin{split}
&{\cal A}_{\gamma \to a}(\tilde{\eta}) =\\
&\quad\frac{g_{a\gamma} B_0 R}{ 2} \Big[ \Big( s_0(\tilde \eta) + s_1(\tilde \eta) \sin (\tilde \eta) + s_2(\tilde \eta) \cos (\tilde \eta) \Big)
+\\
&\quad- i
  \Big( c_0(\tilde \eta) + c_1(\tilde \eta) \sin (\tilde \eta) + c_2(\tilde \eta) \cos (\tilde \eta)
\Big) \Big] \, .
\label{eq:Apoly}
\end{split}
\end{equation}
where $s_i$ and $c_i$ denote the polynomials obtained from the sine and cosine transforms, respectively.\footnote{For $\tilde \eta <1$, one may alternatively Taylor expand the integrand in $\tilde \eta u$, which is numerically preferable for small values of $\tilde \eta$.} The conversion probability is simply found by squaring and summing the real and imaginary parts of the amplitude, as given by equation \eqref{eq:Apoly}.

Alternatively, the conversion probability can be found by first evaluating the magnetic field autocorrelation function. Writing $\ell = L/R$, we have that
\beq
\begin{split}
c_B(\ell)&= B_0^2 R \int_0^{1-\ell} du\, b(u) b(u+\ell) =\\
&=B_0^2 R  \sum_{n,m} b_n b_m \int_0^{1-\ell} du\, u^n (u+\ell)^m \, .
\end{split}
\eeq
Each integral in this double sum can be expressed using  incomplete Euler beta-functions, or alternatively, by expanding the factors of $(u+\ell)^m$ and doing the integral term-by-term. The latter puts $c_B$ in polynomial form:
\beq
c_B(\ell) = B_0^2 R \,\Theta(1-\ell) \sum_{n=0}^{2n_{\rm max}+1} c_n\, \ell^n \, .
\label{eq:cbpoly}
\eeq

From the autocorrelation function, the conversion probability can be found by taking the cosine transform. For a correlation function of the form \eqref{eq:cbpoly}, this involves hypergeometric functions, however, we again use repeated applications of equation \eqref{eq:cosintparts} to simplify the expression and determine its general form. We find that
\beq 
P_{\gamma \to a} = \frac{\g^2 B_0^2R^2}{4}\Big(
p_0(\tilde \eta) + p_1(\tilde \eta) \cos(\tilde \eta)+ p_2(\tilde \eta) \sin(\tilde \eta)
\Big) \, ,
\label{eq:Ppoly2}
\eeq 
where the $p_i(\tilde \eta)$ are polynomials. 
The shape of the conversion probability is then determined by these polynomials and the low-frequency oscillations from $\sin(\tilde \eta)$ and $\cos(\tilde \eta)$, which typically results in a slow-varying conversion probability. 

We now consider a particular example of axion-photon conversion in slowly varying magnetic fields. Recently, reference \cite{LT} adapted a model of the magnetic field in a radio bubble from \cite{Gourgouliatos:2010nu} as a model of the \emph{cluster magnetic field} in the Perseus cluster, suggesting that it provides a limiting case of observationally viable magnetic fields.  We will return to the astrophysical issues of this  non-standard, and somewhat controversial, generalisation in section \ref{sec:astrophys}. Here, we simply show that the methods of this paper can be used to find the  perturbative conversion probability of this model analytically. 

The magnetic field model of \cite{Gourgouliatos:2010nu} was found as a solution to the Grad-Shafranov equation in equilibrium magnetohydrodynamics (MHD). In spherical coordinates, the explicit form of the components of the magnetic field for $r\leq R$ were determined to
\begin{align}
    B_r &= \frac{2 \cos\theta}{r^2} f(r) \\
    B_\theta &= - \frac{\sin \theta}{r} f'(r) \label{eq:BthetaLT} \\
    B_\phi &= \frac{A \sin \theta}{r} f(r) \label{eq:BphiLT} \, ,
\end{align}
where $A$ solves a transcendental equation and is approximately given by  $A= 5.76$. The function $f(r)$ is given by
\beq
f(r) =c_1\Big(
A \cos (\alpha u) - \frac{\sin(A u)}{u}
- u^2\left(
A \cos\alpha - \sin A
\right)
\Big) \, .
\eeq
Here $u=r/R \leq 1$ is the re-scaled radial coordinate, and $c_1$ is a normalisation factor that sets the overall magnitude of the magnetic field. The parameter values adopted in reference \cite{LT} in applying this model to the entire Perseus cluster were $\theta=\pi/4$, $c_1 = -0.060\, \mu$G, and $R=93$ kpc. Since the magnetic field is slowly varying in $u$, this system is easily solved to high accuracy by approximating the components of $B$ by finite-order Taylor expansions.  The main results are summarised in figure  \ref{fig:LT}.

The two paths to the conversion probability are illustrated in figure \ref{fig:LT}: by taking the cosine and sine transforms of the magnetic field components as in equation \eqref{eq:amp-c3}, we obtain the transition amplitudes for linearly polarised photons converting into axions, as plotted in figure \ref{fig:bcsLT}. Taking the squared norm  of the amplitudes give the polarised conversion probabilities, plotted in figure \ref{fig:PLT}. Alternatively, we directly calculate the magnetic autocorrelation function as shown in figure \ref{fig:cbLT}. The cosine transform of these autocorrelation functions again give the conversion probability. 

This example magnetic field model is inconsistent with observations of the Perseus cluster, as we explain in section \ref{sec:magneticmodels}.
However, it is still interesting to interpret it through the Fourier formalism. We have seen that long-ranged magnetic autocorrelations translate into support for rapid-oscillation modes in $P_{\gamma \to a}(\eta)$, however, this does not automatically translate into  oscillations of the full function $P_{\gamma \to a}(\eta)$ at those frequencies; Fourier analysis  decomposes any function into its 
oscillatory components, but of course, this does not mean all functions are oscillatory. The bumpy features of the conversion probabilities of figure \ref{fig:PLT} indicate the broadly  preferred scales of the autocorrelation function, $c_B(L)$. Heuristically, we expect that featured or oscillatory autocorrelation functions at large $L$ correspond to a featured and oscillatory conversion probability at high frequencies. 

\section{The Fourier transform formalism: the massless case}
\label{sec:massless}
The assumption of $m_a^2 \gg \omega_{\rm pl}^2$ used in the preceding section simplifies the phase-factor of the transition amplitude to $\Phi(z') = \tfrac{m_a^2 z'}{2 \omega} = \eta z'$.  The linear appearance of the integration variable $z'$ makes the  connection to Fourier transforms evident. When the plasma frequency cannot be neglected, the analysis is  more subtle. 
In this section, we focus on the next-simplest case of massless axions. We will see that in this case, a change of variable again makes the simple Fourier formulas applicable,  although at a slightly modified form.   

\subsection{The transition amplitude and the conversion probability}
In this section, we set $m_a=0$ and consider $\omega_{\rm pl}^2(z) \neq 0$. The transition amplitude is again  given by
\bea
{\cal A}_{\gamma_x \to a} = -i \int_0^\infty dz' \Delta_x(z') 
e^{-i \Phi(z')}
\label{eq:amp41}
\eea
but now the phase factor reads
\beq 
\Phi(z') = -\frac{1}{2\omega}  \int_0^{z'} dz'' \omega_{\rm pl}^2(z'') = - \lambda \varphi \, ,
\eeq   
where we have defined $\lambda=1/\omega$ and
\beq
\varphi(z') = \frac{1}{2}\int_0^{z'} dz'' \omega_{\rm pl}^2(z'')  \, . \label{eq:varphimassless}
\eeq
Since $\omega_{\rm pl}^2$ is positive definite, $\varphi$
is monotonically increasing with  $z'$, has the range $0 \leq \varphi \leq \infty$, and is a well-defined coordinate, replacing  $z'$. The measure transforms as  $dz' = d\varphi/\omega_{\rm pl}^2$ and
the full amplitude can again be expressed using Fourier cosine and sine transforms:
\beq
{\cal A}_{\gamma_x \to a} =
 -i \int_0^\infty d \varphi \,   \frac{2 \Delta_x}{\omega_{\rm pl}^2} e^{i  \lambda \varphi}  = \hat G_s(\lambda) - i \hat G_c(\lambda)
\eeq
where $\hat G$ denotes the transform of the function
\beq
G(\varphi) =   \frac{2 \Delta_x}{\omega_{\rm pl}^2} = \frac{\g B_x}{\omega_{\rm pl}^2}  \, . \label{eq:fmassless}
\eeq
We emphasise that $B_x$ and $\omega_{\rm pl}$ in this equation are functions of $\varphi$, as defined by \eqref{eq:varphimassless}. Since the $\hat G$ are real, the conversion probability is simply given by
\beq
P_{\gamma_x \to a}(\lambda) = \hat G_s^2 + \hat G_c^2 \, ,
\label{eq:Pmassless1}
\eeq
which is the power spectrum of $G(\varphi)$. Our derivation of the Wiener–Khintchine theorem leading to equation \eqref{eq:Pautocorr} now implies that 
\beq
P_{\gamma_x \to a}(\lambda) =  \hat G^2_s(\lambda) + \hat G^2_c(\lambda) = 2 {\cal F}_c (c_G(\psi) )
\, ,
\label{eq:Pmassless2}
\eeq
where $c_G(\psi)$ denotes the autocorrelation of $G$ in $\varphi$-space, i.e.~
\beq
c_G(\psi) = \int_0^\infty d\varphi\, G(\varphi+ \psi) G(\varphi) \, .
\label{eq:cG}
\eeq
In sum, also in the massless case there are two possible routes to calculate the conversion probability: either by calculating the cosine and sine transforms of $G$,  then taking the sum of squares, or by computing the autocorrelation function of $G$ and taking its cosine transform. 
A novel feature of the massless case is that, for the purpose of axion-photon conversion, distances are more naturally measured in terms of the phase  $\varphi$ instead of the spatial coordinate $z$. Moreover, the Fourier cosine and sine transforms are taken of the function $G$, which includes a  factor of $1/\omega_{\rm pl}^2$. The appearance of this factor is mathematically intuitive: when $\omega_{\rm pl}^2$ is large, the phase in equation \eqref{eq:amp41} varies quickly and the integral tends to wash out. For smaller values of $\omega_{\rm pl}^2$, the oscillations are slower and it is easier to build up a non-vanishing amplitude. When changing integration variable from $z$ to $\varphi$, the phase $\Phi$ grows at a constant rate, by definition. The impact of $\omega_{\rm pl}^2$ on the transition amplitude is instead captured by the explicit factor  of $1/\omega_{\rm pl}^2$ in $G(\varphi)$. This factor makes the impact of the plasma frequency on the transition probability apparent. In particular, this form of the amplitude suggests that fluctuations in the electron density along the direction of propagation can be an important source of axion-photon oscillations. We expect to return to this phenomenon  in   future work.



\subsection{Examples}
We have seen that axion-photon transitions for massless axions proceed similarly to the case of massive axions (cf.~e.g.~equations \eqref{eq:P1cs} and \eqref{eq:Pautocorr} to equations \eqref{eq:Pmassless1} and \eqref{eq:Pmassless2}).  The examples worked out for massive axions  in section \ref{sec:massive_examples}  applies with small modifications to massless axions if the plasma frequency is constant. More generally however, one needs to account for the non-trivial coordinate change from $z'$ to $\varphi$, as we here illustrate by an example. 

\subsubsection{Example 7: Massless axion in an oscillating magnetic field}

\label{sec:analytic_warmup}
Consider an axion with $m_a=0$ in an environment where the electron density decreases like $n_e \sim \sqrt{R/z}$. 
The squared plasma frequency is then given by 
\beq
\omega_{\rm pl}^2(z) = \omega_{\rm pl}^2(R)  \sqrt{\frac{R}{z}} \, ,
\label{eq:omegapl_example}
\eeq
where $\omega_{\rm pl}^2(R) = \frac{4\pi e^2}{m_e}  n_e(R)$. From equation \eqref{eq:varphimassless}, we have that
\beq
\begin{split}
\varphi(z)&= \frac{1}{2}\int_0^{z} dz' \omega_{\rm pl}^2(z') =\\
&=\frac{1}{2}\omega_{\rm pl}^2(R) R \int_0^{z/R} du \frac{1}{\sqrt{u}} = \\
&=\omega_{\rm pl}^2(R) R \left(\frac{z}{R} \right)^{1/2}\, .
\end{split}
\eeq
For this simple form of the electron density, we can invert this expression to find $z(\varphi)$ explicitly:
\beq
\frac{z }{R} =  \tilde \varphi^2 \, ,
\eeq
where we have defined the dimensionless variable $\tilde \varphi = \varphi/( \omega_{\rm pl}^2(R) R)$. 


We consider a damped, oscillatory magnetic field of the form
\beq
B_x(z) = B_R   \Theta(R-z) \, \frac{n_e(z)}{n_e(R)}\, \cos(k z) \, .
\label{eq:Bmasslessexpl}
\eeq
The explicit damping factor mimics a common parametrisation for galaxy cluster magnetic fields (cf.~e.g.~\cite{Bonafede:2010}), but is here mainly motivated by convenience: since $n_e(z)/n_e(R) = \omega_{\rm pl}^2(z)/\omega_{\rm pl}^2(R)$, the function $G$ becomes
\beq
G(\tilde \varphi) =  \frac{\g B_R}{\omega_{\rm pl}(R)^2}   \Theta(1- \tilde \varphi^2) \, \cos(\tilde k \tilde \varphi^2) \, ,
\eeq
where $\tilde k= k R$.

The real  part of the transition amplitude is given by
\begin{widetext}
\begin{align}
&{\rm Re}({\cal A}_{\gamma_x \to a}) = {\cal F}_s(G) = \int_0^\infty d \varphi\, G(\varphi) \sin(\lambda \varphi) 
= 
\g B_R R\int_0^1 d\tilde \varphi\,  \cos\big(\tilde k  \tilde \varphi^2\big) \sin\big(\eta \tilde \varphi\big) \nonumber \\
&= \g B_R R
\sqrt{\frac{\pi }{8\tilde k}} \Bigg[
\left(
C\left(\frac{2 \tilde k-\eta }{\sqrt{2 \pi \tilde k}}\right)+2 C\left(\frac{\eta }{\sqrt{2 \pi \tilde k }}\right)-C\left(\frac{2 \tilde k+\eta
   }{\sqrt{2 \pi \tilde k }}\right)\right) \sin \left(\frac{\eta ^2}{4
   \tilde k}\right)+ \nonumber \\
&+   
   \left(-S\left(\frac{2 \tilde k-\eta }{ \sqrt{2 \pi \tilde k }}\right)-2
   S\left(\frac{\eta }{\sqrt{2 \pi \tilde k}}\right)+S\left(\frac{2 \tilde k+\eta }{
   \sqrt{2 \pi \tilde k }}\right)\right) \cos \left(\frac{\eta ^2}{4 \tilde k}\right) \Bigg] \, ,
  \label{eq:ReAmassless}
%
\end{align}
where we have introduced  $\eta = \lambda \omega_{\rm pl}^2(R) R$, and where $C$ and $S$ denote the corresponding Fresnel integrals. 
The imaginary part of the amplitude is given by
\begin{align}
&{\rm Im}({\cal A}_{\gamma_x \to a}) = -{\cal F}_c(G) = -\int_0^\infty d\varphi\, G(\varphi) \cos(\lambda \varphi) 
= -
\g B_R R\int_0^1 d\tilde \varphi\,  \cos\big(\tilde k  \tilde \varphi^2\big) \cos\big(\eta \tilde \varphi\big) \nonumber \\
&= -\g B_R R
\sqrt{\frac{\pi }{8\tilde k}} \Bigg[\left(C\left(\frac{2 \tilde k-\eta }{ \sqrt{2 \pi \tilde k
   }}\right)+C\left(\frac{2 \tilde k+\eta }{ \sqrt{2 \pi \tilde k}}\right)\right) \cos
   \left(\frac{\eta ^2}{4 \tilde k}\right)+
   \left(S\left(\frac{2 \tilde k-\eta }{ \sqrt{2 \pi \tilde k
   }}\right)+S\left(\frac{2 \tilde k+\eta }{ \sqrt{2 \pi \tilde k }}\right)\right) \sin
   \left(\frac{\eta ^2}{4 \tilde k}\right)\Bigg] \, .
    \label{eq:ImAmassless}
\end{align}
The transition probability is, as usual, obtained from the sum of squares of equations \eqref{eq:ReAmassless} and \eqref{eq:ImAmassless}.
The autocorrelation function of $G(\varphi)$ is readily found to be 
\begin{align}
c_G(\tilde \psi)&=
\frac{\g^2 B^2_R R}{\omega_{\rm pl}(R)^2} 
\int_0^{1-\tilde \psi^2} d\tilde \varphi\, \cos(\tilde k \tilde \varphi^2) \cos(\tilde k (\tilde \varphi+ \tilde \psi)^2) \nonumber \\
&= \frac{\g^2 B^2_R}{\omega_{\rm pl}(R)^4} \Bigg[
\frac{1}{4 \tilde k \tilde \psi} \bigg(\sin( \tilde k \tilde \psi \tilde \psi_2)
- \sin(\tilde k \tilde \psi^2)
)\bigg) + \frac{1}{4} \sqrt{\frac{\pi}{\tilde k}} \left\{
\left(C\left(\sqrt{\frac{\tilde k}{\pi}} \tilde \psi_2\right)-C\left(\sqrt{\frac{\tilde k}{\pi}} \tilde \psi\right)\right) \cos \left(\frac{\tilde k \tilde \psi ^2}{2}\right)
\right.
+ \nonumber \\
&+\left.\left(S\left(\sqrt{\frac{\tilde k}{\pi}} \tilde \psi\right)-S\left(\sqrt{\frac{\tilde k}{\pi}} \tilde \psi_2\right)\right) \sin \left(\frac{\tilde k \tilde \psi^2}{2}\right)
\right\}
\Bigg] \, ,
\end{align}
\end{widetext}
where $\tilde \psi_2= 2+\tilde \psi - 2 \tilde \psi^2$. Using equation \eqref{eq:Pmassless2} and that the cosine transform is its own inverse,~cf.~equation \eqref{eq:invcos}, we note that this expression 
gives the spectrum of oscillations of $P_{\gamma_x \to a}$, as understood through the cosine transform. This spectrum would be non-trivial to obtain explicitly from directly transforming the conversion probability, and similarly, in this case it's challenging to analytically take the cosine transform of the autocorrelation function to obtain the conversion probability.

\section{The Fourier transform formalism:  the general case} 
\label{sec:general}

In general, the axion/photon trajectory may pass through dense regions with $\omega_{\rm pl} > m_a$ as well as dilute regions with $\omega_{\rm pl} < m_a$. The phase $\Phi$ will then increase in some regions and decrease in others, which prohibits a one-to-one, global change of coordinate from $z'$ to $\Phi$.  {\color{dm} Moreover, at points of stationary phase, $d\Phi/dz=0$, the photon and axion are mass degenerate, and can interconvert \emph{resonantly}.  The full amplitude is then the sum of resonant and non-resonant contributions, and in sections \ref{sec:res}--\ref{sec:generalnonres} we discuss how to carefully calculate these contributions  in the perturbative formalism. 

}

Our approach is to split the integration domain into sub-regions in which the phase $\Phi$ is  monotonic, either  increasing or decreasing, or approximately stationary. Points of stationary phase will produce a `resonant' contribution to the amplitude that depends only on the local environment around that point.  
For each region in which  $\Phi$ is monotonic, we can identify a suitable coordinate extension and express the contribution to the amplitude as a Fourier transform. The total amplitude is simply given by the sum over all resonant and non-resonant contributions. Figure \ref{fig:coords} shows a particular example of the general case (there denoted `Case II'): the plasma frequency crosses the axion mass at three points, resulting in four contributions to the non-resonant amplitude (labelled by 1 through 4) and three resonant contributions. 

Our notation for the general case is as follows: we factorise the phase into the (dimensionful) `coordinate' $\varphi$ and its Fourier conjugate $\lambda=1/\omega$ as
\beq
\begin{split}
\Phi(z') & =\int_0^{z'} dz'' [\Delta_\gamma(z'') - \Delta_a(z'')]\approx \\
&\approx  -\frac{1}{\omega} \int_0^{z'} dz''\, \tfrac{1}{2}[\omega_{\rm pl}^2 - m_a^2] = -\lambda\, \varphi(z') \, .
\end{split}
\eeq
We denote the points of stationary phase by $z_i^*$ (for $i=1, \ldots , N$), so that $\varphi'(z^*_i)=0$. We also define a small interval around each stationary point as $z_i^\ell < z^*_i < z_i^u$. As we discuss  in section \ref{sec:res},  these intervals should be chosen to be small enough so that $\varphi$ is well-approximated by a leading-order Taylor expansion around $z_i^*$. For ease of notation, we also define $z_0^u =0$ and $z_{N+1}^\ell =z$. In this notation and without loss of generality, the full amplitude is given by
\begin{equation}
\begin{split}
i {\cal A}_{\gamma_x \to a}(z, \lambda)& = \int_0^z dz' \Delta_x(z') e^{i \lambda \varphi(z')} 
=\\
&=i {\cal A}^{\rm res}_{\gamma_x \to a} + i {\cal A}^{\rm non-res}_{\gamma_x \to a}
\label{eq:amp1}
\end{split}
\end{equation}
where
\begin{align}
i {\cal A}^{\rm res}_{\gamma_x \to a}  &=  \sum_{i=1}^{N}\int_{z_i^\ell}^{z_{i}^u} dz'\,  \Delta_x(z') e^{i \lambda \varphi(z')} \label{eq:Ares}\, , \\
i {\cal A}^{\rm non-res}_{\gamma_x \to a} &=  \sum_{i=1}^{N+1} \int_{z_{i-1}^u}^{z_{i}^\ell} dz' \,  \Delta_x(z') e^{i \lambda \varphi(z')} \label{eq:Anonres}\, .
\end{align}
Equation \eqref{eq:Ares}  picks up the `resonant' amplitudes from stationary points, while equation \eqref{eq:Anonres} collects the non-resonant integrals. We will now discuss these contributions in turn.

\begin{figure*}[t!] 
  \begin{subfigure}[t]{\fourpanelwidth\linewidth}

      \includegraphics[width=\linewidth]{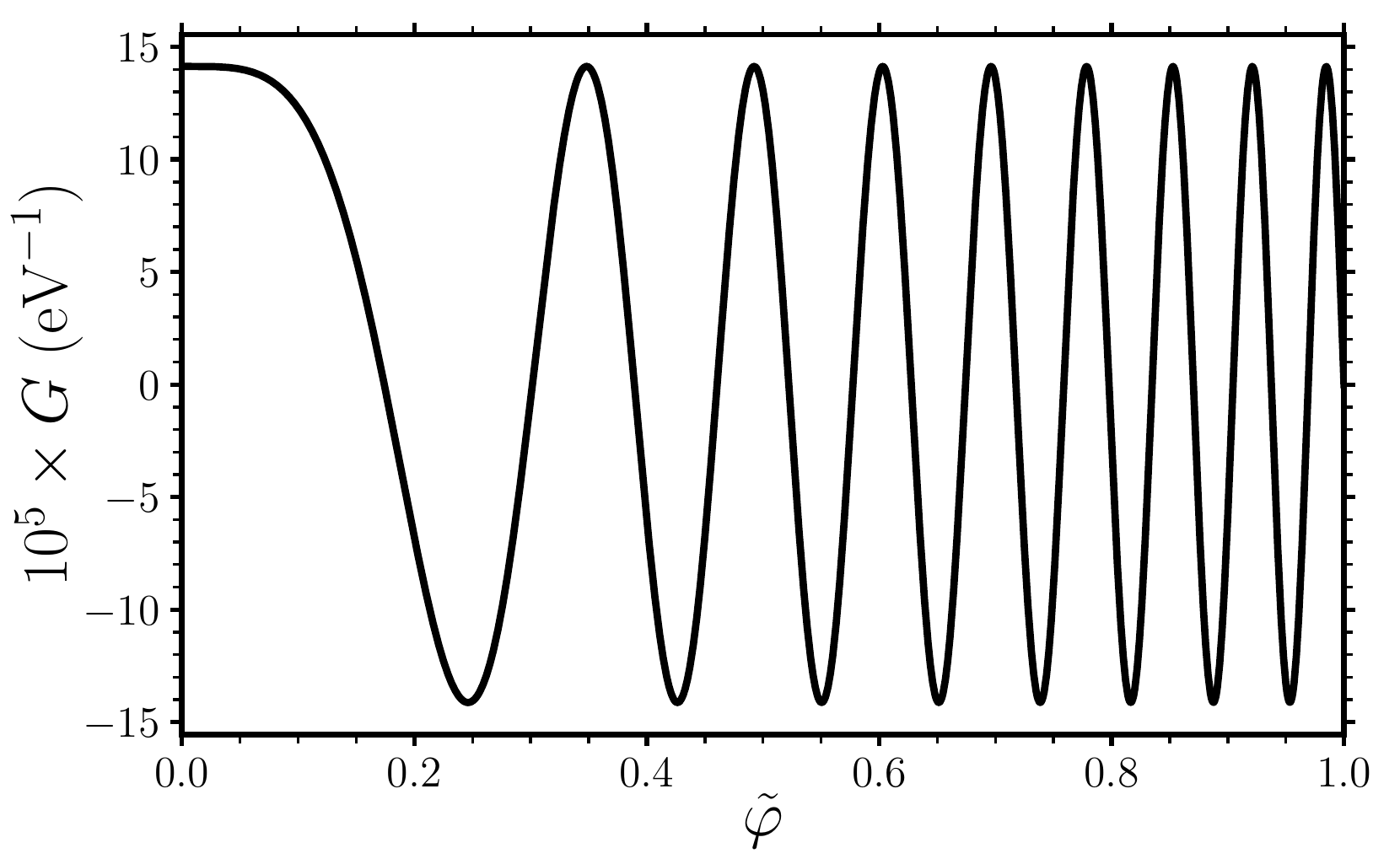} 
        \caption{$G = \g B_x/\omega_{\rm pl}^2$ as a function of $\tilde \varphi= \varphi/(\omega_{\rm pl}^2(R) R)$.}
    \vspace{4ex}
  \end{subfigure}
  \begin{subfigure}[t]{\fourpanelwidth\linewidth}
          \includegraphics[width=\linewidth]{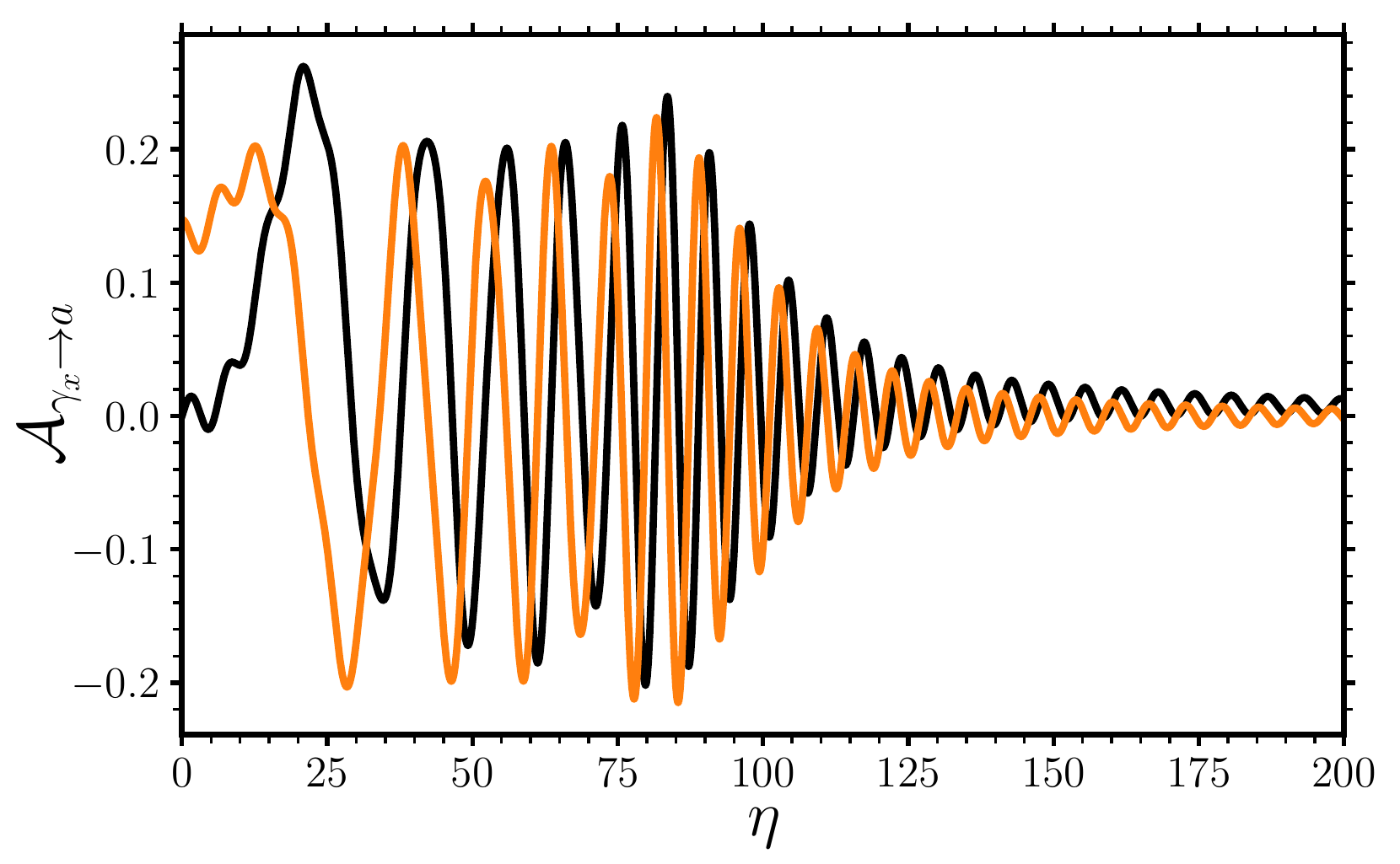} 
        \caption{Re(${\cal A}_{\gamma_x \to a})={\cal F}_s(G)$ (black), and Im(${\cal A}_{\gamma_x \to a})=-{\cal F}_c(G)$ (orange) as functions of  $\eta = \lambda \omega_{\rm pl}^2(R) R$.}
    \vspace{4ex}
  \end{subfigure} 
  \begin{subfigure}[t]{\fourpanelwidth\linewidth}
      \includegraphics[width=\linewidth]{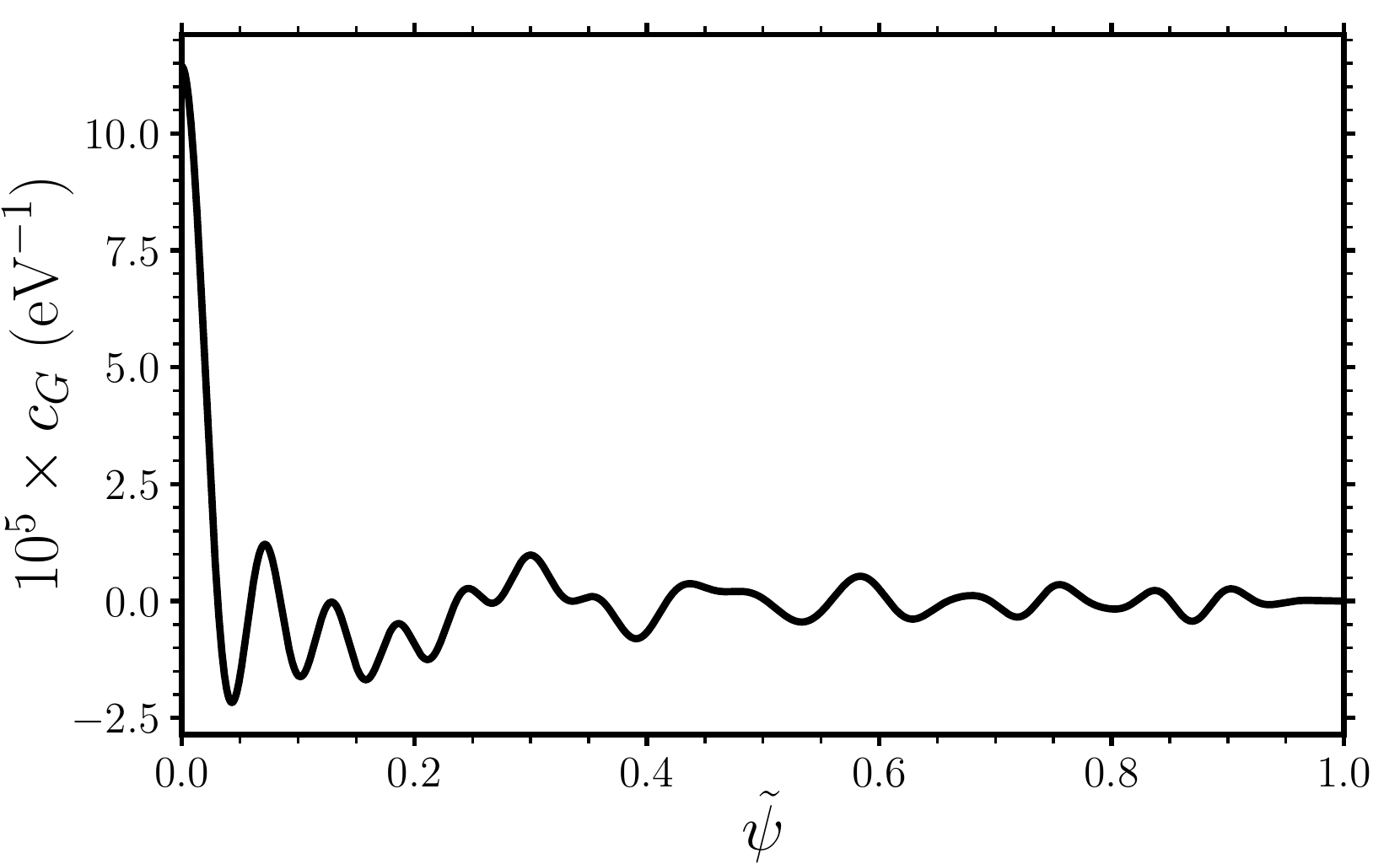} 
        \caption{The autocorrelation function $c_G(\tilde \psi)$ where $\tilde \psi= \psi/(\omega_{\rm pl}^2(R) R)$.}
    \vspace{4ex}
  \end{subfigure}
  \begin{subfigure}[t]{\fourpanelwidth\linewidth}
       \includegraphics[width=\linewidth]{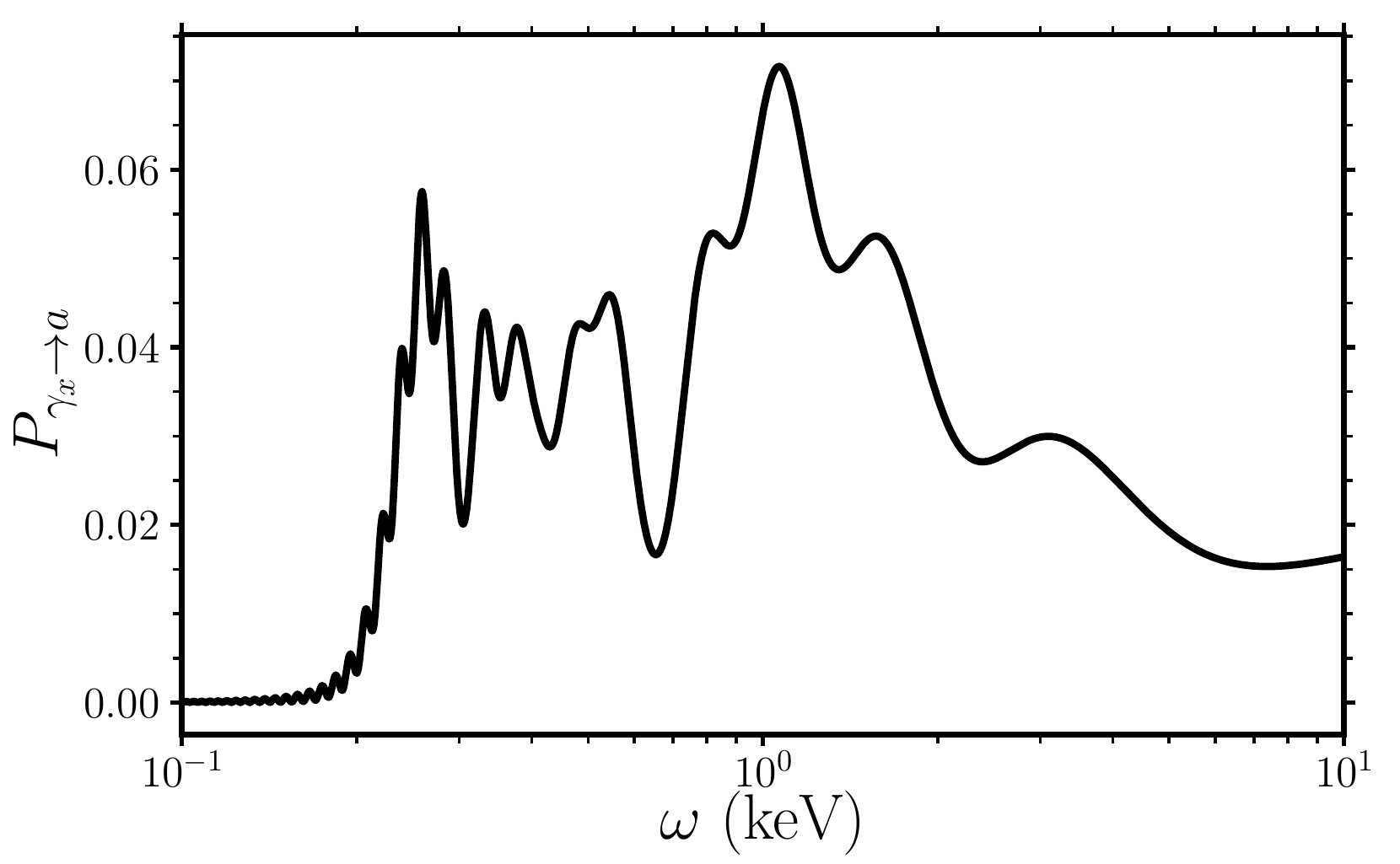} 
        \caption{Conversion probability as a function of energy.}
    \vspace{4ex}
  \end{subfigure} 
    \caption{Conversion probability of photons and massless axions (Example 7, cf.~\ref{sec:analytic_warmup}) with a magnetic field given by equation \eqref{eq:Bmasslessexpl} and a plasma frequency given by \eqref{eq:omegapl_example}. Here $\g=10^{-11}~{\rm GeV}^{-1}$, $B=1~\mu{\rm G}$, $R=50$ kpc, $\tilde k= 33\pi/2$ and $n_e(R)=10^{-3}~{\rm cm}^{-3}$.
    }
    \label{fig:massless_example}
\end{figure*}

\subsection{Resonant contributions: {\color{dm} standard stationary phase approximation} and $\omega/m_a$ enhancement}
\label{sec:res}
{\color{dm}
In this section we provide context for the  general discussion on resonant conversion in section \ref{sec:res2}, and point out the subtleties relating to the use of the stationary phase approximation when evaluating resonant amplitudes, that can lead to large inaccuracies at sufficiently high energies.

The resonance condition, $m_a = \omega_{\rm pl}(z^*)$, implies that the eigenvalues of $H_0$ are degenerate at the resonance point, and the splitting of the eigenvalues of the total Hamiltonian is only due to the off-diagonal interaction $H_I$. This corresponds to an `avoided level crossing' and under certain assumptions, transitions occurring at avoided level crossings can be solved non-perturbatively in $\g$. In our notation, if one assumes that $m_a^2-\omega_{\rm pl}^2(z)$ varies linearly with $z-z^*$, that the magnetic field is constant, and that the boundaries of integration can be taken to $z \to \pm \infty$, then  the conversion probability is described by the  Landau-Zener formula \cite{Landau, Zener}:
\beq
P^{\rm LZ}_{\gamma_x \to a} = 1- e^{-2\pi \gamma} \, ,
\label{eq:PLZ}
\eeq
where $\gamma = \Delta^2_x(z^*)/|\Delta_\gamma'(z^*)|$. When $\gamma \gg 1$, the conversion probability is non-perturbatively large and referred to as `adiabatic'.   In the opposite, `non-adiabatic' regime, $\gamma \ll 1$ and the exponential can be approximated order-by-order in $\gamma$, with the linear-order probability given by:
\beq
P^{\rm LZ}_{\gamma_x \to a} \approx 2 \pi \gamma = \frac{2 \pi \Delta_x^2}{|\Delta_\gamma'|} \, .
\label{eq:LZapprox}
\eeq
This expression is equivalent to what one obtains by 
applying the stationary phase approximation to a perturbative solution of the Schrödinger-like equation, as  we now show. 

}


 The premise of the method of stationary phase is that integrals of rapidly oscillating functions, such as \eqref{eq:amp1}, tend to cancel, and so the dominant contribution is expected from where the phase is approximately constant.  Suppose that  the amplitude $ {\cal A}^{\rm res}_{\gamma_x \to a}$ has a point of stationary phase at $z^*$, so that $\Phi'(z^*)= \lambda \varphi'(z^*)=0$. In the limit of $\lambda \to \infty$, the amplitude is dominated by the resonant contribution, which can be evaluated by Taylor expanding $\varphi(z)$ to second order and  treating the non-exponential part of the integrand as constant:
\begin{equation}
\begin{split}
i {\cal A}^{\rm res}_{\gamma_x \to a} &=
\int_{0}^{z}dz'\,  \Delta_x(z') \, e^{i\lambda \varphi(z')}
\\
&\approx
\Delta_x(z^*)
\int_{-\infty}^{\infty}dz'\, e^{i\lambda \left(\varphi(z^*) + \tfrac{1}{2} \varphi''(z^*) (z-z^*)^2 \right)} \\
&=
  \Delta_x(z^*) L_{\rm res}(z^*)e^{i \left(\lambda \varphi(z^*)  - \tfrac{\pi}{4} {\rm sign}(\varphi''(z^*))\right)} \, ,
\label{eq:ampres}
\end{split}
\end{equation}
where 
\beq
L_{\rm res}(z^*) = \sqrt{\frac{2\pi}{\lambda |\varphi''(z^*)|}}
=
\sqrt{\frac{2\pi  }{|\Delta_\gamma'(z^*)
|}} 
 \, .
\label{eq:Lres}
\eeq
Using the explicit expressions for the matrix elements in \eqref{eq:Deltas}, we can write the resonant conversion lengths as
\beq
L_{\rm res} =
\sqrt{
\frac{2 \pi \omega}{\omega_{\rm pl} \omega_{\rm pl}'}
} = \sqrt{
\frac{2 \pi }{\omega_{\rm pl}'} \frac{\omega}{m_a}} = L \sqrt{ \frac{\omega}{m_a}} \, ,
\eeq
where in the last step we have defined the energy-independent conversion length {\color{dm}
$$L = \sqrt{\frac{2\pi}{|\omega_{\rm pl}'|}} \, ,$$ 
and set $\omega_{\rm pl}(z^*) = m_a$. 
 The conversion probability is given by
\begin{align}
P^{\rm res}_{\gamma_x \to a} &= \left( L_{\rm res} \Delta_x \right)^2  
= \frac{2\pi \Delta_x^2}{|\Delta_\gamma'|} 
= P_0 \frac{\omega}{m_a} \, , \nonumber \\
{\rm for}~~~P_0&= \frac{\g^2 B^2}{4} \frac{2\pi}{\omega_{\rm pl}'} \, ,
\label{eq:Pres}
\end{align}
with all quantities evaluated at the point of stationary phase. Clearly, $P^{\rm res}_{\gamma_x \to a}  = 2 \pi \gamma \approx P^{\rm LZ}_{\gamma_x \to a}$, so that the probability obtained from the perturbative amplitude identical to the Landau-Zener probability in the non-adiabatic limit, cf.~equation \eqref{eq:LZapprox}. This is unsurprising, since these expressions are derived under the same assumptions, although taken in different order. 

An important aspect of equation \eqref{eq:Pres} is that the conversion length is energy-dependent, and can grow large for highly relativistic axions.
As we will now discuss, this signals the breakdown of the stationary phase method at sufficiently high energies, and makes  equation \eqref{eq:Pres}  inaccurate. In this section,  we propose an approximate, simple formula that regulates the probability at high energies, and in section \ref{sec:res2}, we instead carefully re-derive the amplitude and find an analytic formula that is accurate, but more complicated.  }

The factor of $\omega/m_a$ in equation \eqref{eq:Pres} comes from the square-root dependence of the resonance length on the energy, or equivalently the $1/\sqrt{\lambda}$ in equation \eqref{eq:Lres}. The stationary phase approximation is expected to be accurate for $\lambda \to \infty$, but in the relativistic case, we are interested in small $m_a \lambda$, where equation \eqref{eq:ampres} is no longer guaranteed to be applicable. Indeed, the first step of equation \eqref{eq:ampres} approximates the non-exponential part of the integrand, in our case $\Delta_x(z)$, as a constant {\color{dm} over distances large} compared to  $L_{\rm res}$. This is reasonable if $L_{\rm res}$ is sufficiently small, but less so if the resonance length extends over astronomical distances, as is the case when $\omega/m_a$ is sufficiently large. When the magnetic field varies significantly over the resonance length, the amplitude may partially cancel and the stationary phase approximation becomes inaccurate. 

A better approximation of the resonant amplitude for relativistic axion-photon conversion can be obtained by replacing $L_{\rm res}$ by a regulated resonance length, $L_{\rm reg}$. This corresponds to assuming that contributions to the amplitude further away than a distance of $L_{\rm max}/2$ from the resonance point cancels:
\beq
L_{\rm reg}=
\begin{cases}
L \sqrt{ \frac{\omega}{m_a}} & {\rm if ~\omega < \omega_{\rm max}} \\
L_{\rm max} & {\rm if ~\omega \geq \omega_{\rm max}} \,  \label{eq:Lreg} ,
\end{cases} 
\eeq
where $\sqrt{\omega_{\rm max}/m_a}= L_{\rm max}/L$. This expression involves the free parameter $L_{\rm max}$, which can be approximated by the length scale associated with variations in the non-exponential part of the integrand, in our case:  $L_{\rm max}=\Delta_x(z^*)/|\Delta_x'(z^*)|$. We expect equation \eqref{eq:Lreg} to be accurate for $\omega \ll \omega_{\rm max}$. For $\omega \geq \omega_{\rm max}$, using $L_{\rm reg}$ instead of $L_{\rm res}$ in equation \eqref{eq:Pres} for the conversion probability should lead to an improved estimate, though a more detailed calculation is required for accuracy. {\color{dm} Moreover, we expect this simple regularisation to be more accurate than simply discarding contributions where $L_{\rm res}$ becomes large compared to other physical scales of the problem, as done in e.g.~\cite{Mirizzi:2009nq, Witte:2021arp}.}

Two conceptual points emerge from this analysis. First, with all else the same, resonant axion-photon conversion of relativistic axions is more relevant in environments with \emph{slowly} varying magnetic fields.  That is, the resonance length is not only dependent on the plasma frequency as in the standard analysis, but also the magnetic field properties. Second, the spectral shape of the resonant amplitude is in all cases very simple: it increases with $\sqrt{\omega}$ up to a maximal energy that is determined by the spatial variation of the magnetic field, after which it is expected to plateau.

{\color{dm}
To contextualise these results, we close this section with  a brief review of previous work on relativistic axion-photon conversion in astrophysical environments, focussing on the extent to which resonant conversion has been accounted for.

Reference \cite{Yanagida:1987nf} discussed  resonant production of axions in cosmological magnetic fields, considering primarily 
the Cosmic Microwave Background (CMB) as the source.  In the adiabatic limit considered in \cite{Yanagida:1987nf}, the full expression of \eqref{eq:PLZ} need to be used to find the conversion probability, so that the linear dependence of $L^2_{\rm res}$ on the mode energy (cf.~equation \eqref{eq:Pres}) appears in the exponential. The issue of the validity of the Landau-Zener approximation when accounting for the spatial variation of primordial, cosmic magnetic fields was not discussed in \cite{Yanagida:1987nf}. This work was later updated in \cite{Mirizzi:2009nq}, which used the stringent limits on CMB spectral distortions from COBE/FIRAS to constrain resonant axion-photon conversion, now pressed into the non-adiabatic limit. The enhancement of the resonance length  with $\sqrt{\omega/m_a}$ appears in \cite{Mirizzi:2009nq}, but contributions where $L_{\rm res}$ was larger than the estimated coherence scale of the magnetic field were simply discarded.      

At gamma-ray energies, reference \cite{Hochmuth:2007hk} provided an overview of astrophysical environments that could lead to substantial conversion probabilities, considering in particular adiabatic, resonant mixing. More recently, reference  \cite{Fortin:2018ehg} discussed the mixing of relativistic axions from magnetars with hard X-rays, and considered also the resonant case. These references did not explicitly discuss the dependence of $L_{\rm res}$ on the mode energy. 

Axion-photon conversion at X-ray energies in galaxy clusters can be highly efficient, and has been studied by several groups at X-ray and gamma-ray energies. For axion masses that fall in the range between the minimal and maximal plasma frequencies in the cluster, the conversion amplitude will include one or more resonant contributions, in addition to non-resonant contributions. The relevant mass range has been studied in X-ray searches for axions \cite{ Wouters:2013hua, Conlon:2013txa, Berg:2016ese,Marsh:2017yvc,Chen:2017mjf, Conlon:2017qcw, Reynolds:2019uqt,  Reynes:2021bpe} (but not gamma-ray searches, which have focused on the higher-mass region where non-resonant conversion can induce large irregularities). However, many studies have neglected the resonant contribution, either explicitly or implicitly.  In part, this is an immediate consequence of the reliance on cell models, or comparatively coarsely sampled GRF models, for which the mixing equations have been solved numerically. In such models, the resonance condition is typically not satisfied in any patch, and the  resonant contribution is effectively omitted. If by chance the resonance condition is nearly satisfied in a patch, but the cell length is large compared to $L_{\rm res}$,  the resonant conversion probability is overestimated, cf.~\cite{Reynes:2021bpe} for a recent discussion. This issue has been briefly discussed in \cite{Marsh:2017yvc} using the single-domain formula, where it was argued that unless the coherence length of the magnetic field  grows very large near the resonance point, the resonant contribution is negligible. Our discussion below will make this argument much  more precise, but will confirm the general conclusion.    

Finally, a related issue appears when considering the resonant conversion of \emph{non-relativistic} axions and photons. 
The mixing is still governed by the  Schrödinger-like equation, but the resonance length is now enhanced by a factor of $\sim 1/\sqrt{v_a}$. The perturbative amplitude can be evaluated using the standard stationary phase approximation, cf.~\cite{Hook:2018iia, Battye:2021xvt, Witte:2021arp} in the case of dark matter conversion in neutron star magnetospheres. However, the standard stationary phase approximation again suffers analogous theoretical corrections,  as recently discussed in  \cite{Millar:2021}.


}

\subsection{Resonant contributions: general treatment}
\label{sec:res2}
The standard treatment of the resonant amplitude assumes that it suffices to make a quadratic  Taylor expansion of the phase, while extending the limits of integration to $\pm\infty$. This procedure is motivated by the assumption that contributions from regions far away from the resonance point cancel, and erases information about non-resonant contributions. However, non-resonant axion-photon conversion can be non-negligible or even dominate resonant contributions, and a careful calculation requires both types of contributions to be accounted for. We here show how this can be done, and that the width of the finite region attributed to the resonant contribution is arbitrary if taken sufficiently small, and the non-resonant contributions are always important for an accurate result.

To evaluate the resonant amplitude \eqref{eq:Ares}, we treat the non-oscillatory part of the integrand as slowly varying (or constant), and Taylor expand the phase $\lambda \varphi(z)$ to second order around $z_\star$. The integral runs from $z^\ell= z^* - \Delta z$ to $z^u = z^* + \Delta z$. However, to make the build-up of the resonant contribution more apparent, we replace the upper limit by an intermediate  $z$ satisfying $z^\ell < z \leq z^u$. We then have that
\begin{widetext}
\beq
\begin{split}
i {\cal A}^{\rm res}_{\gamma_x \to a} &= \Delta_x (z^*) e^{i \lambda \varphi(z^*) }
\int_{z^*-\Delta z}^{z} dz' e^{i \lambda \frac{\varphi''(z^*)}{2} (z-z^*)^2} 
\\
&=
\frac{1}{2} \Delta_x (z^*)  L_{\rm res}(z^*) e^{i (\lambda \varphi(z^*)- \tfrac{\pi}{4} {\rm sign}(\varphi''(z^*)))}
\left[ {\rm Err}\left(e^{-i\pi/4} \sqrt{\pi}
\frac{ z-z^{*}}{L_{\rm res}}
\right) +{\rm Err}\left(e^{-i\pi/4} \sqrt{\pi}
\frac{ \Delta z}{L_{\rm res}}
%
%
\right) \right]
\\
&\xrightarrow{z\to z^u}
\Delta_x (z^*)  L_{\rm res}(z^*) e^{i (\lambda \varphi(z^*)- \tfrac{\pi}{4} {\rm sign}(\varphi''(z^*)))}\, 
 {\rm Err}\left( e^{-i\pi/4} \sqrt{\pi}
\frac{ \Delta z}{L_{\rm res}}\right)
\, .
\label{eq:staphasefinite}
\end{split}
\eeq
\end{widetext}
The error function  grows linearly for small argument, and performs damped oscillations as the argument increases, and finally asymptotes to 1, consistently with equation \eqref{eq:ampres}. So far, we have kept $\Delta z$ arbitrary, however, we expect that accurate results are obtained when $\Delta z/L \ll 1$ so that $\Delta z/L_{\rm res} \ll 1$ for all relevant energies, and the Taylor expansion of the phase can be trusted. In this limit, the resonant amplitude simplifies further: 
\beq
\begin{split}
i {\cal A}^{\rm res}_{\gamma_x \to a} &= \frac{2 i e^{i \lambda \varphi(z^*)}}{\sqrt{\pi}}
\Delta_x (z^*)\, \Delta z \, .
\end{split}
\eeq
Note that $\Delta z$ is assumed small but is still arbitrary, and varying $\Delta z$ shuffles around contributions to the amplitude between the resonant and non-resonant parts.  This implies that the non-resonant contributions  will generically always be important: 
when insisting on taking $\Delta z$ small enough for the Taylor expansion of the phase to be a good approximation, the  resonant contributions never  dominate the non-resonant contributions.

\subsection{Non-resonant contribution}
\label{sec:generalnonres}
We now turn to the non-resonant amplitudes of equation \eqref{eq:Anonres}. The terms are of the form  
\beq 
I_i= \int_{z_{i-1}^u}^{z_{i}^\ell} dz' \,  \Delta_x(z') e^{i \lambda \varphi(z')} \label{eq:ampex}
\eeq
and $\varphi$ is either monotonically increasing or decreasing in the entire interval $[z_{i-1}^u, z_{i}^\ell]$. To interpret this integral as a Fourier transform, we would like change integration variable from $z'$ to $\varphi$, and extend the range of $\varphi$ to $\pm \infty$. 

A suitable extension, $\varphi_i(z')$, of $\varphi$ can be defined by having $\varphi_i$ coincide with $\varphi$ for $z\in [z_{i-1}^u, z_{i}^\ell]$, and extend linearly outside this range:
\begin{align}
\varphi_i(z_{i-1}^u) &= \varphi(z_{i-1}^u) \, ,\\
\varphi'_i(z') &=
\begin{cases}
\varphi'(z_{i-1}^u) & {\rm for}~z' \leq z_{i-1}^u \\
\varphi'(z') & {\rm for}~z_{i-1}^u \leq z' \leq z_{i}^\ell  \\
\varphi'(z_{i}^\ell) & {\rm for}~ z' \geq z_{i}^\ell \, .  
\end{cases}
\end{align}
With this definition, $\varphi_i$ is continuous and has a continuous first derivative, and an unbounded range for $z'\in (-\infty, \infty)$. Clearly, this is a one-to-one map from $z'$ to $\varphi_i$, which provides the change of coordinates that we seek. We furthermore denote the  even extension of $\Delta_x(z')$  by $\Delta^e_B(z')$, defined for $z'\in (-\infty, \infty)$.  Equation \eqref{eq:ampex} can now be written as
\beq
\begin{split}
I_i&=\int_{z_{i-1}^u}^{z_{i}^\ell} dz' \,  \Delta_x(z') e^{i \lambda \varphi(z')}  =\\
&=\int_{-\infty}^\infty dz' \, W_{[z_{i-1}^u, z_{i}^\ell]}(z') \Delta^e_B(z') e^{i \lambda \varphi_i(z')} \, . 
\end{split}
\eeq
Using $d\varphi_i = \varphi'_i dz'$, this may be written as
\beq
I_i(\lambda)= \int_{-\infty}^\infty d\varphi_i  \,
W_{[\varphi_{\rm min}, \varphi_{\rm max}]}(\varphi_i)
 \frac{\Delta_{B, i}(\varphi_i)}{|\varphi_i'|}  e^{i \lambda \varphi_i} \, ,  \label{eq:Fgen1}
\eeq
where for a monotonically increasing phase we have $\varphi_{\rm min} = \varphi(z_i^u)$ and 
$\varphi_{\rm max} = \varphi(z_{i+1}^\ell)$, and for a decreasing phase we have  $\varphi_{\rm min} = \varphi(z_{i+1}^\ell)$  and $\varphi_{\rm max} = \varphi(z_i^u)$. We've slightly abused the notation to write $\Delta^e_B(z')= \Delta^e_B(z'(\varphi_i))=  \Delta_{B, i}(\varphi_i)$.   Equation \eqref{eq:Fgen1} shows that $I_i(\lambda)$ is, up to a factor of $2\pi$, an inverse  Fourier transform of the function 
$G_i(\varphi_i)$, defined as
\beq
G_i(\varphi_i) = W_{[\varphi_{\rm min}, \varphi_{\rm max}]}(\varphi_i) f_i(\varphi_i) \, , ~~~{\rm for }~f_i(\varphi_i) = 
 \frac{\Delta_{B, i}(\varphi_i)}{|\varphi_i'(\varphi_i)|}   
\, .
\eeq
Given that $\varphi_i$ is now merely a coordinate, we can relabel it by $\phi$, dropping the $i$, and write the total amplitude from \eqref{eq:Anonres} concisely as
\beq
\begin{split}
&i {\cal A}^{\rm non-res}_{\gamma_x \to a} =  \sum_{i=1}^{N+1} \int_{z_{i-1}^u}^{z_{i}^\ell} dz' \,  \Delta_x(z') e^{i \lambda \varphi(z')}
=\\
&=\int_{-\infty}^{\infty} d\phi\,
G(\phi) e^{i \lambda \phi}  = 2 \pi\, \hat G^{-1}(\lambda) = \hat G(-\lambda) \, ,
\label{eq:AFouriergen}
\end{split}
\eeq
where 
\beq
G(\phi) = \sum_{i=1}^{N+1}G_i(\phi) = \sum_{i=1}^{N+1}
W_{[\varphi_{\rm min}^i, \varphi_{\rm max}^i]}(\phi)
 f_i(\phi)  \, .
\eeq
In equation \eqref{eq:AFouriergen}, the circumflex denotes the ordinary, complex Fourier transform, for which we use the convention: 
\begin{align}
    {\cal F}(G) &= \hat G(\lambda) = \int_{-\infty}^{+\infty} d\phi\, G(\phi)\, {\rm exp}(-i\lambda \phi) \, . \\
     {\cal F}^{-1} (G) & = \hat G^{-1}(\lambda)=\frac{1}{2\pi} \int_{-\infty}^{+\infty} d\phi\, G(\phi)\, {\rm exp}(i\lambda \phi) \, \, .
\end{align}
We note that the transform of equation \eqref{eq:AFouriergen} is well-defined for any real $\lambda$, but only positive values have the physical interpretation as inverse energies: $\lambda=1/\omega$. Furthermore, this equation makes it evident that the non-resonant amplitude \emph{in general} can be understood as (an inverse) Fourier transform of $G$, when expressed as a function of the wavelength. 

Clearly, questions about the oscillations in the amplitude, and more specifically its \emph{spectrum}, map directly to questions of the real-space properties of $G$. The spectrum of the amplitude is given by its Fourier transform:
\beq
{\cal F}\left( i{\cal A}^{\rm non-res}_{\gamma_x \to a}\right) 
= 2\pi i\,  \int_{-\infty}^{\infty} d\lambda\,  {\cal A}^{\rm non-res}_{\gamma_x \to a}(\lambda) e^{-i \psi \lambda} =  2\pi \, G(\psi) \, .
\eeq
The coordinate $\psi$ has dual interpretations: it is the frequency of probability-oscillations in wavelength-space, and it's the phase corresponding to a set of positions along the trajectory. The map to position space is one-to-many and given by $\phi \to \varphi_i \to z'$.  
%
%
Thus, the Fourier transformed amplitude reads off, and sums, the values of $\Delta_x/|\varphi'|$ at up to $N+1$ locations along the line of sight. Clearly, if $\psi$ falls outside the range of all window functions in $G$, there is no transition amplitude.

The  total conversion probability from the non-resonant contributions is given by the power spectrum of $G$
\beq
P^{\rm non-res}_{\gamma_x \to a}(\lambda) = \left| \hat G (-\lambda) \right|^2  
= \sum_{i=1}^{N+1} P_i(\lambda) + 2 \sum_{i<j =1}^{N+1} P_{ij}(\lambda) \, , 
\eeq
where in the last line we've defined the individual probabilities from each regions, $P_i= |\hat G_i|^2$, and the interference terms, $P_{ij}= {\rm Re}(\hat G_i \hat G^*_j)$.  
The Wiener–Khintchine theorem can again be applied to express the conversion probability as the Fourier transform of the auto-correlation function of $G$:
\begin{equation}
\begin{split}
P^{\rm non-res}_{\gamma_x \to a}(\lambda) &=  \left| \hat G (-\lambda) \right|^2 = \hat c^{-1}_G(\lambda) \, ,
\end{split}
\label{eq:Pgeneral}
\end{equation}
where
\beq
c_G(\psi)=\int_{-\infty}^{\infty} d\phi\, G(\phi) G(\phi + \psi) \, .
\eeq
Note that $c_G$ is an even function of its argument.


\subsection{Examples}
Fully solvable analytical examples are more sparse in the general case than in the simpler cases of very massive or massless axions. Here, we provide one example in which both the resonant and non-resonant contributions can be calculated exactly (to this order in perturbation theory), and which illustrates the application of the Fourier formalism in the general case.    

\subsubsection{Example 8: Resonant and non-resonant contributions}
\label{sec:generalexample}
We consider a model which is simple enough that some of the approximations of the stationary phase method actually hold exactly, and the resulting amplitude can be calculated in three ways: as a purely resonant contribution; as  an arbitrary  mix of resonant and non-resonant contributions; or as a purely non-resonant contribution. We assume the following plasma frequency profile
\begin{equation}
    \omega_{\rm pl}^{2}= m_{a}^{2}\frac{z}{z^{*}} = m_a^2\, u\, ,
\end{equation}
leading to a resonance $m_{a}=\omega_{\rm pl}$ for $u=z/z^{*}=1$. The resonance point is assumed to lie within a region with a constant magnetic field of magnitude $B_{0}$, which extends up to $u_{\rm max}>1$. 
The (dimensionful) phase coordinate is then given by 
\begin{equation}
    \varphi(z)=\frac{1}{2}\int_{0}^{z}dz'\,(\omega_{\rm pl}^{2}(z')-m_{a}^{2})=|\varphi_*|\left[(u-1)^2 -1\right]\,  ,
    \label{eq:genexvarphi}
\end{equation}
where $\varphi_* = - \frac{z^{*}m_{a}^{2}}{4}$.
Clearly, this phase is a quadratic function of the spatial coordinate around the resonance point. 

We first evaluate the amplitude by using our version of the stationary phase approximation, as derived in section \ref{sec:res2}. The resonance length is given by
\beq
L_{\rm res} = \sqrt{\frac{ \pi \omega}{\omega_{\rm pl}' m_a}} = z^* \sqrt{\frac{\pi}{|\varphi_*| \lambda}} =  z^* \sqrt{\frac{\pi}{\Phi_*}} \, ,
\eeq
where we in the last term have introduced the full phase at the resonance point: $\Phi_* = - \lambda \varphi_*$. The resonant contribution from some small region $[z^{*}-\Delta z,\, z^{*}+\Delta z]$ is now (exactly) given by
\beq 
i {\cal A}_{\rm res} =\frac{\g B_0 z^*}{2} \sqrt{\frac{i \pi}{\Phi_*}} e^{-i\Phi_*} {\rm Err}\left( \sqrt{-i \Phi_*} \frac{\Delta z}{z^*}\right) \, .
\label{eq:Aresgenexpl}
\eeq 
In this simple example where the magnetic field is constant and the phase quadratic in $z$, we can also calculate the full amplitude (i.e.~over $[0, \, z_{\rm max}]$) as a resonant amplitude by using equation \eqref{eq:staphasefinite}:
\begin{widetext}
\beq
i {\cal A}  =\frac{\g B_0 z^*}{2} \sqrt{\frac{i \pi}{\Phi_*}} e^{-i\Phi_*}
\frac{1}{2}\Bigg[
{\rm Err}\left( \sqrt{-i \Phi_*} \frac{z_{\rm max} - z^*}{z^*}\right)
+
{\rm Err}\left( \sqrt{-i \Phi_*} \right)
\Bigg] \, .
\label{eq:genex-fullresA}
\eeq
\end{widetext}
In this example, the standard treatment of the stationary phase approximation, which neglects the Error functions and  uses equation \eqref{eq:ampres}, is a good approximation when $\Phi_* \gg1$, but not otherwise. 

The transition amplitude can also be computed using the Fourier techniques developed in section \ref{sec:generalnonres}. The first step is then  to change coordinate from $z$ to $\varphi$. Inverting equation \eqref{eq:genexvarphi}, gives two branches:
\begin{equation}
 u_{\pm}(\varphi)=\left[1\pm\sqrt{1+\tilde \varphi}\right]\,  ,
\end{equation}
where we have introduced the notation $\tilde \varphi = \varphi/|\varphi_*|$.
Following the procedure outlined in section \ref{sec:generalnonres}, we divide the  integration domain in $u$ into three regions: $[0,\, 1-\delta ]$, $[1-\delta ,\, 1+\delta]$ and $[1+\delta ,\, u_{\rm max}]$. The first and last of these can be evaluated with Fourier techniques as non-resonant amplitudes, and the middle contribution is a resonant amplitude that  evaluates as in equation \eqref{eq:Aresgenexpl}. In the first region $\varphi$ drops from 0 to $-|\varphi_*| + \delta \varphi$ (while $u=u_-$), and in the third region the phase increases from $-|\varphi_*| + \delta \varphi$ to some $\varphi_{\rm fin} = \varphi(u_{\rm max})$ (while
$u=u_+$). To account for the change of measure, we need $d\varphi/dz$:
\beq
\frac{d\varphi_\pm}{dz} = 2\frac{|\varphi_*|}{z^*}(u_\pm-1) = \pm 2 \frac{|\varphi_*|}{z^*} \sqrt{1+ \tilde \varphi} \, .
\eeq
The function $f_i(\varphi_i) = \Delta_x/|\varphi'_i|$ are in this simple case identical between the two regions:
\beq
f(\tilde \varphi) =\frac{\g B_0 z^*}{4 |\varphi_*|} \frac{1}{ \sqrt{1+ \tilde \varphi}} \, .
\eeq
The full function $G(\phi)$ is then given by
\beq
G(\phi) = 
\frac{\g B_0 z^*}{4 |\varphi_*|} \frac{W_{[-1 + \delta \tilde \varphi, 0]}(\tilde \phi) + W_{[-1 + \delta \tilde \varphi, \tilde \varphi_{\rm fin}]}(\tilde \phi)}{ \sqrt{1+ \tilde \phi}} \, .
\eeq
The non-resonant amplitude is now given by the inverse Fourier transform of $G$,  and the result can be expressed using Fresenel's $C$ and $S$ integrals, incomplete Euler gamma functions, or, as we do here, error functions:
\begin{widetext}
\beq
i {\cal A}^{\rm non-res}_{\gamma_x \to a}=\frac{\g B_0 z^*}{2} \sqrt{\frac{i \pi}{\Phi_*}} e^{-i\Phi_*}
\frac{1}{2}\Bigg[
{\rm Err}\left( \sqrt{-i \Phi_*} \frac{z_{\rm max} - z^*}{z^*}\right)
+
{\rm Err}\left( \sqrt{-i \Phi_*} \right)-2{\rm Err}\left( \sqrt{-i \Phi_*}\frac{\Delta z}{z^{*}} \right)\Bigg]\, .
\label{eq:Agenex-nonres}
\eeq
\end{widetext}
To get the full amplitude, we add the resonant contribution from \eqref{eq:Aresgenexpl} to this equation, which simply cancels the last term of equation \eqref{eq:Agenex-nonres}, giving a total amplitude that agree with the resonant-only calculation of   equation \eqref{eq:genex-fullresA}. 

 \begin{figure*}[t!] 
  \begin{subfigure}[t]{\fourpanelwidth\linewidth}
        \includegraphics[width=\linewidth]{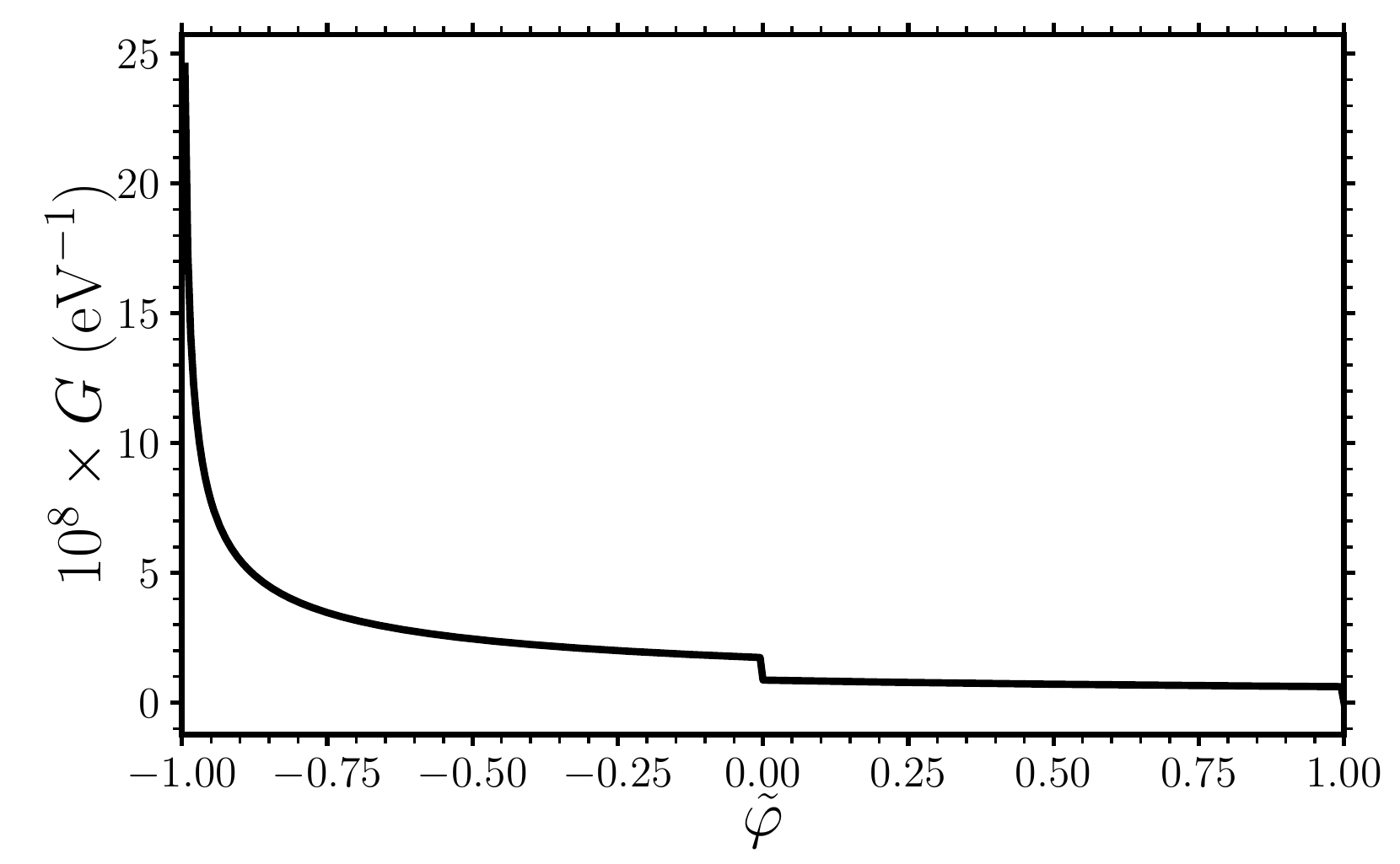} 
        \caption{$G$ as a function of the dimensionless phase $\tilde{\varphi}= \varphi/|\varphi_*|$. The resonance point is at $\tilde \varphi=-1$.}
    \vspace{4ex}
  \end{subfigure}
  \begin{subfigure}[t]{\fourpanelwidth\linewidth}
          \includegraphics[width=\linewidth]{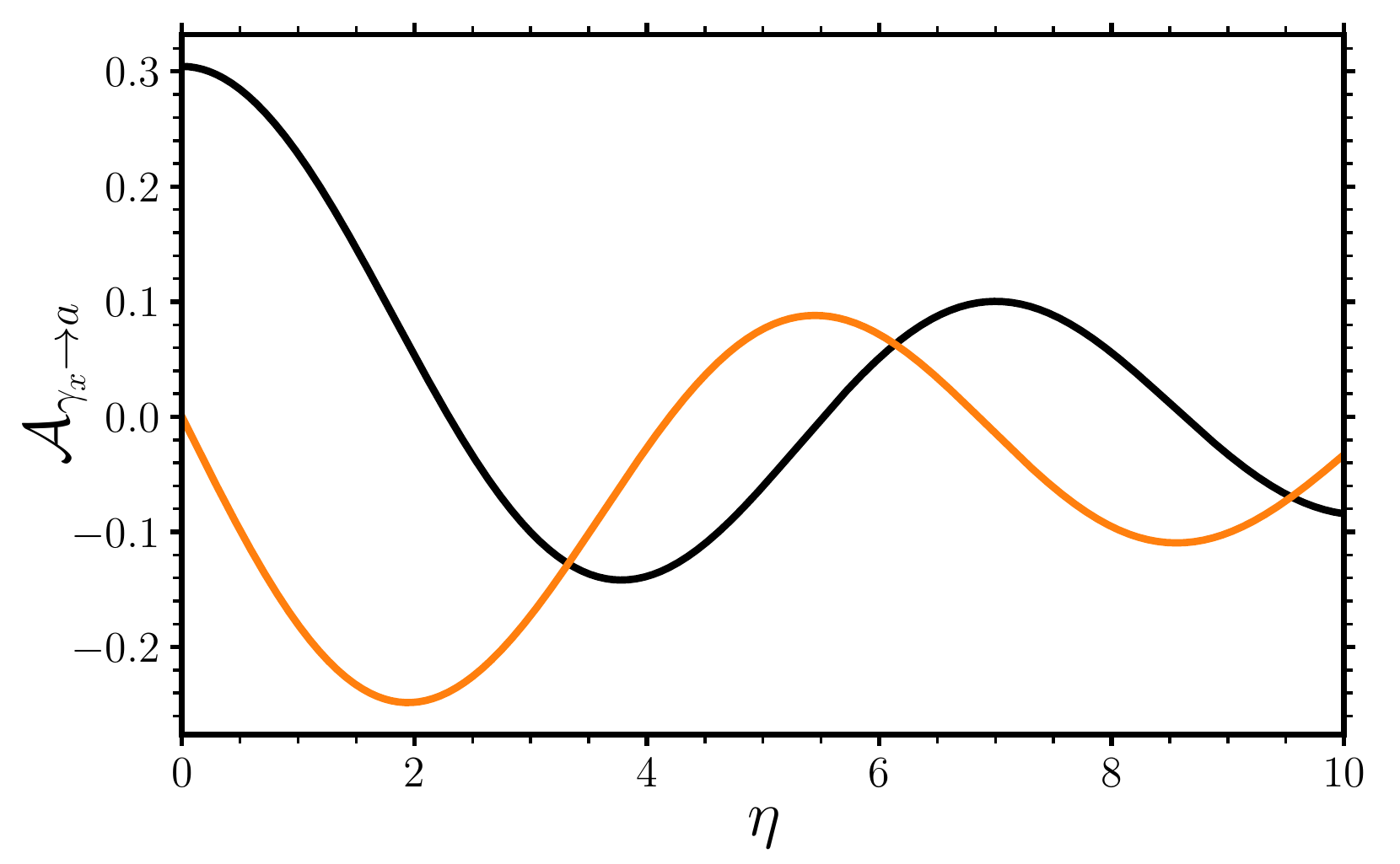} 
        \caption{Re(${\cal A}_{\gamma_x \to a})={\cal F}_s(G)$ (black), and Im(${\cal A}_{\gamma_x \to a})=-{\cal F}_c(G)$ (orange).}
    \vspace{4ex}
  \end{subfigure} 
  \begin{subfigure}[t]{\fourpanelwidth\linewidth}
       \includegraphics[width=\linewidth]{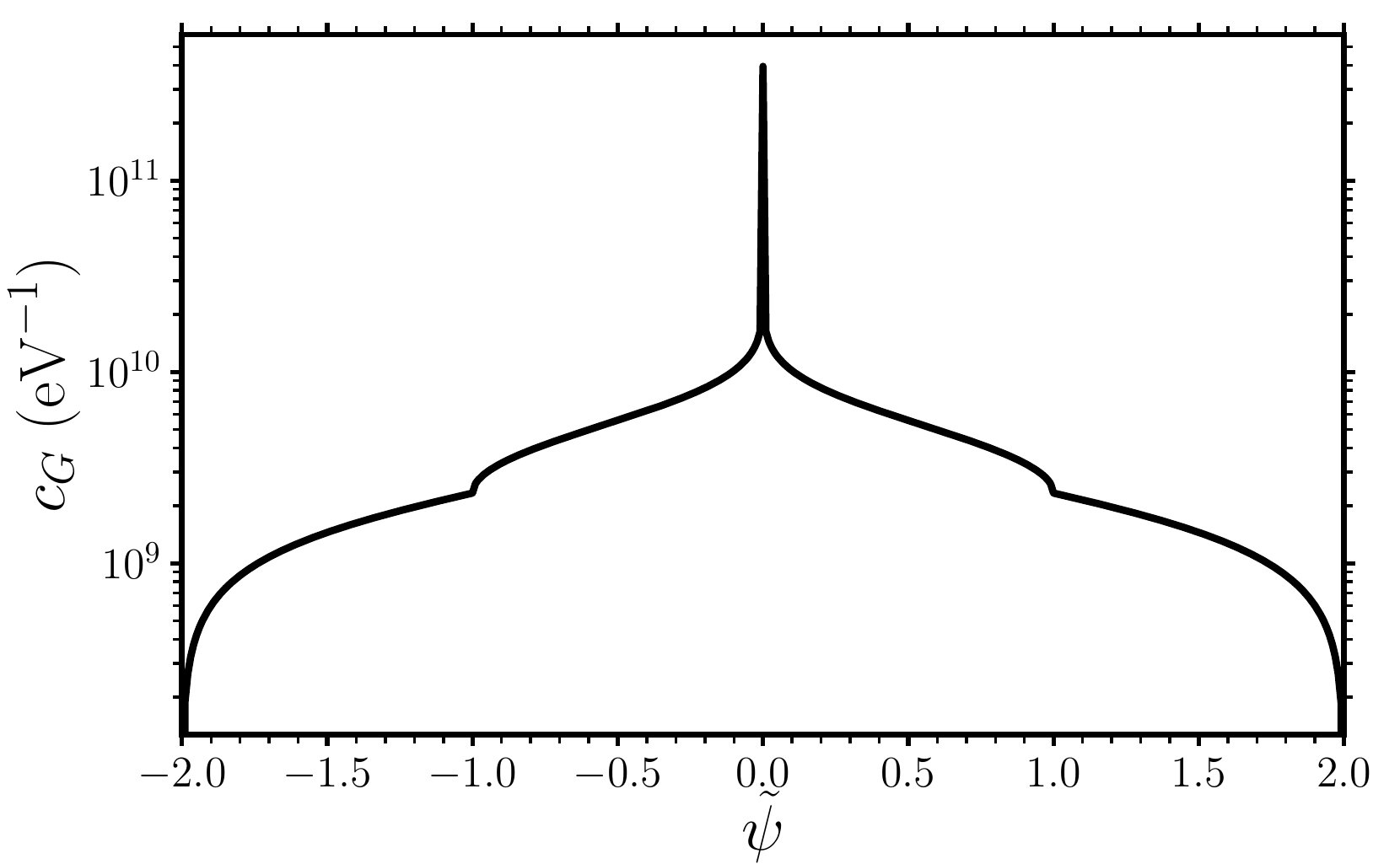} 
        \caption{Autocorrelation function of  $G$.} 
    \vspace{4ex}
  \end{subfigure}
  \begin{subfigure}[t]{\fourpanelwidth\linewidth}
       \includegraphics[width=\linewidth]{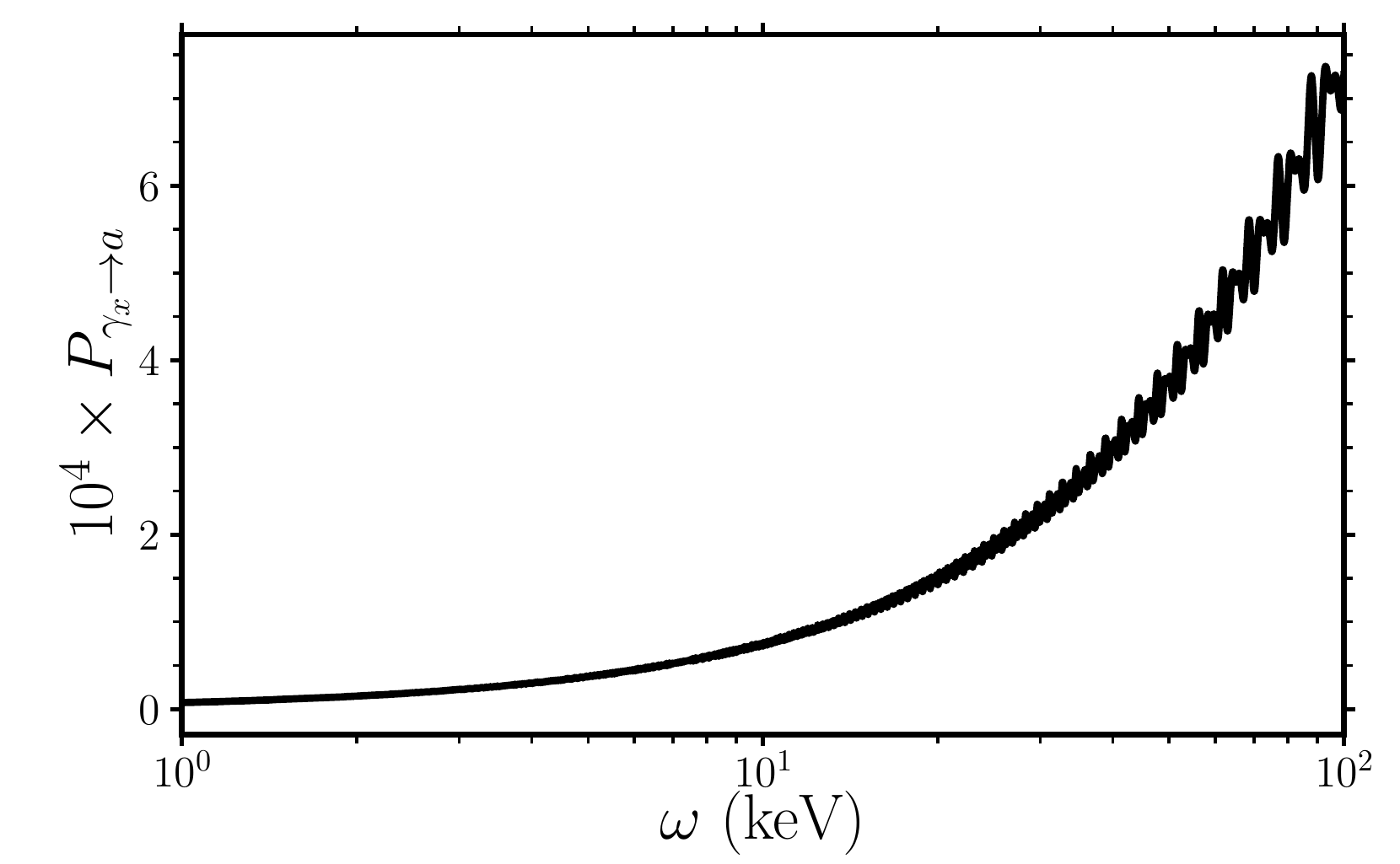} 
        \caption{Conversion probability as a function of energy.} 
    \vspace{4ex}
  \end{subfigure} 
    \caption{The axion-photon conversion probability  (lower right) for $g_{a\gamma}=10^{-11} \GeV^{-1}$, $m_{a} = 0.15$~neV, $\omega_{pl,0}=0.5$~neV and $z^{*}=10$~kpc is calculated for a constant magnetic field $B=1\, \mu G$ (upper left), as discussed in section \ref{sec:generalexample} (Example 8). 
    }
    \label{fig:fig_X2}
\end{figure*}

Finally, by taking $\Delta z \to 0$ so that there is no resonant contribution, the last term of equation \eqref{eq:Agenex-nonres} goes to zero, and the non-resonant amplitude reproduces the full result of equation \eqref{eq:Aresgenexpl}. We conclude that all three ways of calculating the amplitude are in exact agreement, and the conversion probability
$$
P_{\gamma\rightarrow a}=|\mathcal{A}_{\gamma\rightarrow a}^{\rm non-res}+\mathcal{A}_{\gamma\rightarrow a}^{\rm res}|^{2}\, ,
 $$
 is independent of the split between resonant and non-resonant.

 The conversion probability can also be calculated directly from the autocorrelation function of the function $G(\phi)$, taking $\Delta z =0$. The  autocorrelation function is given by
\begin{equation}
\begin{split}
    c_{G}(\tilde{\psi})&=\frac{g_{a\gamma}^{2}B_{0}^{2}(z^{*})^{2}}{8|\phi_{*}|^{2}}\\
    &\Bigg[\sinh^{-1}\left(\sqrt{\frac{1-\tilde{\psi}}{\tilde{\psi}}}\right)+\sinh^{-1}\left(\sqrt{\frac{1}{\tilde{\psi}}}\right)+\\
    &\hspace{3cm}+2\sinh^{-1}\left(\sqrt{\frac{2-\tilde{\psi}}{\tilde{\psi}}}\right)\Bigg]\, , 
    \end{split}
\end{equation}
for $0\le\tilde{\psi}\le1$, and
\begin{equation}
\begin{split}
    c_{G}(\tilde{\psi})&=\frac{g_{a\gamma}^{2}B_{0}^{2}(z^{*})^{2}}{4|\phi_{*}|^{2}}\sinh^{-1}\left(\sqrt{\frac{2-\tilde{\psi}}{\tilde{\psi}}}\right)\, , 
    \end{split}
\end{equation}
for $1<\tilde{\psi}\le2$. Since the autocorrelation function is an even function defined over the entire real line, this completely defines  $c_G$.

    \begin{figure}[t]
        \centering
        \includegraphics[width=\linewidth]{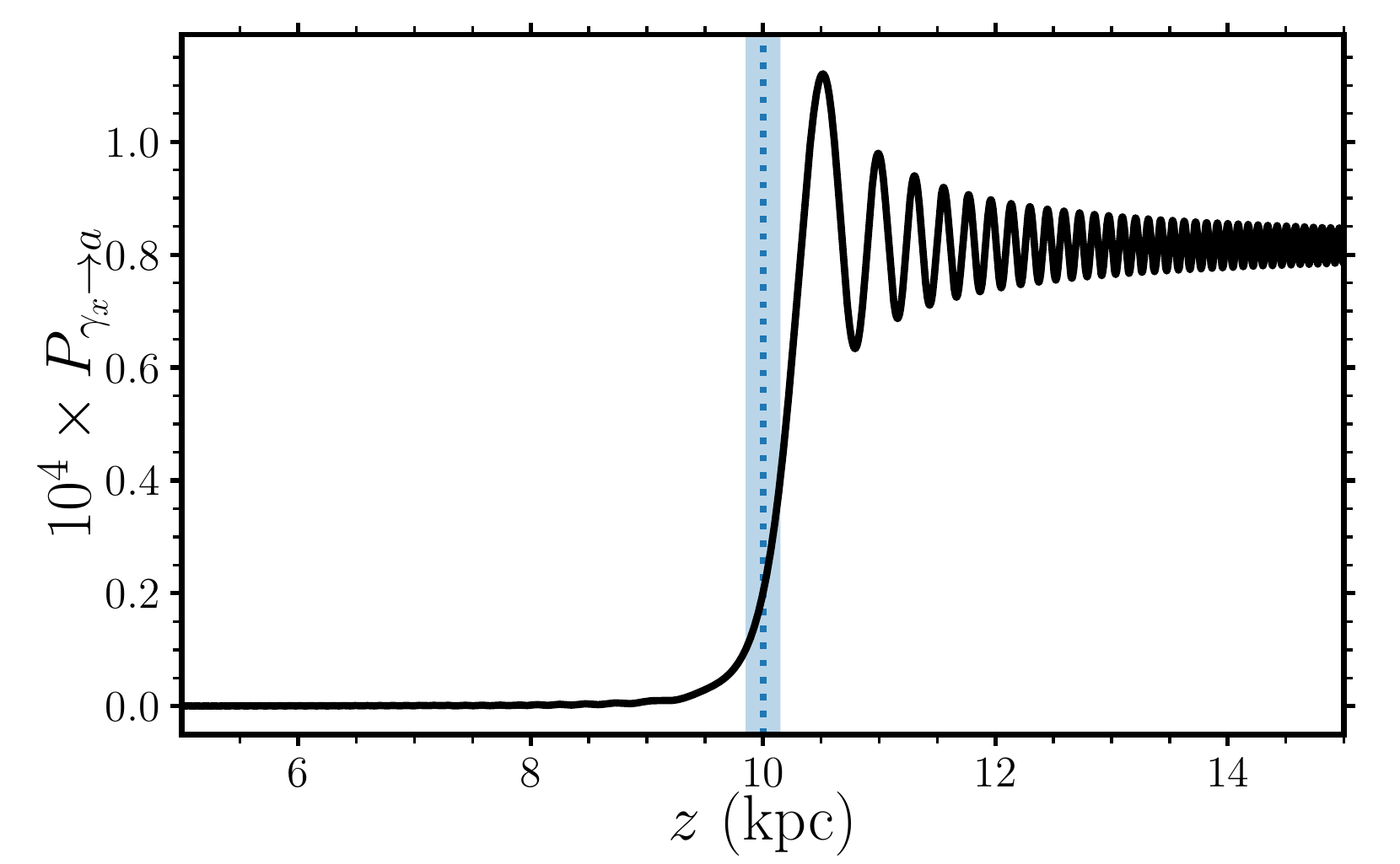} 
        \caption{Conversion probability at $\omega=10$~keV as a function of $z$ for the example of section \ref{sec:generalexample}. The blue dotted line marks the resonance point ($z^* = 10$~kpc) where $m_a=\omega_{\rm pl}$ and the blue shaded region marks the resonance length at this energy: $L_{\rm res}=0.3$~kpc.} \label{fig:general_probz}
    \end{figure}

In figure \ref{fig:fig_X2}, we show the function $G(\tilde \phi)$, the real and imaginary parts of the transition amplitude, the autocorrelation function $c_G$, as well as the resulting conversion probability. A notable feature is the discontinuity in $G(\phi)$ at $\phi=0$ where $G$ goes from having support from two functions ($G_1+G_2$) to just a single non-vanishing function ($G_2$). This is also reflected as a kink in the autocorrelation function of $G$ at $\psi = \pm1$. In figure \ref{fig:general_probz}, we show how the conversion probability builds up with distance across the resonance point. 

\section{Numerical tests of the formalism}
\label{sec:tests}
Expressing  axion-photon mixing as Fourier transforms makes it possible to leverage the highly effective numerical techniques of the Fast Fourier Transform (FFT) to determine the predictions. In this section, we show how FFT techniques can be applied to complex examples with massive ($m_a \gg \omega_p$) and  massless  ($m_a \ll \omega_p$) axions.\footnote{Similar techniques should also be straightforwardly applicable to the general case described in section \ref{sec:general}.}
We use these numerical examples to 
test the applicability of the Fourier formalism, and compare it to the traditional method of solving the Schrödinger-like equation directly. We find that the Fourier-based techniques are much faster than direct simulations (as expected), and that, in our examples, the perturbative expansion holds very well over a wide range of interesting energies and for couplings up to about an order of magnitude larger than the current observational limit. This suggests that numerical Fourier methods are superior to traditional methods when searching for axions using high quality data.

\subsection{The discrete cosine and sine transforms}
\label{sec:dct}
In  sections~\ref{sec:massive}--\ref{sec:general}, we presented a series of examples where the relevant Fourier transforms can be computed analytically. In more complex models for the magnetic field and the plasma density, this may not be possible, and we can instead make use of the discrete cosine and discrete sine transforms (DCT and DST, respectively) \cite{ahmed_dct_1974}.  The DCT and DST can be computed quickly using a modified FFT, and implementations of these algorithms are available as standard in numerical libraries such as scipy \cite{makhoul1980,scipy2020} or the GNU scientific library \cite{gough2009gnu}. We here focus on the DCT, but equivalent equations can be written for the DST. 
There are four different types of the DCT \cite{rao2014discrete}, and here we use the `type-I' definition which simply corresponds to discretising the cosine transform over a finite range, as we now discuss.

The DCT can be obtained by replacing the 
continuous coordinate  ($z$ or $\varphi$) by an evenly sampled list of points: $z_n = n\, \Delta z$ for $n=0, \ldots , N$, so that $z_{\rm max} = N\, \Delta z$. The conjugate variable ($\eta$ or $\lambda$) is  also discretised  with $\eta_m = m\, \pi/z_{\rm max}$ where $m=0,\ldots , N$.  Note that the product $z_n \eta_m = n\, m \, \pi/N$.
The DCT of type I is given by: 
\beq
\hat{f}_m = 
\frac{1}{2} \Big( f_0 + (-1)^m f_N \Big) + \sum_{n=1}^{N-1} f_n \cos\left(\frac{ \pi n m}{N}\right) \, ,
\eeq
which can be understood as the midpoint Riemann sum (divided by $\Delta z$) of the continuous cosine transform. Clearly, the resolution of the conjugate variable is $\Delta \eta = \pi/z_{\rm max}$ and the maximal value is $\eta_N=\pi/\Delta z$. In the context of axion-photon oscillations, obtaining sufficient energy resolution and coverage of the conversion probability, requires  sampling with high enough spatial resolution and out to sufficiently large values of $z_{\rm max}$. From the examples we have considered, a modestly large value of $N=10^5$ gives extremely good resolution in the X-ray band. The value $z_{\rm max}$ is naturally no smaller than the region over which the magnetic field is non-vanishing, but it can be made arbitrarily large by `zero-padding', i.e.~appending the set of $f_n$ with zero-valued data points. 


\begin{figure*}[t!] 
  \begin{subfigure}[t]{\fourpanelwidth\linewidth}
        \includegraphics[width=\linewidth]{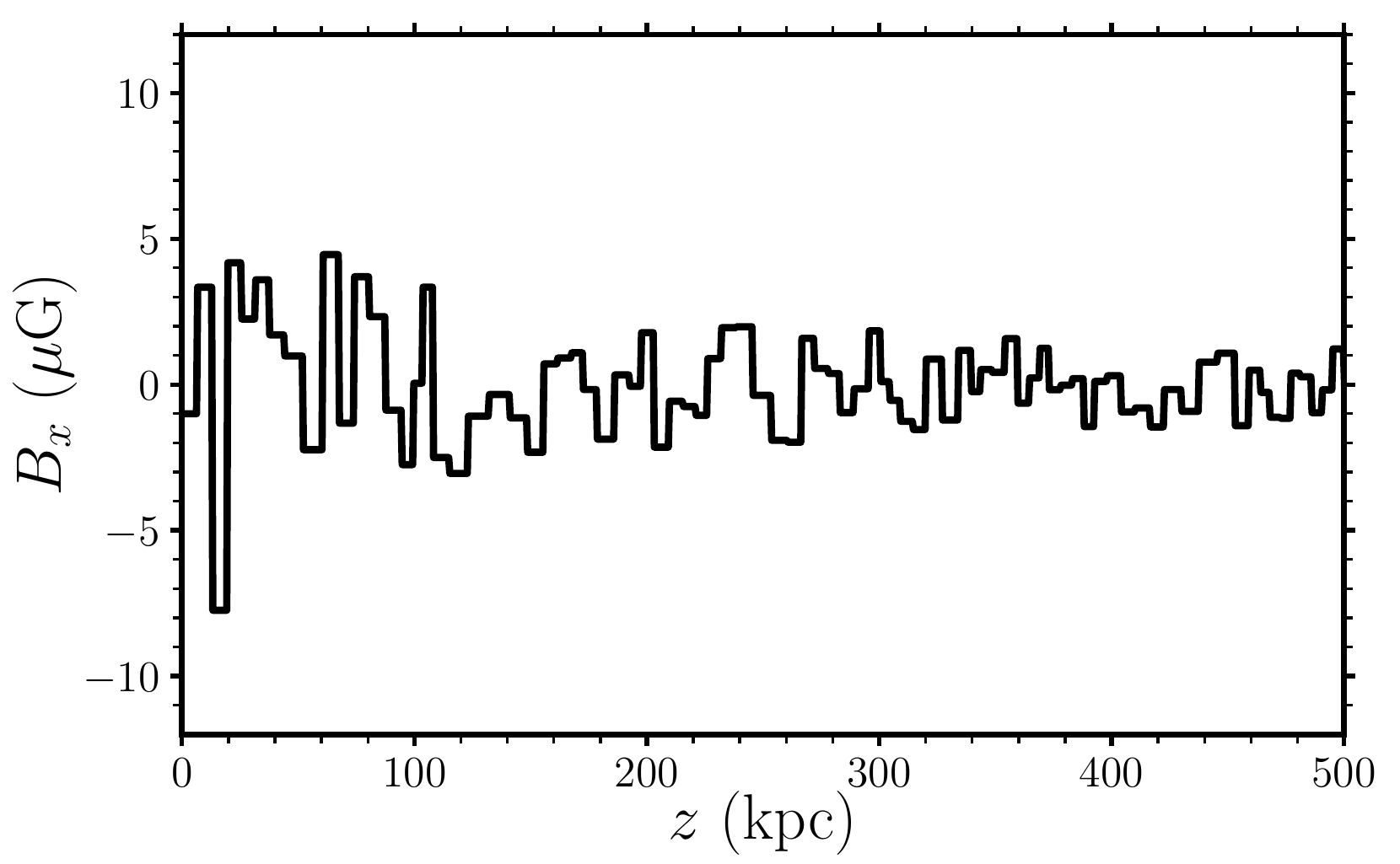} 
        \caption{Magnetic field} \label{fig:1275-bx}
    \vspace{4ex}
  \end{subfigure}
  \begin{subfigure}[t]{\fourpanelwidth\linewidth}
        \includegraphics[width=\linewidth]{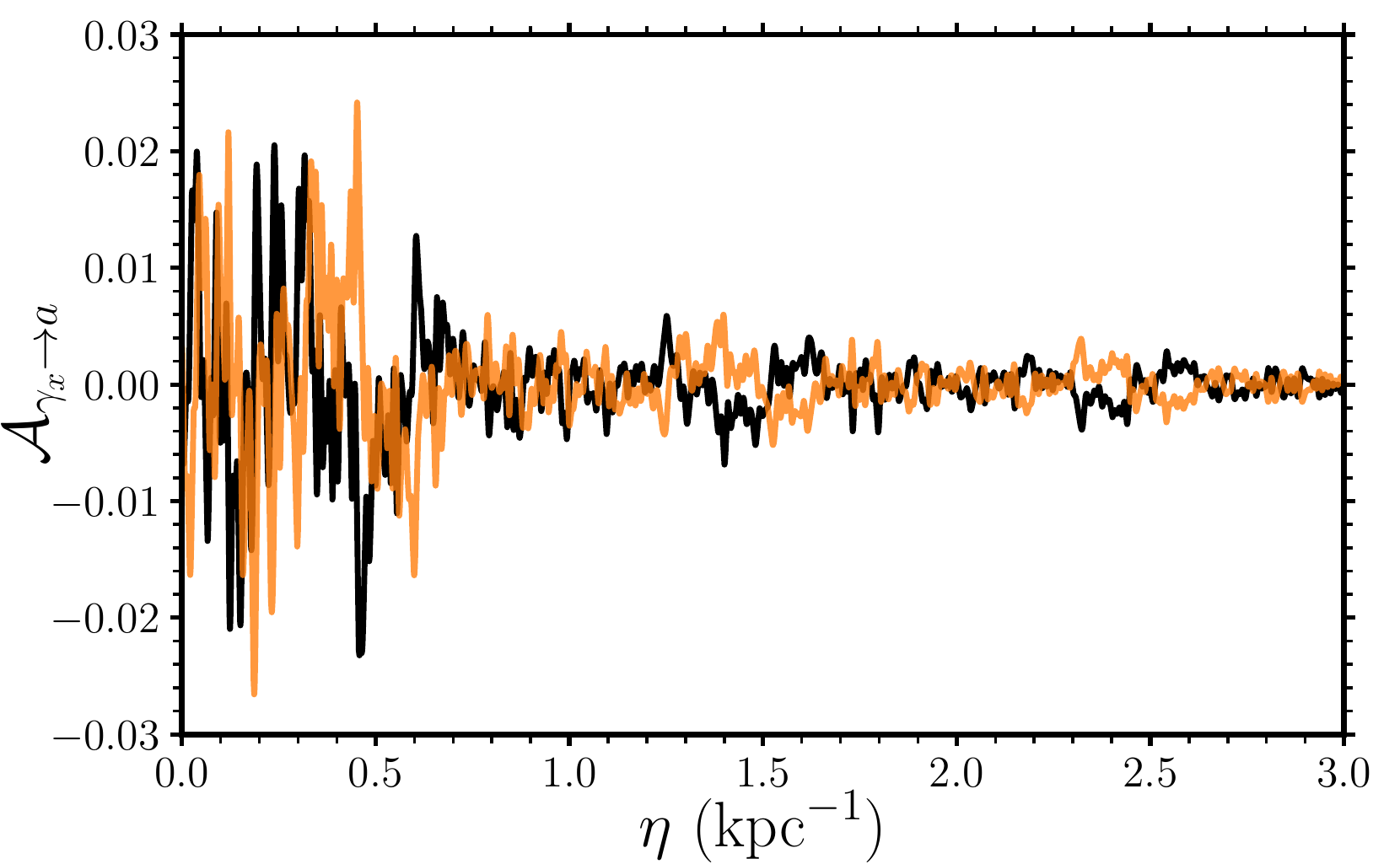} 
        \caption{The real, Re(${\cal A}_{\gamma_x \to a})={\cal F}_s(\Delta_x)$ (black), and imaginary, Im(${\cal A}_{\gamma_x \to a})=-{\cal F}_c(\Delta_x)$ (orange), parts of the amplitude ${\cal A}_{\gamma_x \to a}$ as calculated from discrete transforms.} \label{fig:1275-dct}
    \vspace{4ex}
  \end{subfigure} 
  \begin{subfigure}[t]{\fourpanelwidth\linewidth}
         \includegraphics[width=\linewidth]{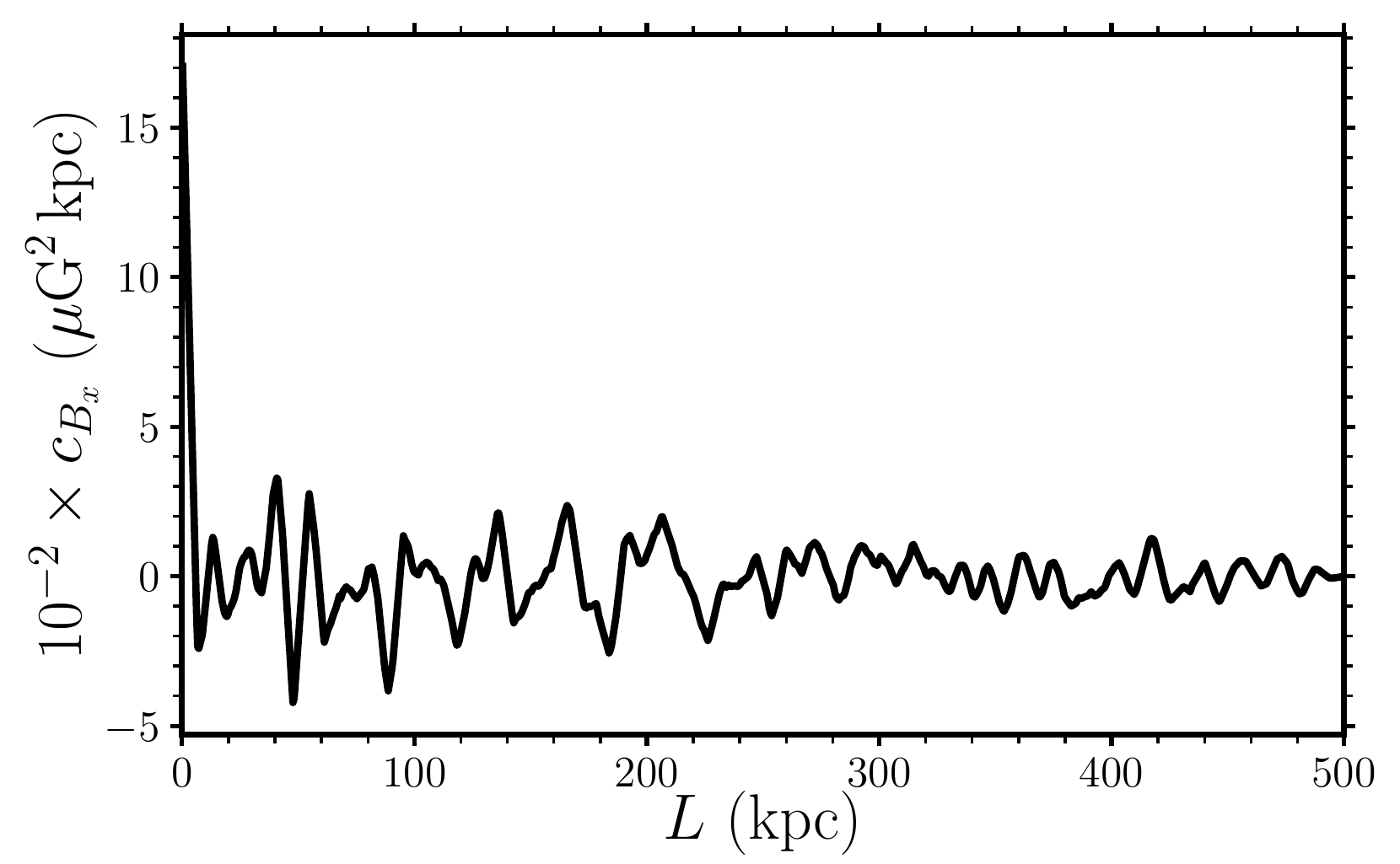} 
        \caption{Magnetic field autocorrelation function.} \label{fig:1275_cb}
    \vspace{4ex}
  \end{subfigure}
  \begin{subfigure}[t]{\fourpanelwidth\linewidth}
         \includegraphics[width=\linewidth]{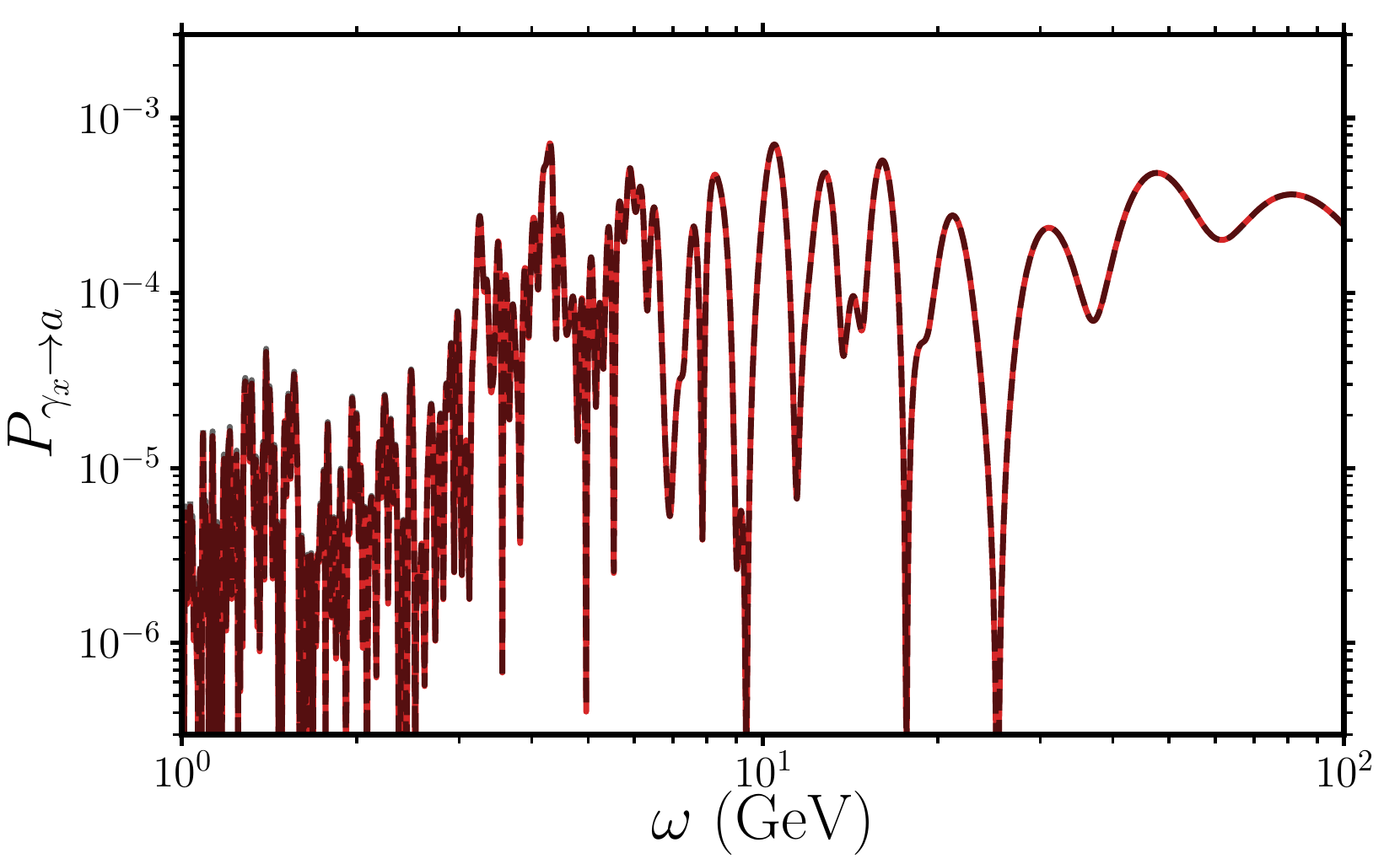} 
        \caption{Conversion probability as calculated from the DCT of $c_{B_x}$, (red solid) compared to a numerical solution of the Schr{\"o}dinger-like equation (black dashed).} \label{fig:1275_pgg}
    \vspace{4ex}
  \end{subfigure} 
 \caption{Numerical cell-model example of section \ref{sec:numcell_massive} for massive axion-photon conversion, with parameter values as in figure \ref{fig:fig_osc}.}
    \label{fig:1275_cell}
\end{figure*}

\subsection{Numerical implementations applied to the Perseus cluster}
\label{sec:numerical}
For our numerical tests, we consider classes of models that have previously been used to study photon-axion conversion in astrophysical settings, adopting magnetic field profiles and density laws appropriate to the Perseus cluster.
We first consider two examples with   massive axions ($m_a \gg \omega_{\rm pl}$): a general cell model, and a turbulent, divergence-free model based on Gaussian random fields. As shown in section \ref{sec:massive}, any such massive model can be solved analytically to leading order in perturbation theory, so here we compare the results of the discrete Fourier transform with direct numerical simulations  of the Schrödinger-like equation. We then consider an example with a massless axion and a `beta-model' plasma density, which illustrates how the discrete Fourier transform can be applied to the phase variable $\varphi$.

In all three examples in this section, we adopt $\g=10^{-13}$\,GeV$^{-1}$ and use a domain size of $500$\,kpc. The DCT is calculated using $2\times10^4$ and $10^5$ samples in the massive and massless case, respectively. We will focus on the gamma-ray regime in the massive ALP case and the X-rays in the massless ALP case, but we stress that the formalism can be applied to a much broader class of problems, and over any energy regime.

\begin{figure*}[t!] 
  \begin{subfigure}[t]{\fourpanelwidth\linewidth}
      \includegraphics[width=\linewidth]{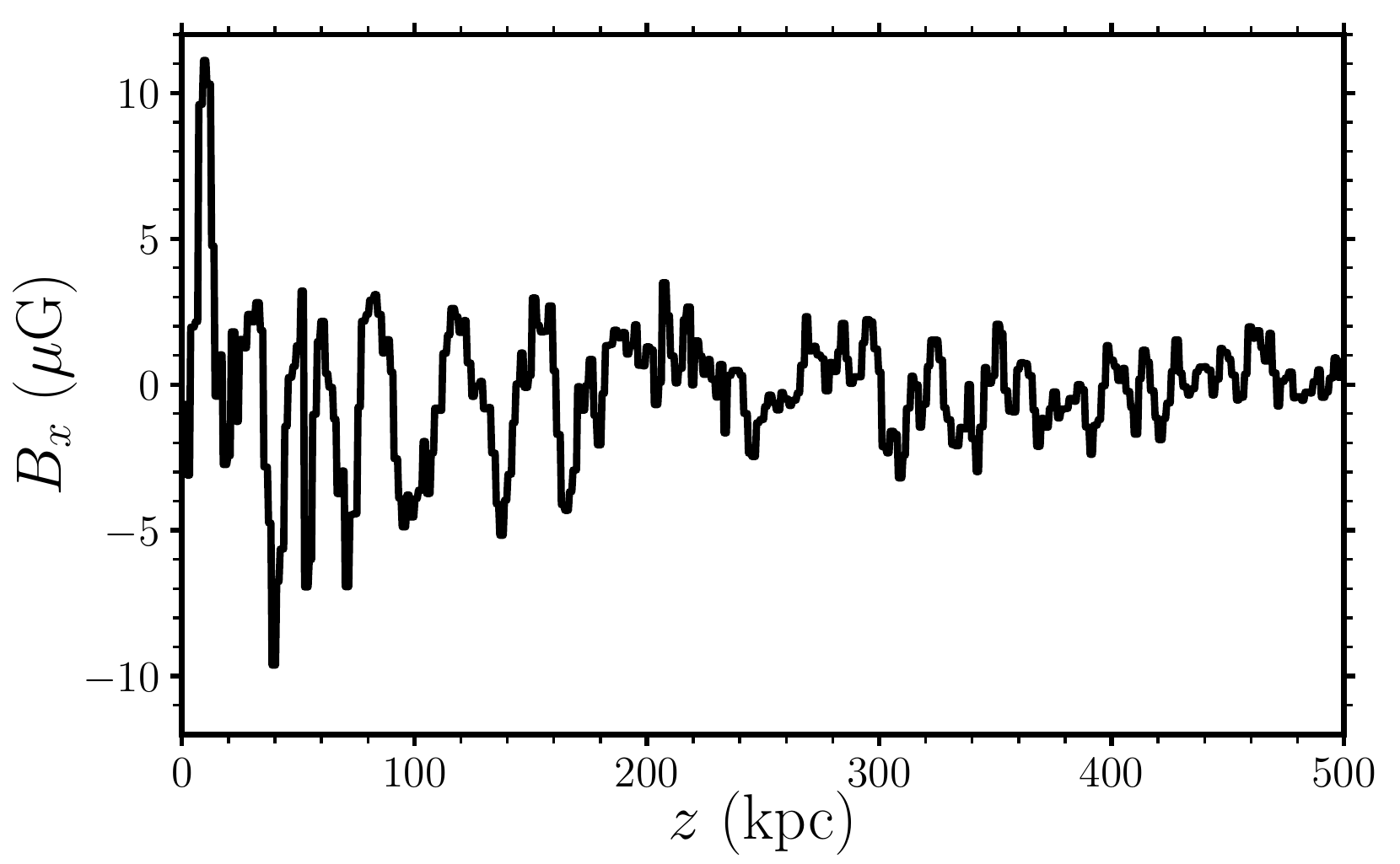} 
        \caption{Magnetic field profile} \label{fig:grf-bx}
    \vspace{4ex}
  \end{subfigure}
  \begin{subfigure}[t]{\fourpanelwidth\linewidth}
        \includegraphics[width=\linewidth]{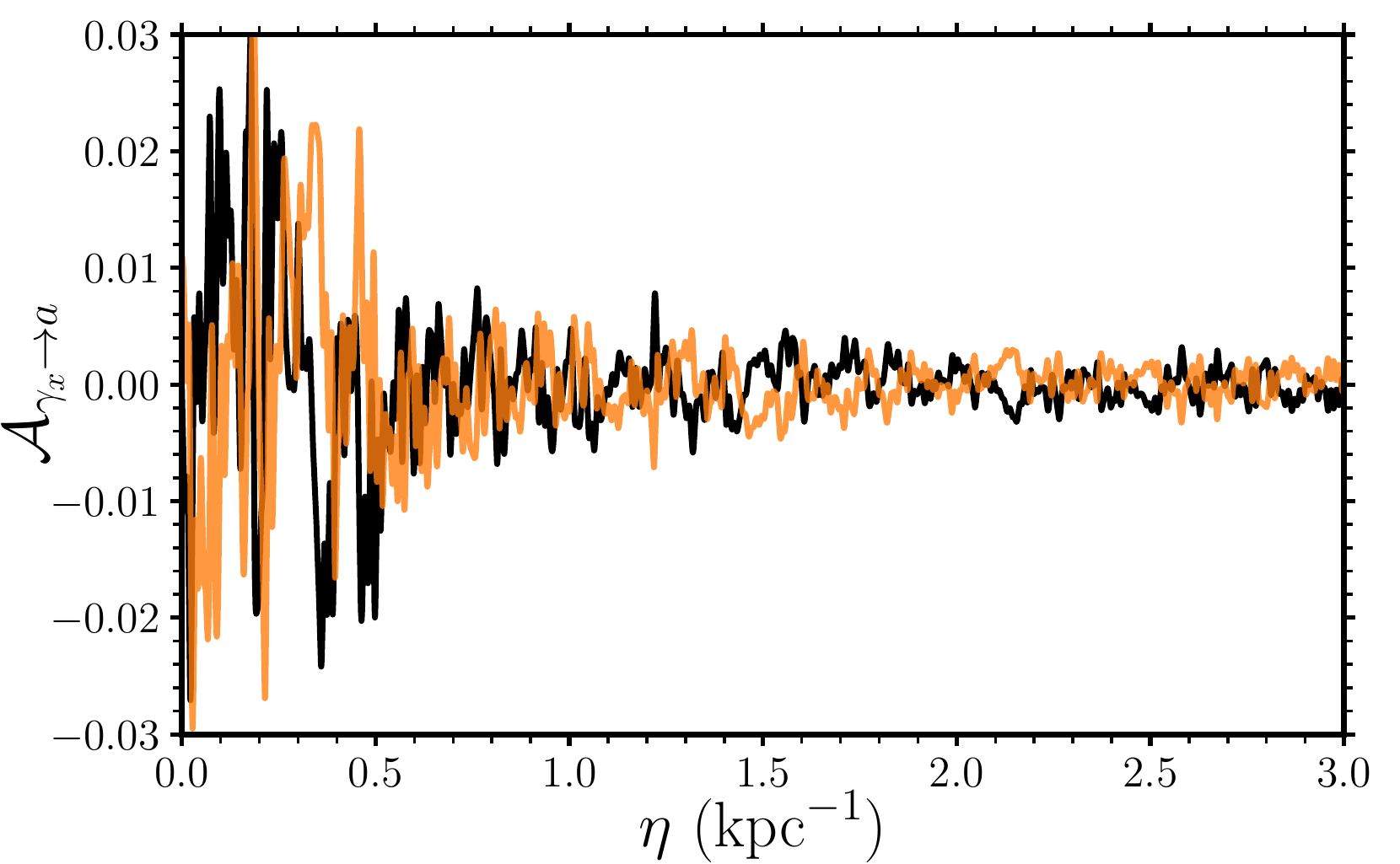} 
        \caption{The real, Re(${\cal A}_{\gamma_x \to a})={\cal F}_s(\Delta_x)$ (black), and imaginary, Im(${\cal A}_{\gamma_x \to a})=-{\cal F}_c(\Delta_x)$ (orange), parts of the amplitude ${\cal A}_{\gamma_x \to a}$ as calculated from discrete transforms.} \label{fig:grf-dct}
    \vspace{4ex}
  \end{subfigure} 
  \begin{subfigure}[t]{\fourpanelwidth\linewidth}
     \includegraphics[width=\linewidth]{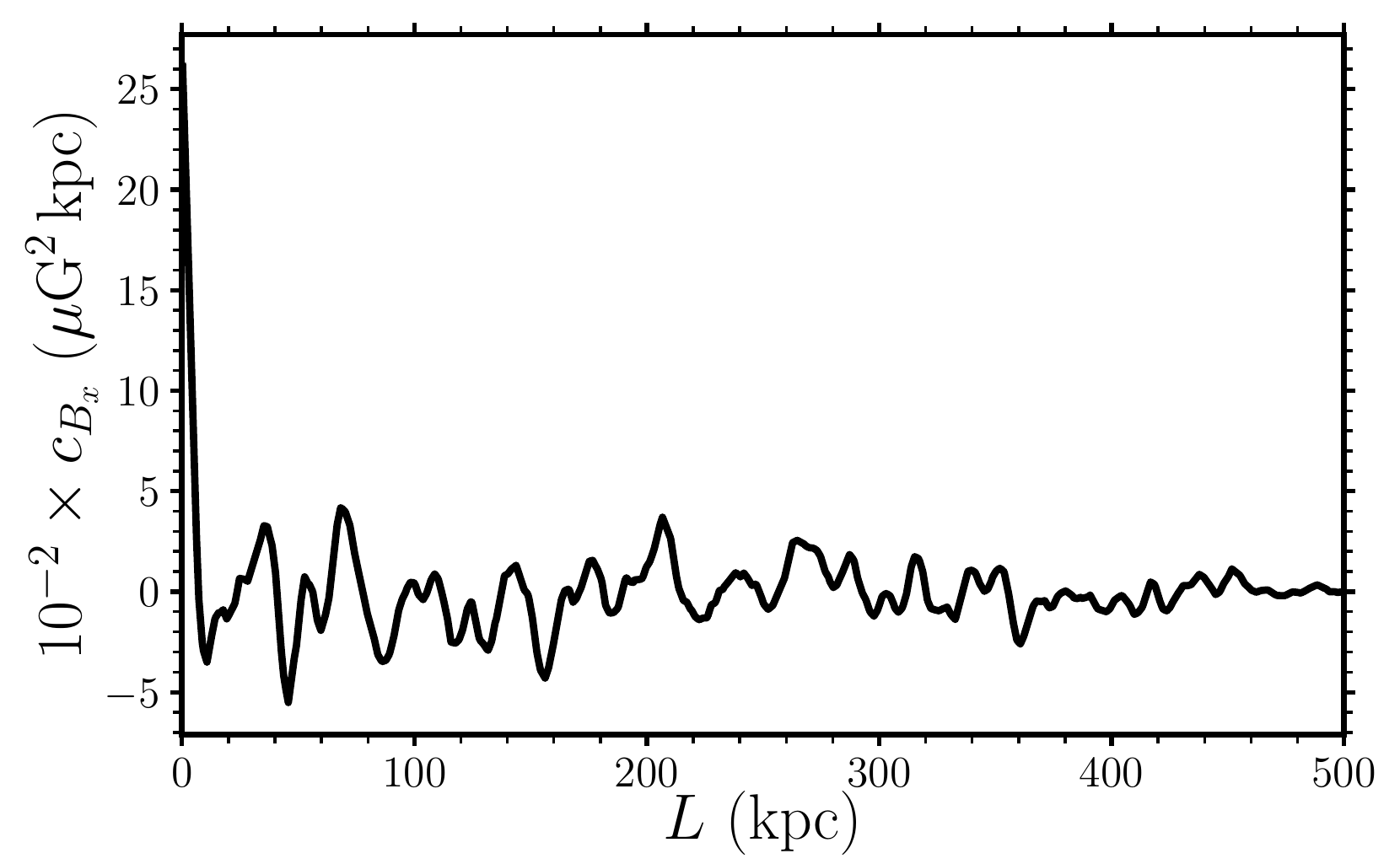} 
        \caption{Magnetic field autocorrelation function.} \label{fig:grf_cb}
    \vspace{4ex}
  \end{subfigure}
  \begin{subfigure}[t]{\fourpanelwidth\linewidth}
     \includegraphics[width=\linewidth]{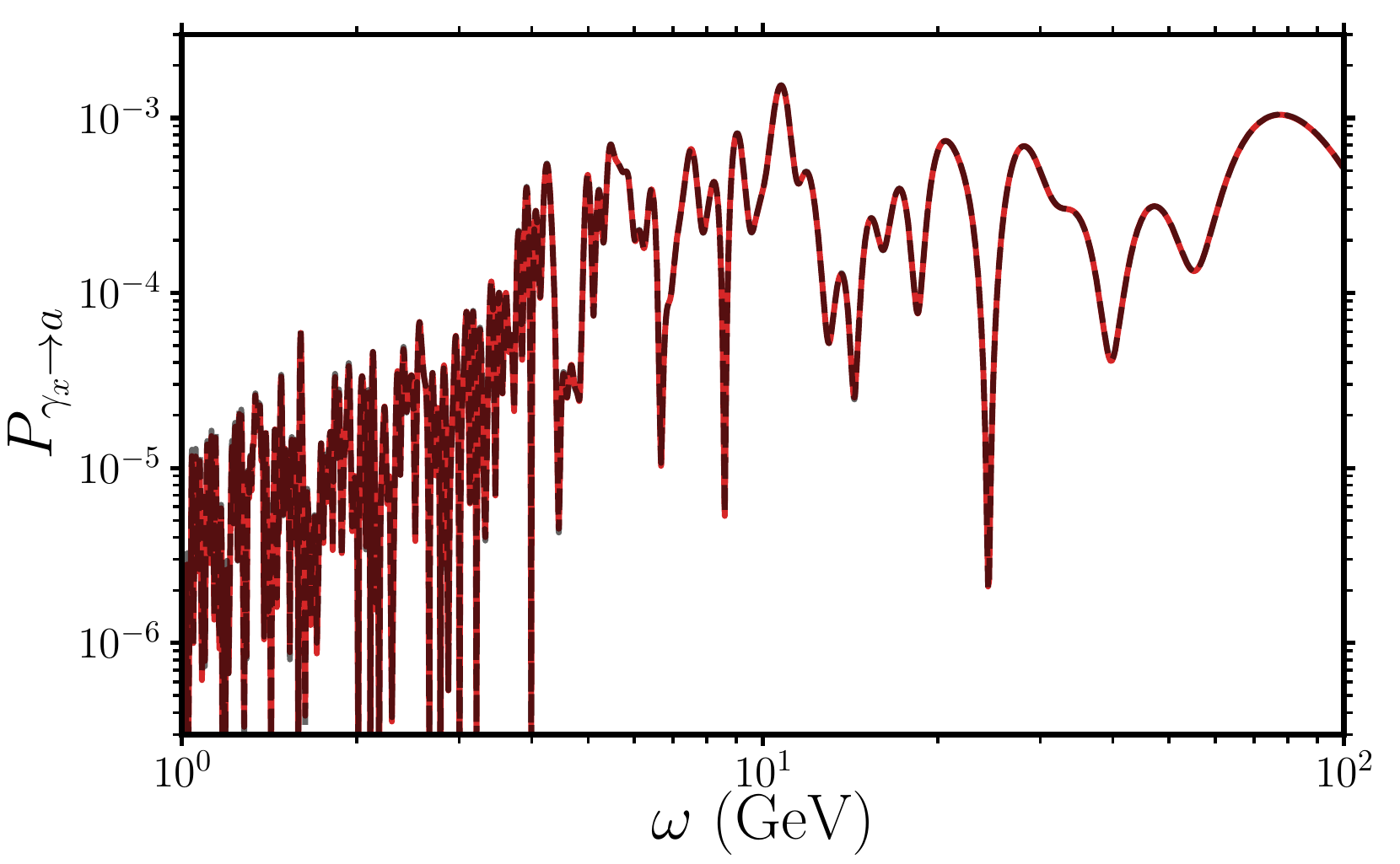} 
        \caption{Conversion probability as calculated from the DCT of $c_{B_x}$, (red solid) compared to a numerical solution of the Schr{\"o}dinger-like equation (black dashed).} \label{fig:grf_pgg}
    \vspace{4ex}
  \end{subfigure} 
    \caption{Numerical GRF example of section \ref{sec:numGRF_massive} for massive axions; parameter values as in figure \ref{fig:fig_osc}.}
    \label{fig:grf}
\end{figure*}

\subsubsection{Cell model, massive axions}
\label{sec:numcell_massive}
We first consider massive axions ($m_a\gg\omega_{\rm pl}$) in a general cell model, cf.~section \ref{sec:general_cell} for the analytical solution. These types of models are commonly used to model photon-axion conversion in complex astrophysical environments, such as galaxy clusters. Specifically, we consider a realisation of the stochastic `Model B' of \cite{Reynolds:2019uqt}, where cells of random size $L$ are generated according to a power-law  probability distribution $p(L) \propto L^{-1.2}$ between 3.5 and 10 kpc, and the magnetic field direction in each cell is random and isotropic. 
Such magnetic domains are assumed to extend to 500 kpc from the centre.
The total magnetic field strength of each cell is assumed to decline with radius as 
\beq
B(z) = 7.5\mu G \left[ \frac{h(z)}{h(25~{\rm kpc})} \right]^{0.5},
\eeq
where the function $h(z)$ is given by a double-$\beta$ law of the form
\beq
h (z) = \frac{3.9\times10^{-2}}{\left[1+ (z/80\, {\rm kpc})^2\right]^{1.8}} + \frac{4.05\times10^{-3}}{\left[1+ (z/280\, {\rm kpc})^2\right]^{0.87}}
\eeq
In \cite{Reynolds:2019uqt}, $h(z)$ corresponds to the electron density, a form originally given by \cite{Churazov:2003hr}, but, in this test, we explicitly set $n_e=0$ everywhere so as to ensure the massive axion case. This cell-based model is a slightly altered version of the model used by \cite{berg_constraints_2017}, and is based on very long baseline array (VLBA) observations of NGC 1275, the central AGN in the Perseus cluster. 

 The magnetic field profile is shown in figure \ref{fig:1275_cell}, together with the associated DCT and DST (giving the real and imaginary parts of the ampllitude), the magnetic autocorrelation function, and the resulting conversion probability. The probability curve obtained from squaring the amplitude or taking the DCT of the magnetic autocorrelation function agrees well with the numerical solution from the Schrödinger-like equation, but is significantly quicker to compute.

\subsubsection{Gaussian random field, massive axions}
\label{sec:numGRF_massive}
Next, we consider the mixing of photons and massive axions in a divergence-free Gaussian random field (GRF) model. The magnetic field profile is generated using a variation of the method of Tribble \cite{tribble_radio_1991}, which has been used in various guises in the astrophysical and axion literature \cite{murgia_magnetic_2004,hardcastle_synchrotron_2013,Angus:2013sua}. Briefly, the field is generated by drawing random Fourier components of the vector potential, from which the real-space  $\boldsymbol{A}$ is obtained through the inverse Fourier transform. The vector potential is then rescaled to account for the decrease of the field strength with radius: specifically we use the scaling of  `Model B' from \cite{Reynolds:2019uqt}, as in the previous subsection. Finally, a divergence-free field magnetic field is generated from the curl of the rescaled vector potential. To generate the field, we have to specify the power spectrum; we assume a Kolmogorov power spectrum such that $E(k) \propto k^{-5/3}$, where $E(k)$ is the energy contained in the interval $(k,k+dk)$ with wavenumber $k$. We use minimum and maximum scale lengths $3.5$\,kpc and $30$\,kpc, respectively, and the field is sampled at $1.75$\,kpc intervals for the solution of the Schrödinger-like equation. As in the previous section, we set $n_e=0$ as we are considering the massive axion regime. 

In figure \ref{fig:grf}, we show $B_x$ of a specific realisation of the magnetic field, the real and imaginary parts of the amplitude (obtained from the DCT and DST), the magnetic autocorrelation function, and the resulting conversion probability. The GRF has significantly more small-scale structure in the magnetic field than the previously studied cell-model, due to the presence of a large  range of Fourier components. This leads to finer structure in the DCT/DST, and the conversion probability. The conversion probability curve calculated from the DCT of $c_{B_x}$ agrees very well with the result from the numerical solution of the equation of motion. 

\begin{figure}
\centering
\includegraphics[width=\linewidth]{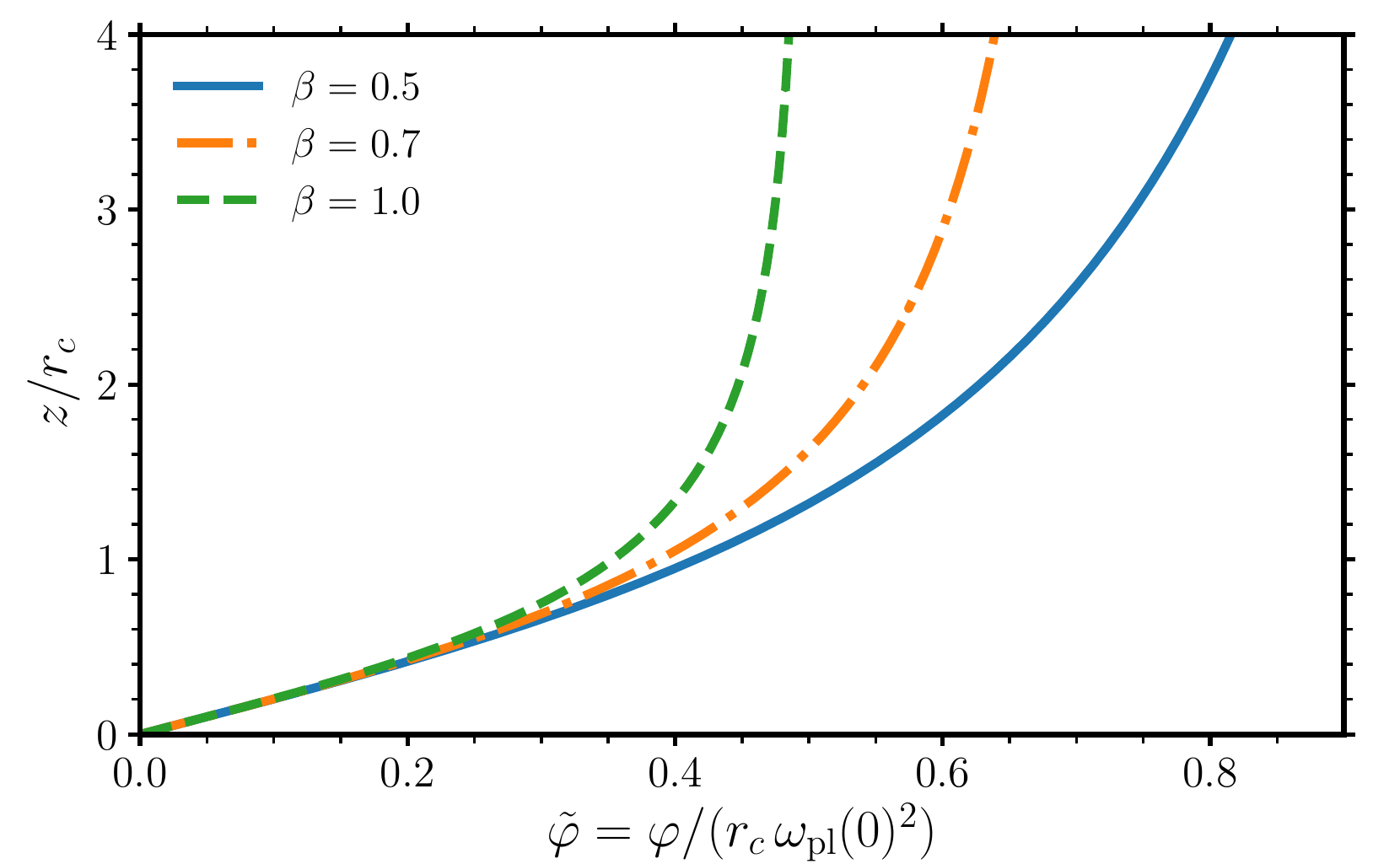} 
\caption{The spatial coordinate $z/r_c$ as a function of $\varphi$ for a $\beta$-law density profile, evaluated up to an assumed cluster radius of $R=4 r_c$ for $\beta=0.5$ (blue), $\beta=0.7$ (orange), and $\beta=1$ (green).}
\label{fig:varphi-massless}
\end{figure}

\subsubsection{Gaussian random field with $\beta$-law density, massless axions}
\label{sec:numGRF_massless}
We now consider the case of massless axions. A non-trivial aspect of the evaluation of the amplitude  is the change of coordinates $z\to \varphi$, which depends on the plasma density. In this example, we consider the famous  class of electron densities described by `$\beta$-models':
\beq
n_e(r) = n_0~\left[1+ \left(\frac{r}{r_c}\right)^2 \right]^{-\frac{3}{2}\beta} \, .
\eeq
Here $n_0$ denotes the electron density at the origin,  and $r_c$  and $\beta$  are positive constants. $\beta$-models and their variants are ubiquitously used as simple approximations of the gas density in galaxy clusters \cite{Cavaliere:1976, Arnaud:2009}.

We consider a source of photons located at the centre of the cluster, and calculate $\varphi$ from  equation \eqref{eq:varphimassless} as:
\beq
\varphi(z) = \frac{2\pi e^2}{m_e} \int_0^z dz'\, n_e(z') = \frac{2 \pi e^2}{m_e}  n_0 r_c \int_0^{z/r_c} du \, (1+u^2)^{-\frac{3}{2} \beta} \, ,
\eeq
where we have made a change of integration variable from $z$ to $u=z/r_c$. The integral  evaluates to the `ordinary', or Gaussian, hypergeometric function: 
\beq
\begin{split}
\varphi(z) 
&=
r_c\, 
\omega_{\rm pl}(0)^2 \tilde \varphi \, , {~~\rm where~~}\\
\tilde \varphi&= \frac{z}{2r_c} \, 
 {}_2F_{1}\left(\tfrac{1}{2}, \tfrac{3 \beta}{2}; \tfrac 3 2; -( \tfrac{z}{r_c})^2 \right) 
\, \, .
\label{eq:varphi2F1}
\end{split}
\eeq
The inverse function, $z(\varphi)$, is readily obtained numerically from this expression, cf.~figure \ref{fig:varphi-massless}.  Four our specific example, we take $\beta=1$, $n_0=0.05~{\rm cm}^{-3}$, $r_c=125$~kpc and a total domain size of $4r_c = 500$kpc, a choice which gives densities and scale lengths comparable to those in typical cool-core clusters. 

\begin{figure*}[t!] 
  \begin{subfigure}[t]{\fourpanelwidth\linewidth}
        \includegraphics[width=\linewidth]{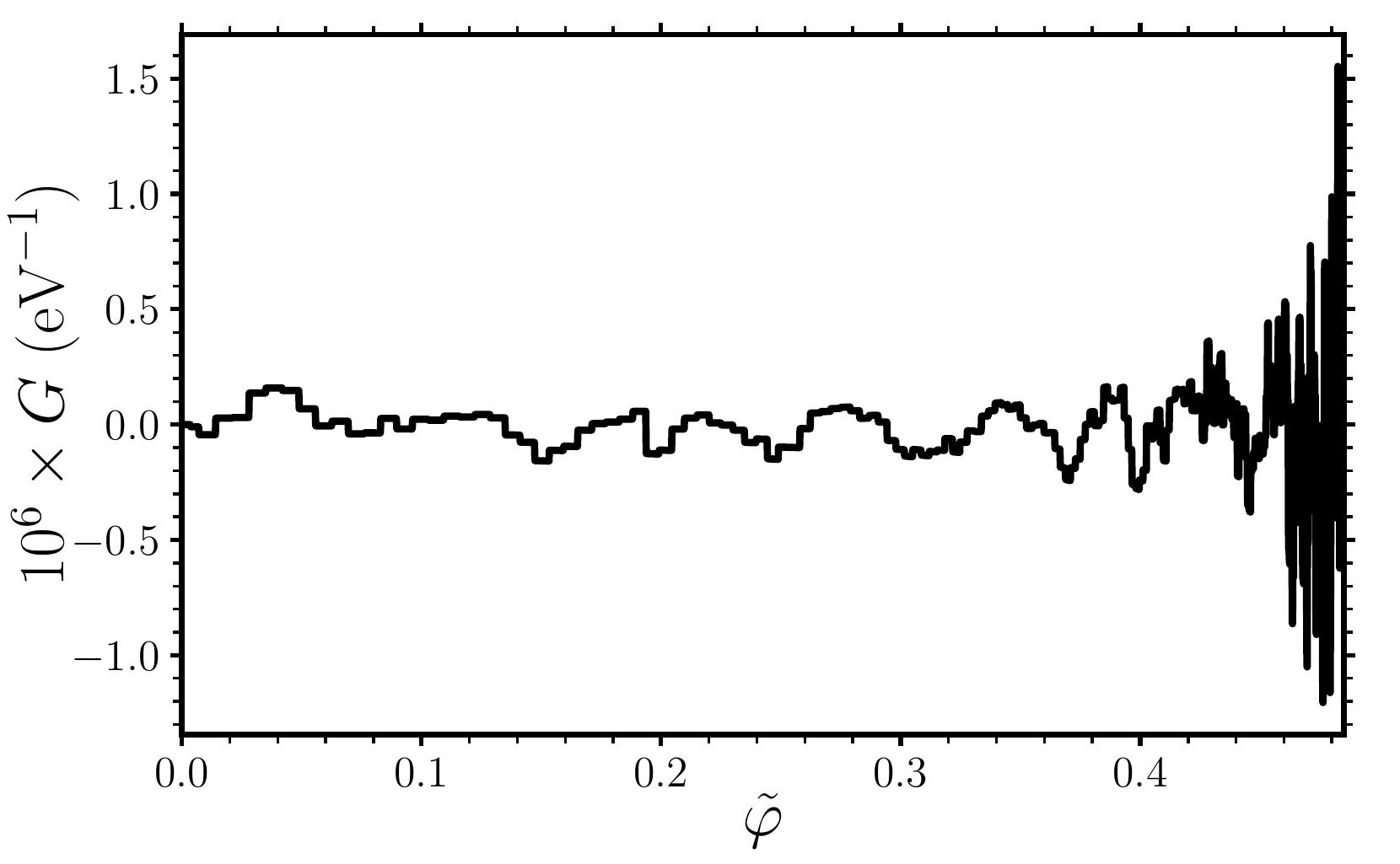} 
        \caption{The function $G(\varphi)=\g B_x / \omega_{\rm pl}^2$.} \label{fig:grf-massless-bx}
    \vspace{4ex}
  \end{subfigure}
  \begin{subfigure}[t]{\fourpanelwidth\linewidth}
       \includegraphics[width=\linewidth]{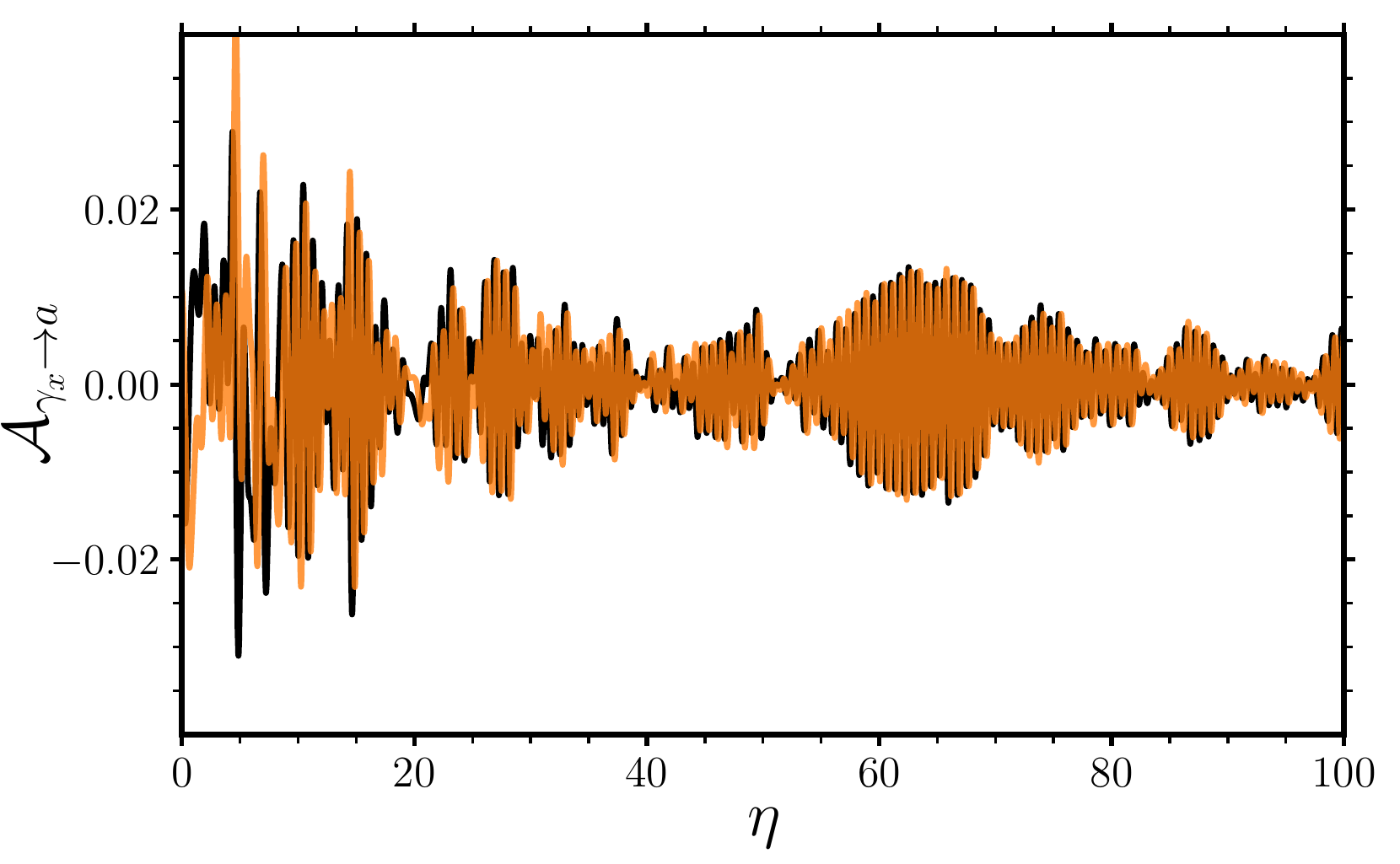} 
        \caption{The real, Re(${\cal A}_{\gamma_x \to a})={\cal F}_s(\Delta_x)$ (black), and imaginary, Im(${\cal A}_{\gamma_x \to a})=-{\cal F}_c(\Delta_x)$ (orange), parts of the amplitude ${\cal A}_{\gamma_x \to a}$ as calculated from discrete transforms.} \label{fig:grf-massless-dct}
    \vspace{4ex}
  \end{subfigure} 
  \begin{subfigure}[t]{\fourpanelwidth\linewidth}
      \includegraphics[width=\linewidth]{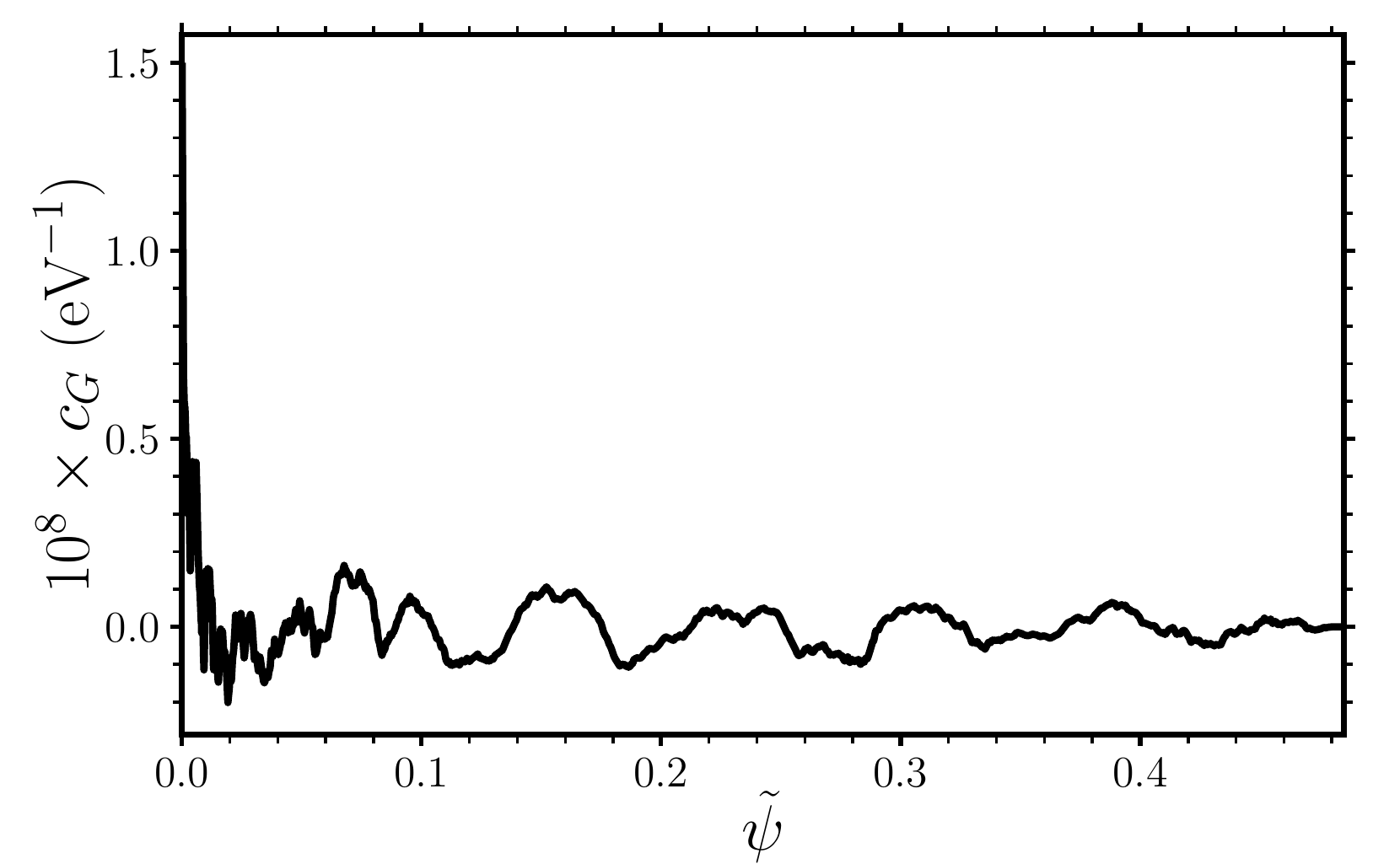} 
        \caption{Autocorrelation function of $G$.} \label{fig:grf-massless-cb}
    \vspace{4ex}
  \end{subfigure}
  \begin{subfigure}[t]{\fourpanelwidth\linewidth}
       \includegraphics[width=\linewidth]{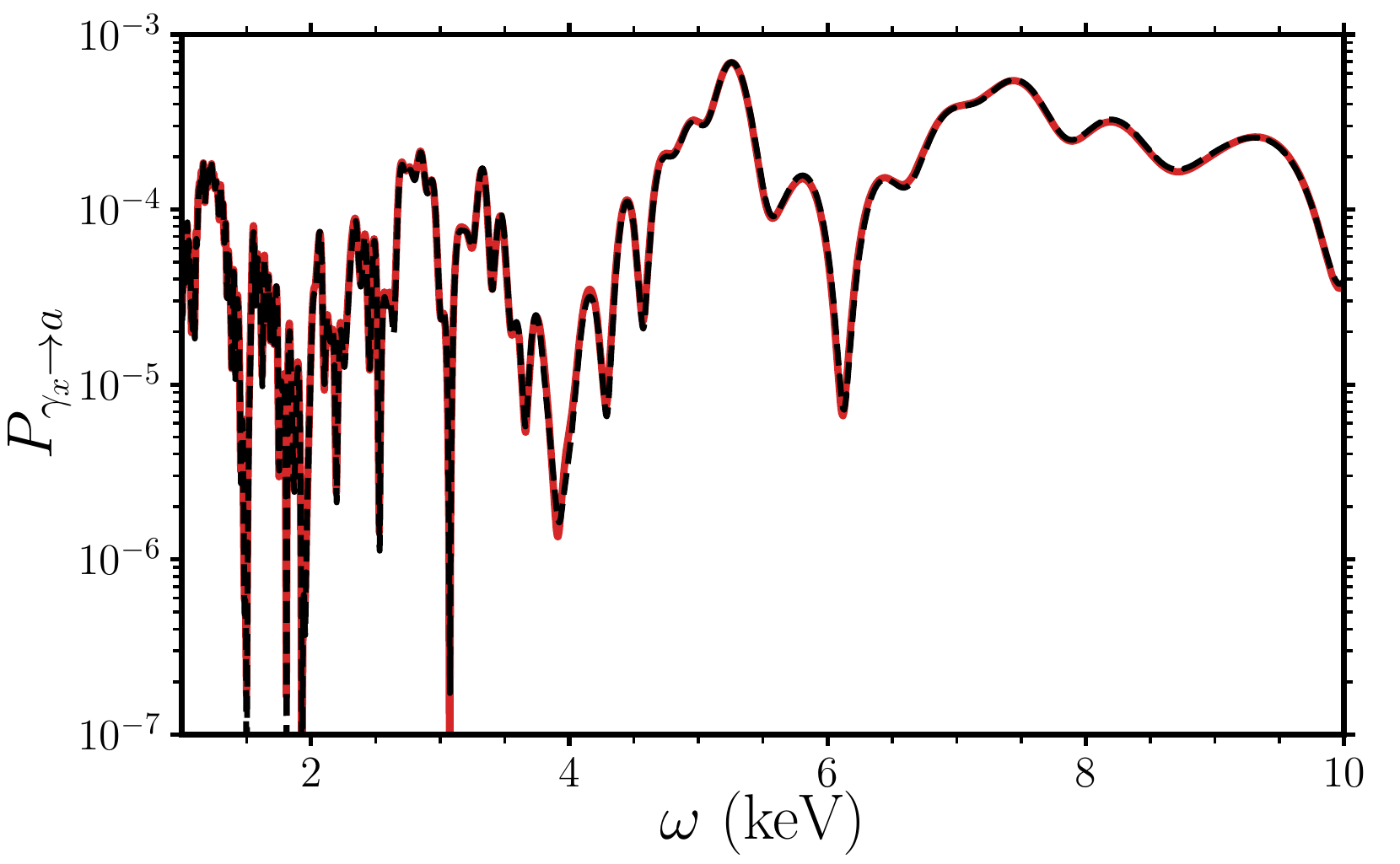} 
        \caption{Conversion probability as a function of energy, calculated from the DCT of $c_{G}(\psi)$, (red solid) compared to a numerical solution of the Schr{\"o}dinger-like equation (black dashed).} \label{fig:grf-massless-pgg}
    \vspace{4ex}
  \end{subfigure} 
    \caption{Numerical GRF example of section \ref{sec:numGRF_massless} for essentially massless axions. Parameters are: $\g=10^{-13}$~GeV$^{-1}$, $m_a=10^{-13}$~eV, $\beta=1$, $n_0=0.05~{\rm cm}^{-3}$, $r_c=125$~kpc and total domain size $4r_c = 500$~kpc.}
\end{figure*}

We assume that the magnetic field is the GRF-example of section \ref{sec:numGRF_massive}. The function $G(\varphi)$ is plotted in figure \ref{fig:grf-massless-bx}, and should be  compared to the magnetic field of figure \ref{fig:grf-bx}: in contrast to the magnetic field that decays with radius, $G(\varphi)$ keeps a rather constant magnitude of oscillations over a large potion of its range, and then increases for large values of $\varphi$. This reflects how the transition probability is more sensitive to the magnetic field in (real-space) regions where the phase $\Phi$ varies more slowly, i.e.~where the plasma density is suppressed. 


The real and imaginary parts of the transition amplitude are plotted in figure \ref{fig:grf-massless-dct}, and the magnetic autocorrelation function, $c_G(\psi)$, is shown  in Fig~\ref{fig:grf-massless-cb}. The conversion probability is shown in Fig~\ref{fig:grf-massless-pgg}, again compared to a numerical solution of the Schrödinger-like equation. Once more, the agreement is extremely good.

\begin{figure}
\centering
\includegraphics[width=\linewidth]{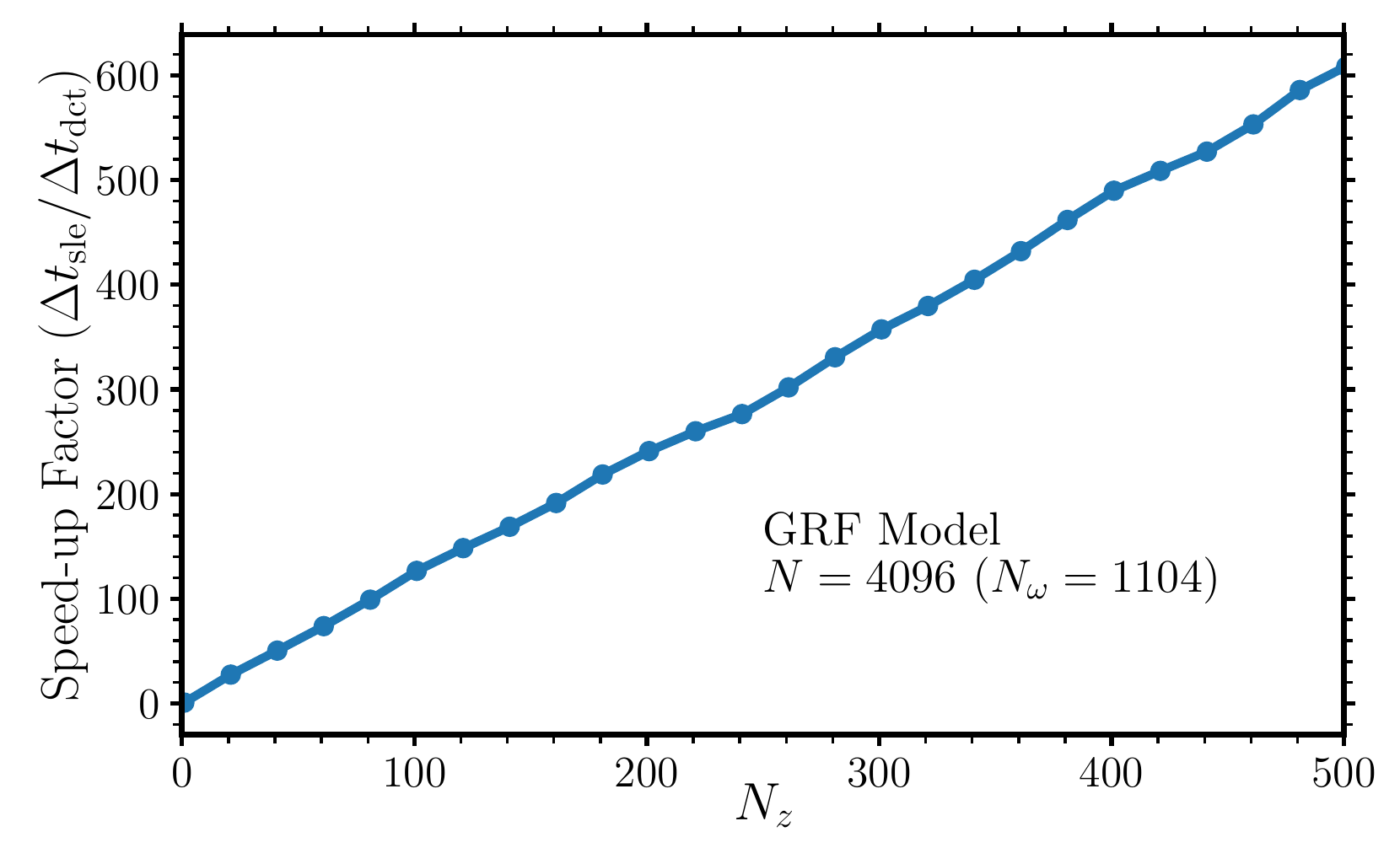} 
\caption{\color{dm} The speed-up factor for the test described in section~\ref{sec:speed} using the same GRF model as in section~\ref{sec:numGRF_massive} truncated in each case at a maximum radius of $z_{\rm max} = N_z \Delta z$, with $\Delta z= 1.75\, $kpc. Here $N_m=N_g=1$.} 
\label{fig:speedup}
\end{figure}

\subsection{Performance Improvement}
\label{sec:speed}
{\color{dm}
To compare the numerical efficiency of the traditional Schrödinger approach to our new, numerical Fourier approach, we consider the following problem. Suppose that we would like to calculate the conversion probability for $N_m$ axion masses and $N_g$ axion-photon couplings in the perturbative regime. We will for simplicity consider massive axions in this example, and assume that the background magnetic field and autocorrelation function are known. Suppose further that to obtain sufficient energy resolution over the specified range, the calculation should include at least $N_\omega$ beam energies, and that axion-photon trajectory needs to be sampled with at least $N_z$ domains for numerical accuracy. We first consider the computational time required  in the traditional approach, before comparing with that of the Fourier  approach. 

To solve this problem in the traditional approach requires calculating the $z$-evolution operator through matrix multiplication of the $N_z$  operators for the uniform domains. This has to be done for each mass, energy and mode energy, so if the calculation time for a single uniform domain is $\Delta t_1$, then the total time to solve this problem is given by
\beq
\Delta t_{\rm sle} = N_g N_m N_\omega N_z\,  \Delta t_1 \, .
\eeq

By contrast, in the perturbative approach, $\g$ is an overall prefactor which trivially rescales the probability (at negligible computational cost), so the discrete Fourier transform can be performed at a fixed coupling. Similarly the mass only appears inside $\eta = m_a^2/(2\omega)$, so that the probability can be calculated for a fixed mass, and then re-interpreted for other masses (again at negligible numerical cost).  Moreover, a single discrete Fourier transform samples all spatial points and generates the probability at all mode energies. The number of points that the discrete Fourier transform runs over, $N$, must then satisfy both $N \geq N_z$ and $N\geq N_\omega$. If $N\geq N_\omega > N_z$, taking the Fourier transform requires `zero-padding' as described above. If $N \geq N_z > N_\omega$, the Fourier transform tends to give a higher energy resolution than the minimal requirement.  Given the usual scaling in the DCT algorithm of $\mathcal{O} (N \log_2 N)$ \citep{makhoul1980}, we expect the total calculation time to scale as 
\beq
\Delta t_{\rm dct} \sim N \log_2 N\, \Delta t_2
\eeq
where $\Delta t_2$ is now a single multiplication in the DCT algorithm, and $\Delta t_2 \ll \Delta t_1$.

The fractional numerical gain can in this example be estimated as
\beq
\text{speed-up} = \frac{\Delta t_{\rm sle}}{\Delta t_{\rm dct}} = \frac{N_g N_m N_\omega N_z }{N \log_2 N} \frac{ \Delta t_1}{ \Delta t_2}\, .
\eeq
This performance improvement is model-dependent but appreciable for all relevant values of the parameters, and tends to be maximised when $N_z \approx N_\omega$, so that the Fourier approach does not lead to `superfluously' high spatial or energy sampling.  

We now consider a concrete example where we numerically compare the actual computational time. We take $N_m = N_g=1$, and set $\g=10^{-13}$\,GeV$^{-1}$ and $m_{a} =5\times10^{-9}$~eV.  For the magnetic field, we consider the model of GRF model described in section~\ref{sec:numGRF_massive}. To illustrate the dependence on $N_z$, we `truncate' the magnetic field at the maximal radius $N_z \Delta z$ with $\Delta z = 1.75$\,kpc and vary $N_z$ from $1$ to $501$ at intervals of $20$. Thus, the parameter $N_z$ in this case parametrises the size of the magnetised region. To ensure sufficient energy resolution in the band  $1-100$\,GeV, we take $N_\omega=1104$. This requires the number of discrete Fourier transform samples to be $N=4096$ for the chosen value of the mass. We note that this example is rather conservative since $N_\omega \gg N_z$, and the Fourier approach samples the spatial profile more densely than the minimal requirement. } 
{\color{dm}

The speed-up factor is shown in Fig~\ref{fig:speedup} as a function of $N_z$. The speed improvement is linear in $N_z$, because $\Delta t_{\rm sle} \propto N_z$ and $\Delta t_{\rm dct}$ is constant for constant $N$ and $N_\omega$. The speed-up factor quickly exceeds $100$ even for modest numbers of cells, and this particular test provides quite a conservative estimate of the performance improvement. For broader energy ranges, the speed-up is even more significant since more solutions to the Schrödinger-like equation must be calculated (the factor $N_\omega / N$ increases). The speed-up is further increased by several orders of magnitude when $N_m$  and $N_g$ are taken to be sufficiently large to ensure a dense sampling of the axion parameter space. 

Overall, our results show that the Fourier formalism is significantly quicker -- often by orders of magnitude -- for models with more than a few cells, and therefore also for the types of magnetic field models commonly used for astrophysical ALP searches.

}

\begin{figure*}
    \centering
    \begin{subfigure}[t]{\fourpanelwidth\textwidth}
        \centering
        \includegraphics[width=\linewidth]{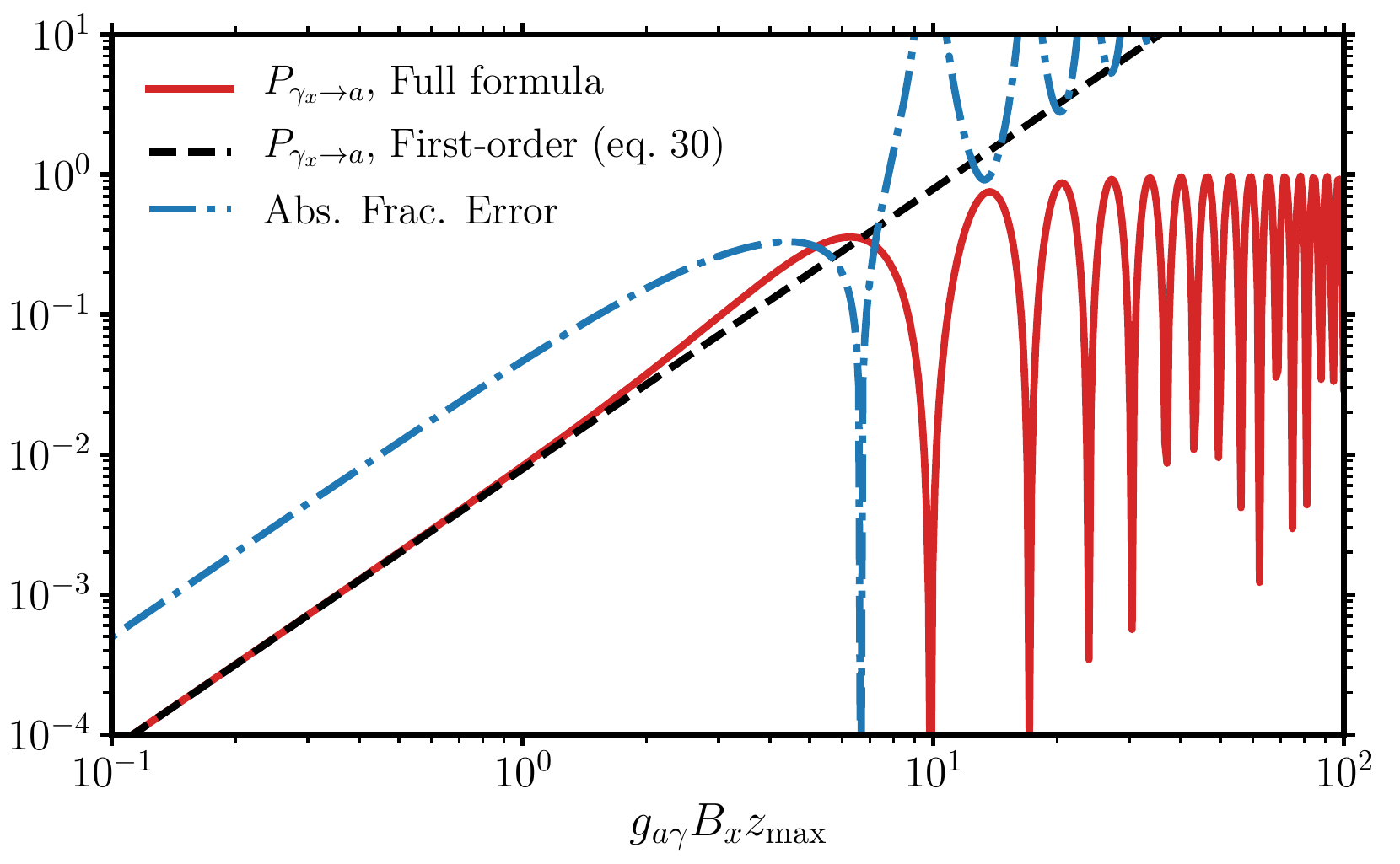} 
        \caption{Analytic calculation for a single domain of size $z_{\rm max}$; calculations from equation \eqref{eq:Psingledomain}, accurate to leading order, are compared to the fully accurate analytic formula, as a function of the dimensionless quantity $\g B_x z_{\rm max}$.} 
        \label{fig:perturb_analytic}
    \end{subfigure}
    \begin{subfigure}[t]{\fourpanelwidth\textwidth}
        \centering
        \includegraphics[width=\linewidth]{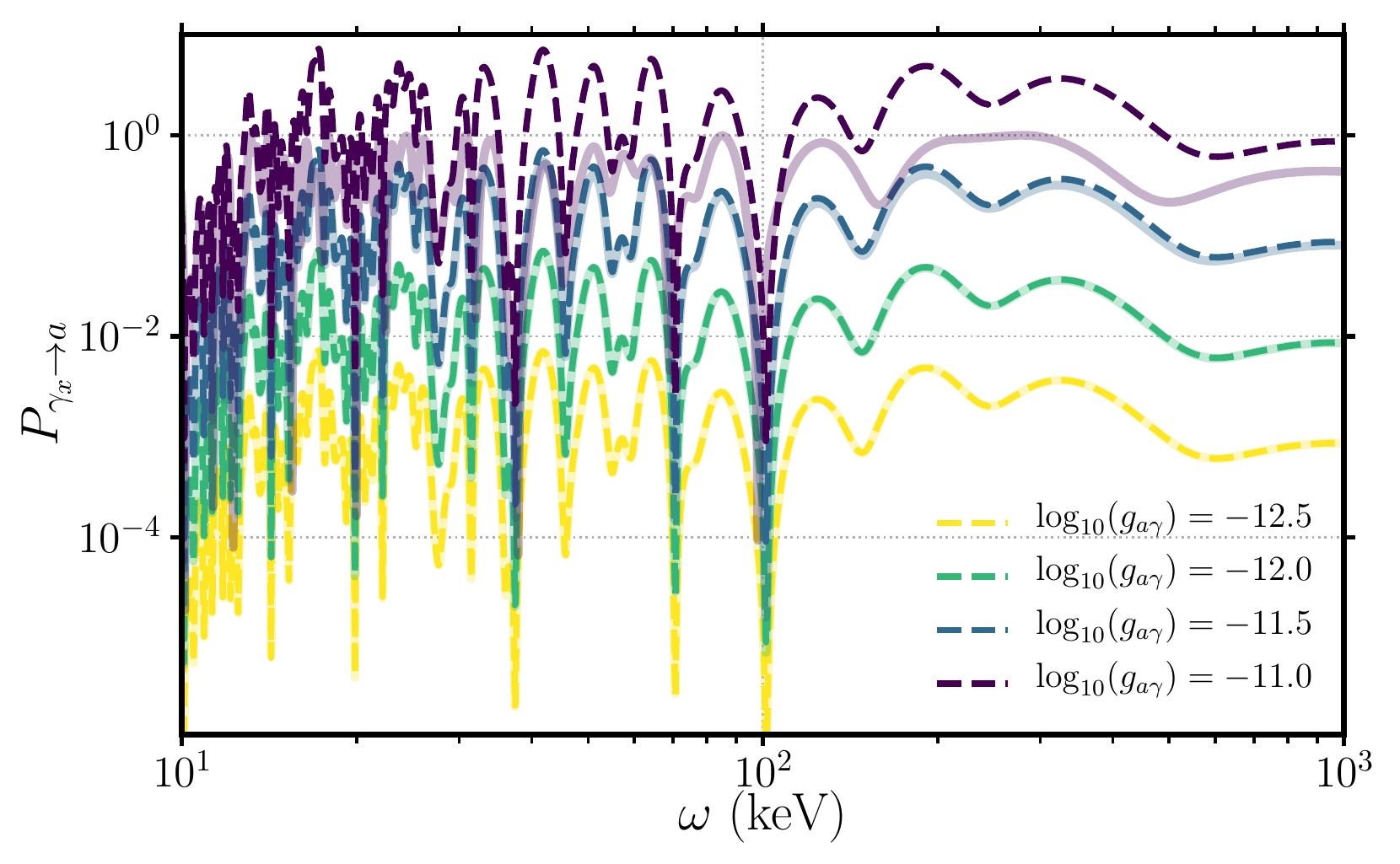} 
        \caption{Numerical calculation using a cell-based model for the Perseus cluster. Conversion probabilities calculated from a DCT (dashed lines) are compared to a full numerical solution to the Schr{\"o}dinger-like equation of motion (translucent solid lines), as a function of $\g$ for the realisation of Model B already discussed and shown in figure \ref{fig:1275_cb}.} 
        \label{fig:perturb_modb}
    \end{subfigure}
    \caption{An illustration of the breakdown of the perturbative treatment (cf.~section \ref{sec:perturbativity}) for a single domain with uniform magnetic field (left) and  a cell-based model for the Perseus cluster (right). Both models use $m_a=10^{-11}\,$eV. Discrepancies are only significant once the amplitude of the probability exceeds a few $\%$.
    }
    \label{fig:perturb}
\end{figure*}

\subsection{Perturbativity}
\label{sec:perturbativity}
A central aspect of our analysis in this paper is perturbation theory, where we have consider the leading-order contributions to the conversion probability (i.e.~${\cal A} \sim \g$ and $P_{\gamma \to a} \sim \g^2$). {\color{dm} As discussed in section \ref{sec:conv}, parametrically, the perturbative expansion should be good when conversion probabilities are small or moderately small, which includes a large fraction of all cases of phenomenological interest.} In this section, we elaborate on this general argument through two examples: the simplest case of a constant magnetic field, and a complex example with a turbulent magnetic field.  In both cases, we seek to identify  the breakdown of the perturbative expansion when the mixing grows large.

First, we consider the single domain of uniform $B$ and length $z_{\rm max}$ discussed in section~\ref{sec:single_cell}. This is one of a very small set of examples in which both the full and the perturbative conversion probabilities can be calculated analytically. The ratio between the leading-order and full conversion probabilities is:
\beq
\frac{P_{\gamma \to a}^{\rm LO} }{P_{\gamma \to a}^{\rm full}} = \Bigg(
1 + 4 \frac{\Delta_x^2}{\eta^2}
\Bigg) \frac{\sin^2 \left(\frac{\eta z_{\rm max}}{2} \right)}{\sin^2 \left(\frac{\eta z_{\rm max}}{2} \sqrt{1 + 4 \frac{\Delta_x^2}{\eta^2}}\right)} \, .
\eeq
The perturbative expansion in $\g B$ can be interpreted as an approximation using  $4\Delta_x^2/\eta^2 \ll1$. Clearly, when $\eta z_{\rm max}$ is sufficiently small to allow for a Taylor expansion of the sines, 
\beq
P_{\gamma \to a}^{\rm LO}/P_{\gamma \to a}^{\rm full} =1 + {\cal O}\left(\frac{\eta z_{\rm max}}{2} \sqrt{1 + 4 \frac{\Delta_x^2}{\eta^2}}\right)^3 \, .
\eeq
 An interesting special case is when $4\Delta_x^2/\eta^2 \ll1$ but $\eta z_{\rm max}$ is sufficiently large that  $\Delta_x^2 z_{\rm max}/\eta \gtrsim \pi$. In this case,  the oscillations of the conversion probability are very fast, and the LO approximation is out-of-phase with the exact result. This discrepancy is not always observable however as detector resolution or Doppler broadening limits the detectability of high frequency oscillations, and the cycle averaged predictions at LO agree with that of the full model. 

In figure \ref{fig:perturb_analytic} we show a comparison of the first-order, Fourier-like formula, which can be calculated from the cosine and sine transforms, to the full formula accurate to all orders. The calculation is conducted for a beam energy of $10$\,keV, with $m_a = 10^{-11}$\,eV, $z_{\rm max}=10$\,kpc and $B=10\,{\rm \mu G}$. We show the conversion probabilities as a function of $\g B_x z_{\rm max}$, together with the absolute fractional error. The relative error is at the few $\%$ level or lower until the conversion probability exceeds $\sim 0.01$; beyond this point errors can be significant and approach unity or higher --- moreover the probability exceeds $1$ at very large $\g$. 

Second, we consider a more complex cell model with many domains, taking the example show in figure \ref{fig:1275-bx} for concreteness (but now calculated using $m_a = 10^{-11}$~eV). Here, we compare the result obtained from a numerical solution of the Schrödinger-like equation to a numerical DCT of the autocorrelation function of the field, $c_{B_x}(L)$. The conversion probabilities are shown as a function of energy for a few different values of gradually increasing $\g$, in figure \ref{fig:perturb_modb}. Overall, the outcome is similar to the single-cell case, with good agreement until the amplitude of the conversion probability exceeds a few $\%$, where second-order effects start to imprint themselves on the curve. In fact, the values of $\g$ above which the first-order predictions from the DCT deviate substantially from the more detailed calculation are already ruled out in X-ray searches using similar models \cite{Reynolds:2019uqt}. The tests presented here are specific and not exhaustive, so caution should be exercised when applying our scheme to any situation whether the conversion probability exceeds a few percent.

\section{Astrophysical modelling of galaxy clusters for axion-photon conversion}\label{sec:astrophys}

\subsection{Magnetic Fields in Clusters}
\label{sec:Bcluster}

The formalism developed in this paper provides new insights into the dependence of axion-photon conversion probabilities on the structure of the magnetic field along the particle trajectory. Given the special role that galaxy clusters play in these studies, it is appropriate to review our current understanding of the structure and strength of the magnetic field in the ICM.

To state the current paradigm up-front --- multiple lines of evidence suggest that the ICM is a turbulent magnetized plasma that is, almost everywhere, in approximate hydrostatic equilibrium in the gravitational potential of the cluster's dark matter halo. In the bulk of the ICM, the magnetic pressure appears to be approximately 1\% of the thermal pressure and the magnetic field is tangled on scales of a few-kpc, consistent with characteristic spatial scales in the turbulent power spectrum. There is little evidence for large-scale regular magnetic field structures in the central regions of clusters. In the rest of this subsection, we summarize in brief the key evidence underpinning this paradigm.

Direct observational evidence for magnetic fields in the ICM comes principally from radio observations \citep{carilli:2002}.  Faraday rotation, i.e. the characteristic rotation of the polarization angle as a function of wavelength due to propagation through a magnetized plasma, is clearly seen in the radio band towards the radio-emitting lobes of AGN embedded in the ICM.  The corresponding Rotation Measure (RM) provides a direct measure of the line-of-sight magnetic field $B_z$ integrated along the path from the source to observer. For a given RM, the inferred typical magnetic field strength depends upon the number of field reversals along the path and hence the field coherence length.  In a number of clusters, the RM mapping across large radio lobes reveals the typical coherence length of the field to be a few kpc \cite{taylor:1993,vogt:2003,Ensslin:2003}. Adopting these in-plane coherence lengths as being typical of the line-of-sight structure allows typical field strengths to be estimated; we find $B\sim 1-10\mu{\rm G}$ in the cores of cool-core clusters, corresponding to a ratio of the thermal-to-magnetic pressures of $\sim 100$.  

Independent support for this picture comes from observations of radio halos and minihalos in the cluster cores and radio relics in cluster outskirts \citep{vanweeren:2019}. The radio emission from these diffuse structures is synchrotron radiation from a population of relativistic electrons gyrating in the magnetic field of the ICM (with the relativistic population likely arising from a central AGN for the central minihalos, and shock acceleration for the radio relics). Measurements (or upper limits) on the X-ray inverse Compton scattering of the relativistic electron population can be combined with measurements of the synchrotron emissivity to estimate (or set a lower limit on) the magnetic field strength.  An example of such a study for the radio halo in the Perseus cluster is provided by \citep{Sanders:2005jx}; note that the reported detection of the inverse Compton X-rays in this work was later discovered to be due to an error in the {\it Chandra} mirror calibration.  Thus the reported measurements of the magnetic field strength in \citep{Sanders:2005jx} should be taken as strict lower bounds, implying $B>3\mu{\rm G}$ in the core of Perseus.  The polarization of these structures provides information about the configuration of the magnetic field. Synchrotron radiation in a highly ordered magnetic field can provide extremely high levels of polarization (with polarization fractions exceeding 50\%).  By contrast, the observed polarization fractions of these structures is observed to be significantly lower ($<10$\%), especially for the radio minihalos in the cores of relaxed clusters \citep{bonafede:2011}. This is interpreted as the effect of magnetic field tangling on a scale smaller than the resolution of the radio observations, i.e. beam depolarization.

The kpc-scale tangling of the magnetic field implied by the radio observations is intimately linked to turbulence in the ICM plasma.  Again, there are multiple lines of evidence suggesting that the ICM in even relaxed clusters is turbulent with typical velocity fluctuations $\delta v\sim 0.1-0.2c_s$, where $c_s$ is the sound speed.  The density fluctuations associated with turbulence directly translates into fluctuations of the X-ray surface brightness, and these have become a powerful tool for constraining ICM turbulence. For example, \citep{zhuravleva:2014} employ the particularly high-quality {\it Chandra X-ray Observatory} data for the Virgo and Perseus clusters, showing that both have a turbulent ICM with a Kolmogorov-like spectrum of fluctuations extending to scales well below 10kpc.  The most direct detection of turbulence in the ICM of a cluster came from measurements of the Doppler broadening of X-ray emission lines by the Soft X-ray Spectrometer (SXS) on the {\it Hitomi} observatory \citep{hitomi:2016}\footnote{The Perseus cluster was the first science target for the X-ray microcalorimeter on the newly launched Hitomi observatory in Feb-2016. Shortly after taking these data, the satellite was destroyed due to a command-and-control error, making Perseus the only galaxy cluster to be observed by an X-ray microcalorimeter at the present time.}. The (1-d) velocity dispersion across most of the core was found to be 100-150\kmps \citep{hitomi:2018}, to be compared with the sound speed of $c_s\approx 1000\kmps$ and in good agreement with the findings of the surface brightness fluctuation analysis. This suggests that the turbulent energy density is a few percent of the thermal energy density, and so very similar to the magnetic energy density.

From a fluid dynamics perspective, the existence of MHD turbulence in the ICM is entirely expected. While the microphysics of viscosity in ICM-like plasmas is still an area of active work \citep{kunz:2014}, it is clear that it is significantly suppressed below the n\"aive textbook value \citep{spitzer:1982}, making the ICM atmosphere a moderate-to-high Reynolds number (Re) system. Dynamics in the ICM atmosphere is driven by a variety of external processes including merging sub-clusters which induces ICM sloshing \citep{zuhone:2010}, the orbiting galaxies \citep{ruszkowski:2010}, and jet-activity from the central AGN that drives ICM shocks and inflates ICM cavities \citep{fabian:2012}. There are also purely internal processes within the ICM that can drive dynamics, including conduction-driven buoyancy instabilities \cite{balbus:2010} and small-scale kinetic dynamos \citep{schekochihin:2005}. Turbulence is a natural outcome of such driving in a high-Re atmosphere.

\subsection{Galaxy cluster magnetic field models in axion searches}
\label{sec:magneticmodels}
Having discussed the evidence that the ICM is a turbulent magnetized plasma, here we briefly review the specific magnetic field models that are used for axion-photon conversion in galaxy clusters. These range from simple cell-models to more sophisticated treatements of turbulent fields. We critically discuss a recent proposal to model galaxy cluster magnetic fields as regular \cite{LT} for the purpose of axion-photon conversion, and point out that this model is inconsistent with observations. 

\begin{itemize}
    \item {\bf General cell models.} The most commonly used class of magnetic field models for axion searches in galaxy clusters is the general cell model, which we discussed (and solved) in section \ref{sec:general_cell} for massive axions. An attractive feature of this class of models is its close link to the single-domain model (cf.~section \ref{sec:single_cell}), which is commonly used to build intuition around axion-photon mixing also in complex environments. In a general cell model, the sizes of the domains can be generated stochastically as in section \ref{sec:numcell_massive}, which allows mimicking some properties of smooth,
    `multi-scale', turbulent magnetic fields.
    Moreover, with a general cell-model, it is easy  (though not very fast) to numerically solve the Schrödinger-like equation.
    
    However, as an astrophysical model, cell models are extremely simplified, and do not satisfy some  of the most basic properties of magnetic fields, such as being  continuous and divergence free. Indeed, the non-vanishing divergence of cell models has been argued to provide a route to rule them out observationally through its impact on Faraday rotation measurements \cite{Ensslin:2003}. Furthermore, while qualitative agreement has been found between the predictions for axion-photon conversion in cell models and in more sophisticated, turbulent models \cite{Angus:2013sua}, there is no clear dictionary between these models that would allow an unambiguous translation of assumptions and constraints. See also \cite{Galanti:2018nvl, WangLai, Bu:2019qqg, Carenza:2019vzg} for a discussion of variants of cell models.

    \item {\bf Gaussian random field models.} The second class of magnetic field models commonly used for axion searches are the `GRF' models that we described in section  \ref{sec:numGRF_massive}. These are smooth and divergence free, and since
    the power spectrum is used when generating the vector potential,  it is easy to connect these models to other theoretical and observational studies of magnetic field structure. GRF models have been used numerically for axion searches in e.g.~\cite{Horns:2012kw,Angus:2013sua, Meyer:2014epa,  Fermi-LAT:2016nkz}. 
    
The magnetic fields generated from a GRF are rather featureless and lack large-scale coherent filamentary structures arising from intermittency and cooling instabilities which may relate to cold filaments that are often observed in galaxy cluster cores. Thus, the GRF models are still too simple to fully capture the complex structure of galaxy cluster magnetic fields.  Moreover, axion-photon conversion in GRF models has not been well-understood analytically (though our discussion in section \ref{sec:turbulent} addresses this issue for massive axions). Finally, GRF models are  slightly more involved to simulate than cell models.

      \item {\bf MHD models.}  Magnetohydrodynamic (MHD) dynamos transfer mechanical energy of the plasma into magnetic energy, and is believed to amplify small seed fields into  galaxy cluster magnetic fields of ${\cal O}(\mu{\rm G})$ strength. Cosmological MHD simulations indicate that such magnetic fields are tangled, can feature turbulence-induced non-Gaussian structures, and extend over several decades in Fourier space (cf.~\cite{Govoni:2004as} for a review). To date, there has been no study of axion-photon conversion in MHD magnetic fields. We note that our formalism will be very useful in developing an understanding of how axion-photon conversion in MHD magnetic fields differs from GRF models. In particular, coherent structures that are clearly evident in MHD simulations are completely erased when the phases of the (exponential) Fourier modes are randomised \cite{Maron:2001}. However, as we've discussed in this paper, axion-photon mixing is only sensitive to the magnetic field through the  autocorrelation function, which is independent of the phases. This suggests that 
     magnetic field models used for axion searches can be visibly different from more realistic magnetic fields, but still produce the same predictions.  

      \item {\bf The `regular model' of \cite{LT}. } Outflows from AGN at the centre of galaxy clusters provide an additional source of energy and complexity to the intracluster medium (ICM). In particular, jets extending from AGNs can inflate bubbles that rise through the ICM by buoyancy, and can reach large radii before `pancaking' or dispersing. Such bubbles, or cavities, are visible in high-resolution images   of galaxy clusters as regions with suppressed X-ray emission, indicating that the thermal ICM has been displaced from the cavity.  For example, in the  Perseus cluster, a number of such cavities have been identified  in various directions within 100 kpc from the central galaxy, NGC 1275 \cite{Fabian:1981, Branduardi-Raymont:1981, Boehringer:1993, Fabian:2000, Sanders:2005jx}. 
            In particular, reference \cite{Sanders:2005jx} used deep \emph{Chandra}  
     observations to identify a high-abundance ridge, or `shell', around 93 kpc from the centre of the cluster (coincident with the edge of the radio mini-halo), and interpreted this ridge as a relic of a collapsed radio bubble.  
     
     Recently, reference \cite{LT} appealed to the evidence for X-ray cavities to propose a galaxy cluster model with slowly varying, `regular' fields in  Perseus, and then used this model to argue that the uncertainties in gamma-ray and X-ray limits on axions had been  underestimated, possibly by several orders of magnitude.  Here, we critically re-examine the motivation and consistency of the astrophysical model proposed in \cite{LT} (see section \ref{sec:regular} for the analytic, perturbative solution of axion-photon mixing in  this magnetic field model).
     
     The central assumptions of {\color{dm} the model of }\cite{LT} are that:
     \begin{enumerate}
         \item The centre of the Perseus cluster consists of an X-ray cavity of radius 93 kpc.
         \item   The cavity can be modelled using the analytical solution of \cite{Gourgouliatos:2010nu} 
     (that we discussed in section \ref{sec:regular}), and the  electron  density is given by the deprojection analysis of \cite{Churazov:2003hr}.
     \end{enumerate}
     However, both these assumptions are problematic: 
     \begin{enumerate}
         \item The Perseus cluster is the brightest cluster in the X-ray sky. The azimuthally averaged surface brightness peaks towards the centre of the cluster \cite{Churazov:2003hr}, and so, the central 93 kpc of Perseus is certainly not an X-ray cavity. As mentioned above, observations by a number of groups have identified several localised cavity regions, including the high-abundance ridge at 93 kpc from the centre \cite{Sanders:2005jx}, but these do not encapsulate the cluster or even fall along the line of sight from earth to the centre of the cluster. Thus, the central assumption of \cite{LT} is in stark conflict with observations. 
         \item A key prediction of the analytical, axisymmetric solution of \cite{Gourgouliatos:2010nu}\footnote{Which strictly only applies to an \emph{unmagnetised} ambient environment beyond the cavity.} is that the gas pressure dips inside the cavity, and attains its peak value at the centre and at the boundary. This prediction is incompatible with the radially decreasing gas pressure found from \emph{Chandra} observations of Perseus in \cite{Schmidt:2002}. Moreover,  the cavity model of \cite{Gourgouliatos:2010nu} assumes that the electron density inside the cavity is smaller than the surrounding ICM, while \cite{LT} implicitly assumes that it is filled by the ICM when using the observational electron density from \cite{Churazov:2003hr}.

     \end{enumerate}
        We conclude that  the regular magnetic field model of \cite{LT} is incompatible with observations, which should be taken into account when examining its consequences for axion-photon mixing.
      
\end{itemize}
We note in closing that magnetic field modelling is currently the dominant systematic in studies of axion-photon mixing, and further theoretical, numerical and observational studies will be critical to further improving the sensitivity to axions.

\subsection{Outlook}
\label{sec:outlook}
In sections \ref{sec:massive}--\ref{sec:general}, we discussed the mathematical and conceptual consequences of the formalism that we have developed, and in section \ref{sec:tests}, we demonstrated how it allows for drastic improvements of numerical simulations of axion-photon mixing. In this short section, we briefly outline how our formalism suggests entirely new analysis methods for axion searches. 

The current paradigm for studying axion-photon mixing involves specifying the magnetic field and the plasma density, which are then used to numerically solve the modified Klein-Gordon and Maxwell equations to find the conversion probability. In many astrophysical applications, this process must be repeated many times as one scans over the allowed magnetic field configurations. 
The key message of this paper is that the perturbative axion-photon conversion probability is given by the magnetic power spectrum (in some cases rescaled by $1/\omega_{\rm pl}^2$), or equivalently, as a Fourier-type transform of the corresponding  magnetic autocorrelation function. The autocorrelation function contains less information than the magnetic field, and visibly distinct magnetic fields can share the same autocorrelation function. This suggests that scanning over magnetic fields is redundant, and that more efficient methods could be developed based directly on the autocorrelation function. Here we consider three possibilities along these lines. 

First, when defining an ensemble of magnetic fields (e.g.~using one of the classes discussed in section \ref{sec:magneticmodels}), one implicitly defines an ensemble of magnetic autocorrelation functions.  This suggests that instead of generating magnetic field profiles along the axion/photon trajectory, one may directly generate the magnetic autocorrelation function. In simple cases, the distribution of autocorrelation functions may be known analytically. In more complex cases, machine learning techniques based on Gaussian processes are likely to be very suitable for learning the distribution. An autocorrelation function generated in real space need only to be Fourier transformed to yield the conversion probability. In some cases, it may be possible to generate the Fourier coefficients of the autocorrelation function directly, which simply corresponds to writing down the conversion probability, without numerically solving the equations of motion. 

Second, whenever high quality radio observations of extended radio halos are available, it may be possible to dispense of explicit magnetic field modelling all together,  and move directly from radio observations to predictions for the axion-photon conversion probability at high energies. In the case of Gaussian, isotropic magnetic turbulence, maps of RMs of an extended radio background source can be used to reconstruct the statistics  of the magnetic autocorrelation tensor, $M_{ij}= \langle \delta {\bf B}_i({\bf x}) \, \delta {\bf B}_j({\bf x + r}) \rangle$, and \cite{vogt:2003}:
\beq
M_{ij} = M_N(r) \delta_{ij} + \left(M_L(r) - M_N(r) \right) \frac{r_i r_j}{r^2} + M_H(r) \epsilon_{ijk} r_k \, ,
\eeq
Here $M_L$, $M_N$ and $M_H$ respectively denote  the longitudinal, normal and helical components of the tensor. We note however that  $M_H$  is irrelevant for both  RMs or axion-photon mixing. The remaining components, $M_N$ and $M_L$, can be reconstructed from the spatial distribution of the observed RMs, and determine the statistics of axion-photon conversion along any trajectory through the Faraday screen. In this context,  the conversion probabilities could be obtained as Fourier transforms of autocorrelation functions generated from the observationally inferred tensor $M_{ij}$. 

Third, photons `disappearing' into axions lead to distortions of astrophysical spectra. These distortions will appear as oscillatory residuals in fits that don't take axions into account. Recently, reference \cite{Conlon:2018iwn} suggested that it may be beneficial to search for these imprints by expressing them as functions of their wavelength, and then performing a Fourier transform to hopefully isolate the oscillatory features. Our formalism suggests that this procedure is a map from the observational residuals to an approximation of the magnetic autocorrelation function (up to corrections due to noise and the fitting procedure). For unpolarised fluxes, the relevant conversion probability is the superposition of the polarised conversion probabilities, cf.~equation \eqref{eq:Punpol}, and the relevant correlation function will be the sum of two components of the autocorrelation tensor. This method opens up the possibility of using mathematical, physical and observational properties of the autocorrelation function to improve the sensitivity of axion searches.      

\section{Conclusions}\label{sec:conclusions}
We have considered the mixing of relativistic axions and photons to leading order in the coupling, $\g$, and asked what properties of the  magnetic field are reflected in the observationally interesting conversion probability, $P_{\gamma \to a}$. Our conclusion is that $P_{\gamma \to a}$ is given by the magnetic power spectrum, or equivalently, by the Fourier transform of the magnetic autocorrelation function. This result has notable conceptual, calculational, numerical and methodological consequences. 

\emph{Conceptually}, our formalism maps questions about the oscillatory spectrum of the conversion probability into questions about real-space magnetic correlations. Moreover, the mature framework of Fourier analysis and  autocorrelation theory lead  to new, interesting identities for axion-photon conversion.

\emph{Calculationally}, we have demonstrated in a series of examples that large classes of magnetic fields models, that have previously only been studied numerically, can be solved analytically. This is particularly striking in the case of  axions more massive than the plasma frequency (which is of direct relevance for gamma-ray searches for axions), where one of our examples comprises the general solution to axion-photon mixing in any magnetic field.

\emph{Numerically}, the current paradigm for computing axion photon mixing is to  simulate the equations of motion for each magnetic field realisation. Our formalism allows one to circumvent  numerical simulation, and  the techniques of the Fast Fourier Transform can be leveraged to efficiently  determine the predictions of each model.

\emph{Methodologically}, our formalism suggests new approaches for generating the predictions of the model, tightening the connection with radio observations of rotation measures, and for analysing observational data that may carry spectral imprints from axions.

Moreover, in the light of our new formalism, we have critically reviewed the types of magnetic field models used in axion searches in the literature. We have argued that the commonly used cell models qualitatively agree well with more sophisticated, turbulent GRF models, though with certain differences that can be explained by the zigzaggy  nature of the cell-model autocorrelation function. We have also pointed out some inconsistencies in a recently proposed regular model for the magnetic field in the Perseus cluster. 

The focus of this paper is the interconversion of photons and relativistic axions, e.g.~as those that may be produced from high-energy astrophysical fluxes. Our formalism extends straightforwardly to non-relativistic axions, such as axion dark matter, in the case of a constant plasma density. The general problem of interconversion of dark matter axions and photons is more subtle, and not discussed in this paper. Similarly, we do not discuss the opposite regime of extremely energetic particles in strong magnetic fields, where the QED birefringence contribution cannot be neglected. 

Our results draw on the quantum mechanical analogy for axion-photon mixing developed in \cite{Raffelt:1987im}, and in the context of quantum mechanical perturbation theory, it is well-known that the leading-order, infinite-time  transition  amplitude is given by a Fourier transform of the interaction Hamiltonian. To the best of our understanding, the equivalent Wiener–Khintchine form in which the transition probability is given by the Fourier transform of the autocorrelation of the interaction Hamiltoinian, is less explored in the literature.

Finally, our approach is based on perturbation theory and will be most useful when searching for spectral irregularities that are sufficiently small. The precise breaking point of perturbation theory depends on the magnetic field model, but our analysis indicate that our approach could consistently be used to search for  axions with couplings of about an order of magnitude larger than the current observational limit, using 
currently available precision X-ray data. 
Thus, our first-order formalism can already be used for large portions of the parameter space considered in axion searches. Moreover, our method is even more relevant to future studies with the {\sl Athena} mission \cite{nandra_athena_2013} and the gamma-ray \emph{CTA} observatory \cite{CTAConsortium:2017dvg}.

\section*{Acknowledgements}
DM is supported by the European Research Council under Grant No. 742104 and by the Swedish Research Council (VR) under grants  2018-03641 and 2019-02337. CSR thanks the UK Science and Technology Facilities Council (STFC) for support under the Consolidated Grant ST/S000623/1, as well as the European Research Council for support under the European Union’s Horizon 2020 research and innovation programme (grant 834203). JHM acknowledges a Herchel Smith Fellowship at Cambridge. The work of PC is partially supported by the Italian Istituto Nazionale di Fisica Nucleare (INFN) through the ``Theoretical Astroparticle Physics'' project and by the research grant number 2017W4HA7S ``NAT-NET: Neutrino and Astroparticle Theory Network'' under the program PRIN 2017 funded by the Italian Ministero dell'Universit\`a e della Ricerca (MUR).

\bibliographystyle{apsrev4-1}
\bibliography{Xrayspec}

\end{document}